\documentclass[%
 aip,
 amsmath,amssymb,
 reprint,%
]{revtex4-1}

\usepackage{graphicx}
\usepackage{dcolumn}
\usepackage{bm}

\usepackage[utf8]{inputenc}
\usepackage[T1]{fontenc}
\usepackage{mathptmx}
\usepackage[colorlinks=true,citecolor=blue,linkcolor=blue,urlcolor=blue]{hyperref}

\begin{document}

\preprint{AIP/123-QED}

\title[HighFieldmKSTM]{Millikelvin scanning tunneling microscope at 20/22 T with a graphite enabled stick-slip approach and an energy resolution below 8$\mu$eV: Application to conductance quantization at 20 T in single atom point contacts of Al and Au and to the charge density wave of 2H-NbSe$_2$}


\author{Marta Fern\'andez Lomana}
\author{Beilun Wu}
\author{Francisco Mart\'in-Vega}
\author{Raquel S\'anchez Barquilla}
\author{Rafael \'Alvarez Montoya}
\affiliation{
Laboratorio de Bajas Temperaturas y Altos Campos Magn\'eticos, Unidad Asociada (UAM/CSIC), Departamento de F\'isica de la Materia Condensada, Instituto Nicol\'as Cabrera and Condensed Matter Physics Center (IFIMAC), Universidad Aut\'onoma de Madrid, E-28049 Madrid,
Spain
}
\author{Jos\'e Mar\'ia Castilla}
\affiliation{
Departamento de F\'isica de la Materia Condensada, Universidad Aut\'onoma de Madrid, E-28049 Madrid,
Spain
}
\author{Jos\'e Navarrete}
\author{Juan Ram\'on Marijuan}
\affiliation{ 
SEGAINVEX, Universidad Aut\'onoma de Madrid, E-28049 Madrid,
Spain
}
\author{Edwin Herrera}
\author{Hermann Suderow}
\author{Isabel Guillam\'on}
 \email{isabel.guillamon@uam.es.}
\affiliation{ 
Laboratorio de Bajas Temperaturas y Altos Campos Magn\'eticos, Unidad Asociada (UAM/CSIC), Departamento de F\'isica de la Materia Condensada, Instituto Nicol\'as Cabrera and Condensed Matter Physics Center (IFIMAC), Universidad Aut\'onoma de Madrid, E-28049 Madrid,
Spain
}

\date{\today}

\begin{abstract}
We describe a scanning tunneling microscope (STM) that operates at magnetic fields up to 22 Tesla and temperatures down to 80 mK. We discuss the design of the STM head, with an improved coarse approach, the vibration isolation system, and efforts to improve the energy resolution using compact filters for multiple lines. We measure the superconducting gap and Josephson effect in Aluminum and show that we can resolve features in the density of states as small as 8\,$\mu$eV. We measure the quantization of conductance in atomic size contacts and make atomic resolution and density of states images in the layered material 2H-NbSe$_2$. The latter experiments are performed by continuously operating the STM at magnetic fields of 20 T in periods of several days without interruption.
\end{abstract}

\maketitle

\section{Introduction}

The scanning tunneling microscope (STM) is one of the most useful and informative experimental tools to probe and manipulate surfaces. Since its invention in 1986 by Binnig and Rohrer\cite{Binnig1982}, many different STM designs have been successfully realized. A sharp metallic tip is approached to a flat sample, until a tunneling current $I(V)$ flows under the applied bias voltage $V$. The tip is then scanned over the surface. One of the most interesting aspects of STM is that the tunneling conductance $dI/dV$ is proportional to the local electronic density of states $N_s(E)$ at the energy $E=eV$, smeared by temperature\cite{Voigtlander,wiesendanger_1994}. Mapping $N_s(E)$ as a function of the position addresses directly many topical problems in the physics of superconductors and quantum materials. For example, a superconducting or a charge density wave (CDW) gap is readily observed in $N_s(E)$. $N_s(E;x,y)$ maps provide then neat images of the CDW pattern on the surface\cite{PhysRevB.37.2741,Hamidian2016}. After applying a magnetic field, the superconducting vortex lattice can be observed and studied\cite{PhysRevLett.62.214,RevModPhys.79.353,Suderow2014}. Very strong magnetic fields lead to Landau quantization with sharp peaks in $N_s(E)$\cite{PhysRevLett.94.226403,Jeon2014}. Furthermore, high magnetic fields allow destroying superconductivity and study the density of states of the normal phase at low temperatures, benefiting from the large energy resolution required to identify low energy features in $N_s(E)$. Other high magnetic field phenomena where $N_s(E)$ is key in their understanding include re-entrant superconductivity in ferromagnets, Ising superconductivity or quantum phase transitions\cite{doi:10.7566/JPSJ.88.022001,Lu1353,Saito2016,Paschen2021}.

Resolving fine features in $N_s(E)$ requires reducing temperature smearing by cooling the STM and applying a well defined bias voltage $V$. State of the art STMs are operated in dilution refrigerators at mK temperatures and within a considerably filtered environment to obtain an energy resolution often of the order of several tens of $\mu$eV. Values of order or below that range are needed to study various tunneling phenomena, such as the Josephson effect in superconductors\cite{Ast2016,Rodrigo2004a,PhysRevLett.87.097004}.

Recent advances in the design and construction of STM at high magnetic fields include the notable achievement of STM imaging at magnetic fields as high as 34 T using a resistive coil, with the considerable difficulty of the vibrations induced by water cooling \cite{Tao2017}. These experiments have been made, however, at liquid Helium temperatures and only on graphite (4.2 K). Other reports in dilution refrigerator temperatures include STM designs using superconducting coils and obtaining magnetic fields up to 15 T \cite{Song2010,Misra2013,Singh2013,Assig2013,Roychowdhury2014} and in one case up to 17.5 T\cite{Machida2018}. Here we describe a dilution refrigerator STM that is operated at magnetic fields up to 22 T in a superconducting coil. We provide a detailed description of the laboratory and the STM head, describing solutions to improve vibration isolation on the support of the set-up, to obtain a more compact STM head with an improved stick-slip approach and to increase energy resolution with new filters. We discuss results obtained with our microscope, including a thorough characterization of the energy resolution, vortex lattice imaging and atomic scale studies requiring un-interrupted operation of the STM at 20 T during several days.

\begin{figure*}[htbp]
		\centering
		\begin{center}
			\includegraphics[width = 2\columnwidth]{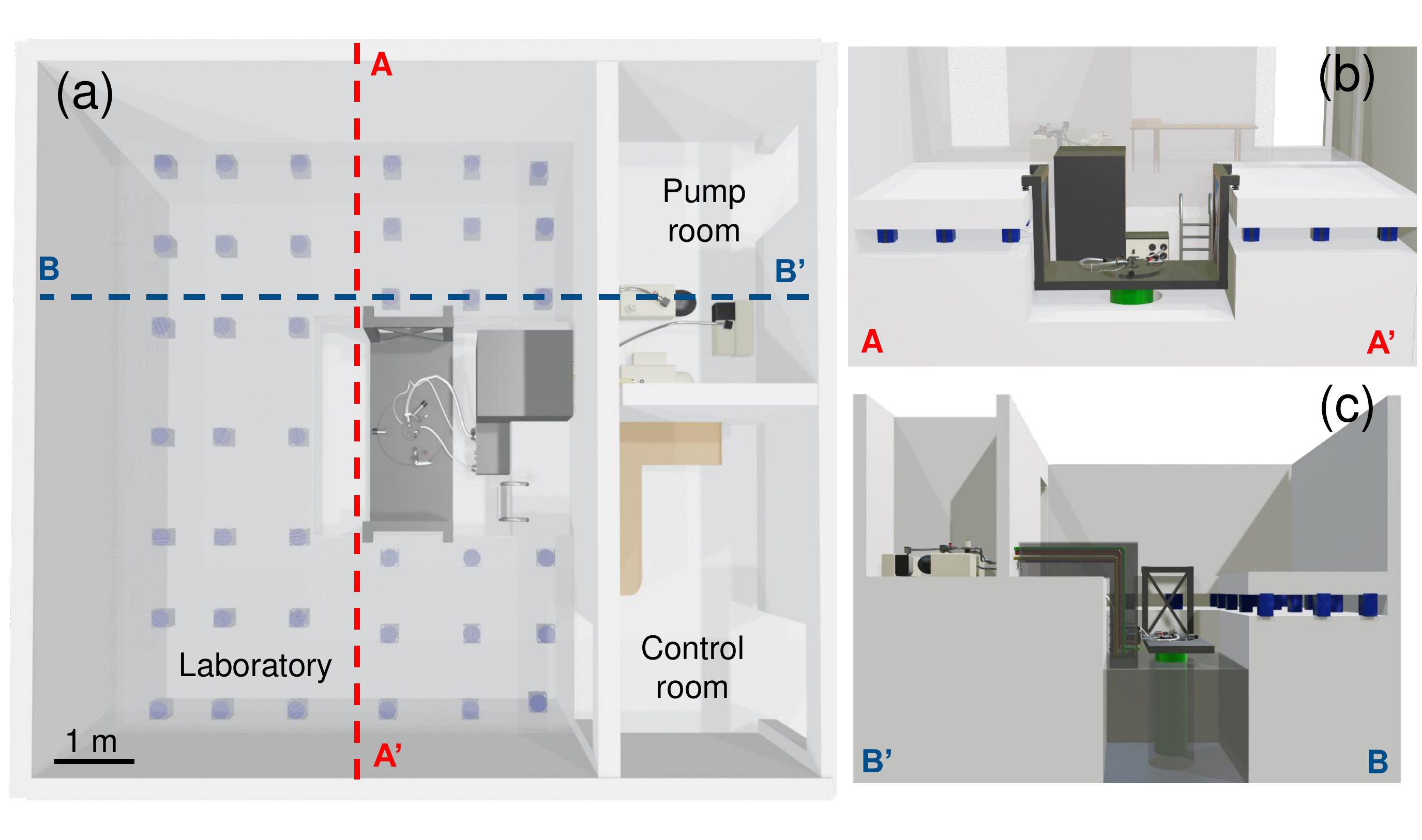}
		\end{center}
		\caption{(a) Schematic of the set-up. The cryostat with the high field magnet is located within the room called laboratory. The control room serves to operate the STM and the pump room holds the pumps and other ancillary equipment. The light blue squares are springs on which the floor of the laboratory is suspended. The cut A-A' is shown in (b). In dark grey color we show a steel structure which holds the support for the cryostat (cylinder in green) on the floating floor. The springs of the floating floor are shown in blue. The cut B-B' in (a) is shown in (c). The tubing from the pumps to the cryostat are located in an enclosure filled with sand which is firmly fixed to the main building. The connection to the cryostat with the STM is through flexible tubes.}\label{FigureLab}
		\end{figure*}

\section{Laboratory arrangement and vibration isolation}

We start by discussing the overall construction of the laboratory, schematically shown in Fig.\,\ref{FigureLab}. The STM holds together tip and sample within a compact STM head. The STM head can eventually vibrate as a whole. However, the distance between tip and sample $\delta z$ should be stable. Thus, the design of the STM head, which determines the connection between tip and sample is the most important element. If we consider the STM head as a damped oscillator with a resonance frequency $\omega_0$, we see that the STM itself damps vibrations for $\omega<\omega_0$, with damping going as $\left(\frac{\omega}{\omega_0}\right)^2$ for low frequencies\cite{ParkStroscio,Voigtlander}. We can take for $\omega_0$ the resonance frequency $\omega_P$ of a piezo tube, which is needed to scan the tip over the sample and for which $\omega_P\approx 10^4$ Hz. We can then consider a noise spectrum with floor vibrations which remains flat up to $\approx 10^2$ Hz and then decreases. We see that floor vibrations are damped by at least $10^{-4}$ simply by building a tip-sample system which is more rigid than the piezo tube. Typical vibration levels of a floor with pumps and other laboratory equipment are, in the worst case, of the order of a few tens of nm, in a bandwidth from DC up to a few $10^2$ Hz. This leads to tip-sample distance variations $\delta z$ at the hundreds of fm level. The role of the vibration damping system is to decrease this level at least by two orders of magnitude, below a fm.

If we take a simple STM head posed on a simple vibration isolation system consisting of springs, it is not very difficult to achieve the required damping\cite{ParkStroscio,Voigtlander}. However, cryogenic equipment required to reach mK temperatures strongly connects vibration sources such as pumps with the STM head. Furthermore, the STM is suspended on the bottom of a long cryostat, holding a heavy weight from a large magnetic field coil, which eventually amplifies floor vibrations at certain frequencies. Thus, vibration isolation must be carefully considered in high magnetic field cryogenic environments\cite{doi:10.1063/1.5111989}.

\begin{figure}[htbp]
		\centering
		\begin{center}
			\includegraphics[width = 1\columnwidth]{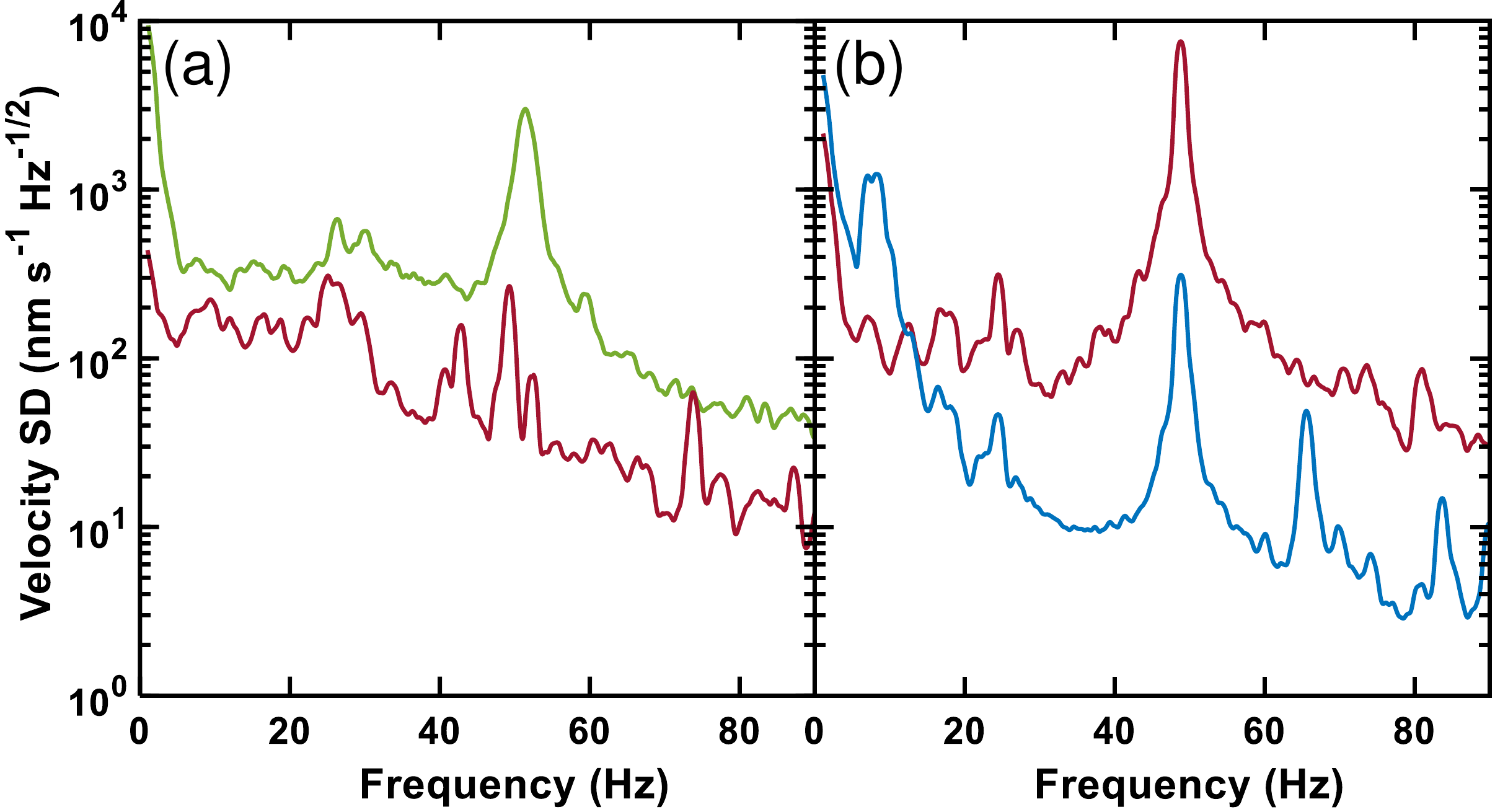}
		\end{center}
		\caption{We show the vibration levels as the velocity vs frequency measured using a geophone, as explained in the text. The actual velocity is obtained by multiplying with the square root of the data recording time, which is 10\,s. (a) We show as a green line the level of vibrations in a usual faculty building. As a red line we show the vibrations obtained in a separate building, where we located the laboratory. (b) We show as a red line the level of vibrations within the building with a running pump, where we removed the usual vibration isolation rubber support. The blue line is the level of vibrations measured on the floating floor.}\label{Geophones}
		\end{figure}

To reduce the vibrations on the cryogenic equipment, we located the STM set-up within a building separate from usual faculty activity. To characterize vibrations we used SM-24 geophones from SENSOR Netherlands. The geophone consists of a magnetic mass suspended on a spring inside a coil. Vibrations cause motion of the mass and induce a voltage which is proportional to the velocity of the suspended mass\cite{hoffmanphd}. To measure the voltage we use two EGG 5113 preamplifiers and a data acquisition system\cite{WinSPM}. We measure vibrations up to $\approx$ 100 Hz. The geophone is a damped harmonic oscillator with a resonance frequency of 20 Hz. We thus correct the measured voltage with the transfer function, which is provided by the manufacturer.

We see in  Fig.\,\ref{Geophones}(a) a comparison between measurements taken on a building with usual faculty activity to the separate building where we have installed the STM. We see a decrease by approximately an order of magnitude in the level of vibrations. Assuming harmonic vibrations, we can write for the amplitude of vibrations $x_0$ as a function of the amplitude of the velocity $v_0$, $x_0=\frac{v_0}{\omega}$, at frequency $\omega$ (for low frequencies as compared to $\omega_0$, see Ref.\,\onlinecite{Voigtlander}). From this, we find that the amplitude of vibrations in the new location is of order of a few nm at a few Hz.

To further isolate the equipment, we use a large concrete floor separated from the building by a set of springs. The floor and the arrangement of 39 springs is shown in Fig.\,\ref{FigureLab}(a). The springs have a spring constant of about $1 \times 10^{3}$ $\frac{N}{m}$ which gives a resonance frequency of the order of a Hz (the concrete floating floor weighs 29 ton). To characterize the effect of this system, we removed the usual rubber vibration isolation from a pump and positioned it on the floor of the building close to the floating floor. The vibrations induced on the floor of the building are shown by the red line in Fig.\,\ref{Geophones}(b). At the floating floor, we observe a reduction of one order of magnitude (blue line in Fig.\,\ref{Geophones}(b)) above 10 Hz. There are a few resonances, at about 65 Hz and 85 Hz and at low frequencies, due to the vibration modes of the concrete floor.

The location of the cryostat is described in Fig.\,\ref{FigureLab}(b). Mounting the STM microscope and changing the sample requires removing the dilution unit from the cryostat and the magnet. This requires a large height. We thus located the magnet and the cryostat well below the floating floor inside a pit. This also brings the magnet very far from the working floor, so that even at the highest applied magnetic fields, the magnetic field on the floor remains at small values. Furthermore, the magnet is not affected by the iron of the springs that hold the floating floor or by the structural steel of the building. To position the cryostat inside the pit we constructed a stainless steel support (grey in Fig.\,\ref{FigureLab}(b,c)). The stainless steel support is located on damped vibration isolators with a resonance frequency of 12 Hz to eliminate internal vibration modes of the floating floor.

The pumps, magnet power supply and cables are connected to the stainless steel support through flexible tubes. Whenever possible (in the He recovery and pumping lines), we used flexible silicone tubes, instead of stainless steel bellows. As shown in Ref.\onlinecite{doi:10.1063/1.4905531}, this reduces the transmission of vibrations. All tubes are located inside a large sand box which is posed on the main building floor. Pumps are located in a separate room and the system is operated from another separate room (Fig.\,\ref{FigureLab}(a)). The door between the laboratory and the control room is acoustically insulated. The walls of the experimental room are also acoustically insulated using fiberglass. In addition, we manufactured a 8 mm thick sound absorbent curtain with which we can cover the pit. This turns out to be a simple and useful method, as it is an effective damper for very low frequency sound transmitted through the air.

\begin{figure*}[htbp]
		\centering
		\begin{center}
			\includegraphics[width = 1.8\columnwidth]{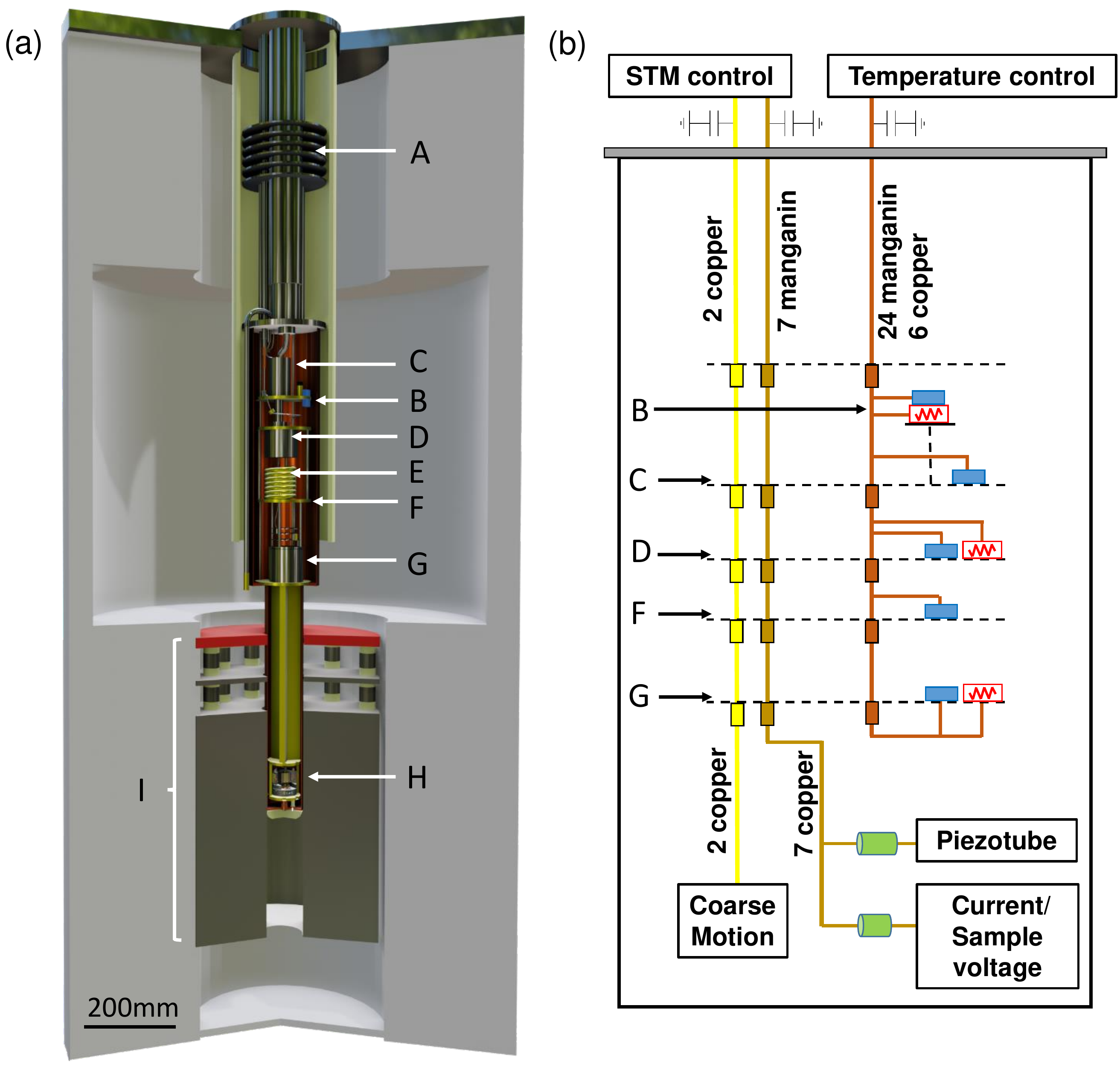}
		\end{center}
		\caption{Schematic of the dilution refrigerator cryogenic system (a) and of the wiring (b). A: radiation shields, B: carbon sorption pump, C: 1K pot, D: still, E: heat exchanger, F: cold plate, G: mixing chamber, H: STM, I: magnet. For the wiring (b), we insert in the dilution unit a set of wires to control the dilution unit (brown) and the STM (yellow and ochre). We use separated stainless steel capillaries all over the length. The temperature control is made using 24 manganin wires thermalized at the different stages and filtered at the upper part of the cryostat using feedthrough capacitors (thermometers are shown by blue rectangles). Heaters (red rectangles with a resistor inside) are operated with copper wires. The STM unit needs two copper wires for the coarse approach motor and seven manganin wires to operate the piezo tube (five), measure the current and apply the voltage (two). The latter two sets of wires are inserted into separated stainless steel capillaries and filtered using the high frequency filters described in Fig.\,\ref{FigFilter} (green cylinders). \label{FigDilFridge}}
		\end{figure*}

\section{Dilution unit and cryogenic STM support}

The STM experiment is mounted in a commercial Kelvinox\textsuperscript{\textregistered}400HA dilution refrigerator provided by Oxford Instruments\textsuperscript{\textregistered}. Fig.\,\ref{FigDilFridge}(a,b) show a scheme of the dilution system and wiring. The STM requires nine wires, among which five are used to send voltage to the piezo tube to control the lateral and vertical position of the tip during scanning, two are used to apply the bias voltage to the sample and to carry the tunneling current, and two are used to operate the coarse vertical movement of the tip. Wires are inserted into stainless steel capillaries and thermalized at each temperature stage. Thermalization is made by wrapping wires tightly on copper cylinder heat sinks. The temperature control is made using a thermometer and a heater located slightly below the mixing chamber. From that point on to the STM, only copper wires are used to enhance thermal conductance between the thermometer and the STM. The dilution refrigerator reaches a base temperature below 50 mK. 

An additional stainless steel wire, used to move the sample holder, is thermalized at each stage. We use the method described in Ref.\onlinecite{Suderow2011}. The movable sample holder allows to prepare the tip by indentation on a pad of the same material as the tip, following Ref.\,\onlinecite{Rodrigo_2004}. At the same time, it is used to cleave the sample in-situ in cryogenic vacuum and obtain a fresh and clean surface\cite{Herrera2021}.

\begin{figure}[htbp]
		\centering
		\begin{center}
			\includegraphics[width = 1\columnwidth]{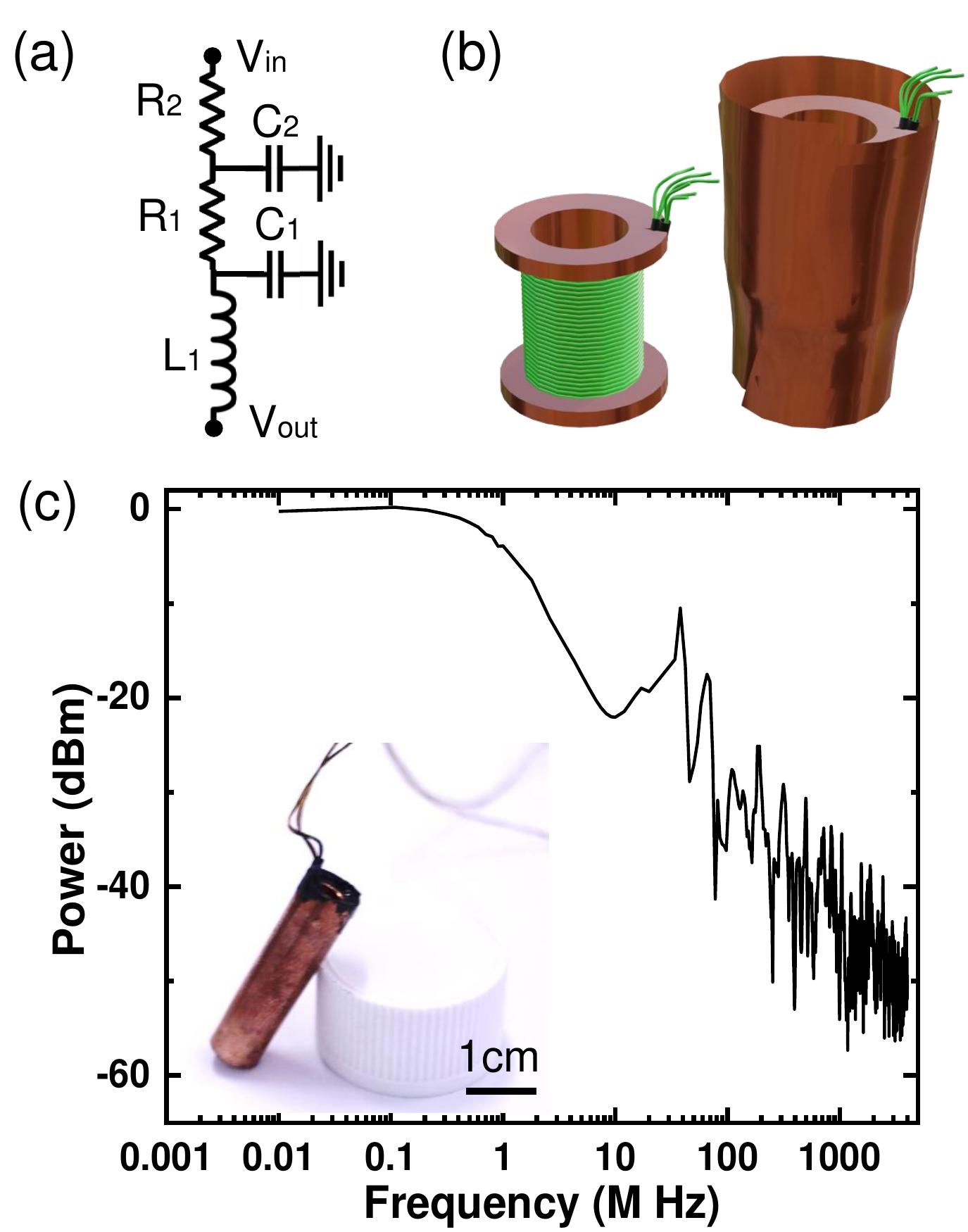}
	\end{center}
		\caption{(a) Scheme of the filtering. We first filter using resistor $R_2$ and a feedthrough capacitor $C_2$. The long wiring builds up a second filter, made of the resistance of the wires $R_1$ and their distributed capacitance to ground $C_1$. In addition, wires are wound on a coil, which produces an inductance $L_1$. In (b) we show a schematic of the filtering coil used at the mixing chamber with a measured inductance of $L_1 = 16$ $\mu$H. The resistance of the wire in the coil is $R_1 = 170$ $\Omega$. We wind a manganin wire twisted to a copper wire that is grounded. This gives a distributed capacitance of about $1$ nF to ground. In (b) we show the wire wound on a copper support which serves for thermalization. To build the filter, we have encapsulated the coil inside eccosorb\textsuperscript{\textregistered} of epotek \cite{eccosorb} and surrounded the whole with a copper sheet. In (c) we show the response of the whole system to an AC signal inserted on the top of the cryostat. Similar curves are obtained for all lines. We see that there is an effective shielding of high frequency radiation. In the inset we show a photograph of the finished cryogenic filter.}\label{FigFilter}    		
\end{figure}

A superconducting magnet provided by Oxford Instruments\textsuperscript{\textregistered}
is used to apply magnetic fields up to 20 T in normal helium cooling conditions, and up to 22 T when the magnet is further cooled down to 2.2 K by pumping the $\lambda$-fridge\cite{Magnet}.

\section{High frequency filtering and electromagnetic environment}

Tunneling phenomena in superconductors are highly sensitive to pair breaking by high frequency radiation. To avoid deleterious effects of high frequency radiation on the measurement of tunneling conductance, one can use low pass filters. These should remain efficient up to very high frequencies. It has been previously shown in several experiments that low pass filters are needed to measure the sharp features that characterize the tunneling conductance of superconductors\cite{GUILLAMON2008537,Rodrigo_2004,Assig2013,Machida2018,Schwenk2020,Roychowdhury2014}. Often, low pass feedthrough capacitors or $\pi$ filters are used at the entries to the cryostat\cite{Machida2018}. These should be complemented by cryogenic filters\cite{Lukashenko2008}. Cryogenic high frequency filters mimic a network of RC filters by using a capacitance that is distributed over long wires. In addition, the inductance is increased to produce a voltage divider for high frequencies. To this end, several reported filters use a conducting wire which is surrounded by magnetic metal powder mixed with epoxy\cite{Bladh2003,Milliken2007,Lukashenko2008,Thalmann2017}.

In our experiment, we use room temperature feedthrough capacitors at the input to the cryostat. Wires are located such that high voltage signal cables are well separated from the rest of the set-up. The capillaries are wound at several stages to ensure thermalization.

At the final stage, close to the STM, we use a new cylinder-shaped cryogenic filter, shown in Fig.\,\ref{FigFilter}(b). To build the cryogenic filter, we first twist 1 m-long 0.1 mm wires together with a 0.1 mm bare (un-insulated) copper wire. We then wind the twisted wires around a copper cylinder which has a central hole to act as a heat sink. We cover the lateral surface of the cylinder with silver epoxy and then cure the whole at 65 \textdegree{}C. Through the silver epoxy and thanks to the bare copper wire, we make a distributed ground plane that is very close to the wires carrying the signal. The ends of the wires come out through a notch at the edge of the cylinder (green wires in Fig.\,\ref{FigFilter}(b)). We then cover the whole block with a copper sheath and pour magnetically loaded silicone rubber Eccosorb CFS-8480 inside the copper sheath, while it is hot (65 \textdegree{}C). We take care to do this process slowly enough to avoid the formation of any bubbles. Finally, we let the whole block cure in the furnace for 12 h at 70 \textdegree{}C. 

The advantage of this design is that we can include many wires in a compact filter. As long as there is no risk of cross-talk (which can be anyhow minimized by twisting pairs of wires together), multiple wires can be efficiently filtered using a small amount of space. We measured the transfer function of the whole set-up using a Picoscope\textsuperscript{\textregistered} oscilloscope for frequencies up to 20 MHz and a Windfreak\textsuperscript{\textregistered} SynthNV v3.0 for frequencies between 35 MHz and 4.5 GHz. The result is shown in Fig.\,\ref{FigFilter}(c). The attenuation we achieve is similar to other low-pass metal-powder filters\cite{Bladh2003,Milliken2007,Lukashenko2008}. The most important aspect is however to obtain clean tunneling conductance in superconductors. This is discussed below.

\section{Design of the STM head}

 \begin{figure*}[htbp]
		\centering
		\begin{center}
			\includegraphics[width = 0.8\textwidth]{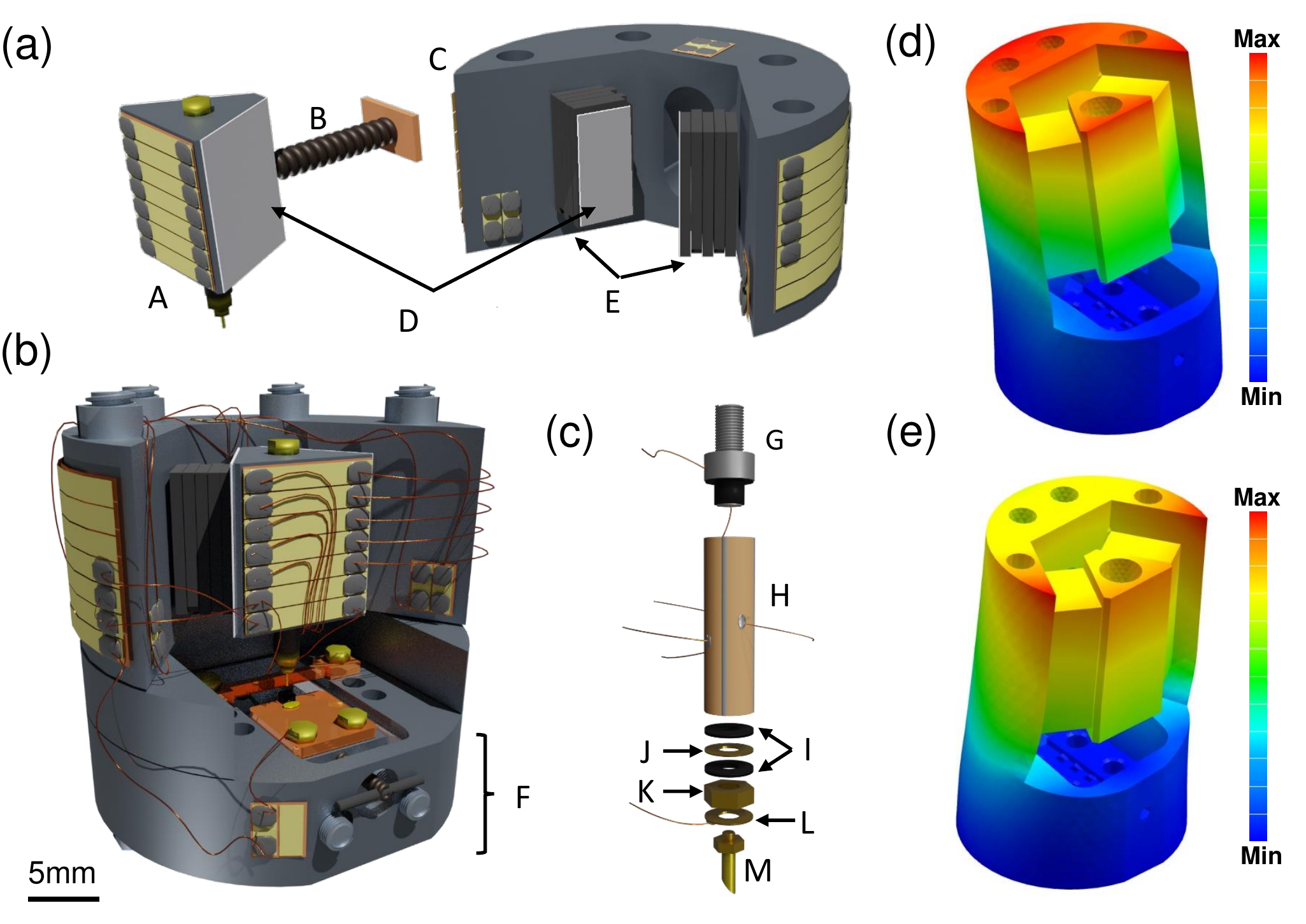}
		\end{center}
		\caption{In (a) we show the mobile part of the STM (A), a spring (B) that holds it to the support (C). (D) are alumina sheets, polished at one end. It is quite important to cover the alumina sheets with graphite, rubbing a pencil on the alumina sheets (D). The piezostacks are shown as (E). The assembled system is shown in (b), where we also highlight (F) the piece where we have the sample holder. A detailed view of the piezo tube is given in (c). The piezo tube (H) is glued to electrical insulation layers (I), a grounded copper washer (J), the tip holder (M1 nut, K) and a washer which we use to electrically connect the tip (L). In (d) and (e) we show the deformations obtained using finite element analysis of our microscope, for the first two resonance frequencies, located at 12.4 kHz and 14.3 kHz respectively.}\label{FigureSTM}
		\end{figure*}

In Fig.\,\ref{FigureSTM}(a-c) we show several relevant aspects of the STM head. The piezo tube is mounted in a vertical configuration and approaches the sample using a piezo driven coarse approach along the z-axis. The system works by moving a prism (A in Fig.\,\ref{FigureSTM}(a)) which holds the piezo tube and the tip with respect to a large body which holds shear piezos that drive the coarse approach. The prism is firmly fixed using a spring (B in Fig.\,\ref{FigureSTM}(a)) to the large body (C  in Fig.\,\ref{FigureSTM}(a)). There are alumina plates (D in Fig.\,\ref{FigureSTM}(a)) glued on the prism and on the shear piezos (E in Fig.\,\ref{FigureSTM}(a)). For the piezos, we use five 5 mm $\times$ 10 mm long shear piezo plates glued together in such a way as to obtain a shear along their long axis when applying a voltage to the surface of the plates. The prism A and the large body C are joined together through polished alumina plates (D in Fig.\,\ref{FigureSTM}(a)). The alumina plates are carefully covered with graphite. The graphite is applied using a pencil and covers the whole surface of the polished alumina plates.

The coarse motion is driven by a saw tooth signal sent to the piezos which is polarized in both directions, i.e. each piezo plate is polarized between +V and -V. This produces consecutive stick-slip situations between the prism and the large body. The operation of stick-slip motors is usually explained by the difference between static and dynamic friction between two bodies\cite{Voigtlander,act6010007}. When the motion is slow (slow part of the saw tooth signal), the two bodies stick together due to the frictional force. The prism A then moves with respect to the large body C, following the deformation of the shear piezo stacks. When the motion is fast (fast ramp in the sawtooth signal), the two surfaces are slipping with respect to each other. The surface of the piezo stacks returns to the initial position, but the prism A remains at the same place. While the slow (stick) motion is easy to understand, as it just relies on the frictional force between the top parts, the rapid (slip) motion is much more involved\cite{roch2021velocitydriven}. The detachment of both parts during slip implies fracturing the joints that are established between two surfaces when these contact each other.

Controlling the slip motion is usually notoriously difficult. We have tested stick-slip between metals, alumina not covered with graphite, two pieces of glass and a piece of glass and a metal, but the best results are obtained with polished alumina covered with graphite. To obtain best performance, we make sure that the two alumina surfaces (D in Fig.\,\ref{FigureSTM}(a)) are parallel to each other. To this end we glue the alumina to their support on the same set-up (Fig.\,\ref{FigureSTM}(a)). Using polished alumina helps reducing the surface roughness to such a level that the graphite applied to the alumina fills all voids and allows for a smooth contact between the two surfaces. The contact between the two surfaces is then made through graphite flakes.

\begin{figure}[htbp]
		\centering
		\begin{center}
			\includegraphics[width = 1\columnwidth]{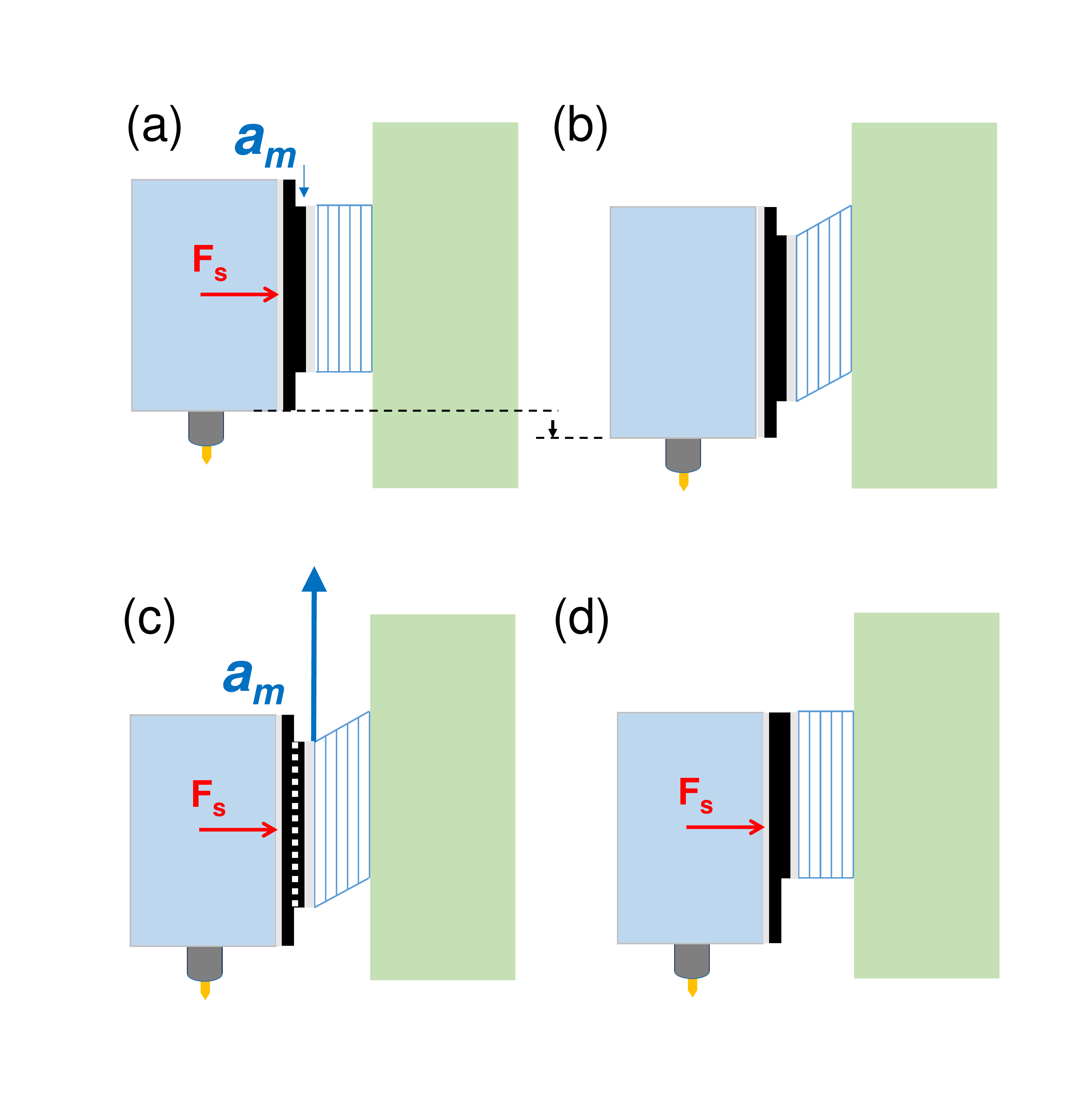}
		\end{center}
		\caption{Scheme of the steps for stick-slip motion. The shear piezoelements are shown by blue rectangles. The alumina plates are shown as grey rectangles. The graphite is shown (with an exaggerated thickness) in black. The mobile part is shown by the light blue rectangle and the fixed part by the green rectangle. In (a) we show the slow motion. The piezos move with a small acceleration $a_m$. The force $F_s$ (applied by the spring which is not shown in this figure) is holding the two parts together. (b) After the piezos have reached the maximum elongation, the mobile has advanced by a certain distance (black arrow). (c) The piezos are then retracted, but with such a large acceleration $a_m$ that the bonds of the graphite layer separating both parts break (white dashed line). (d) Thus, during piezo retraction, the mobile part (blue) stays at the same position. The cycle can the start again.}\label{FigureScheme}
		\end{figure}

During slip, the two alumina pieces move at a relative speed of about 0.15 m/s (Fig.\,\ref{FigureScheme}(c)), whereas motion during stick is made at a speed five orders of magnitude below that value (Fig.\,\ref{FigureScheme}(a)). The estimated acceleration used in the slip regime is of order of $a_M\approx 10^3 \frac{m}{s^2}$. Taking the mass of the prism, we find a force $F_m$ of several N associated to this acceleration. This force is used to fracture the bonds between graphite flakes (white dotted line in Fig.\,\ref{FigureScheme}(c)). The interlayer shear strength of graphite is of the order of 100 MPa\cite{Liu2012}. This suggests that the slip occurs by sliding over each other flakes of sizes of the order of a hundred $\mu$m. This also corresponds well to the sizes of flakes we observe on the alumina surface using optical microscopy. During slip, the prism suffers gravitation. However, the path covered by the prism, assuming free fall, is of the order of an \AA\,during slip (at most 5$\mu$s). In all, and thanks to slip, one obtains a net relative motion between both parts (Fig.\,\ref{FigureScheme}(a,d)) and the cycle can be repeated.

The acceleration between the two alumina plates (D in Fig.\,\ref{FigureSTM}(a)) plays a prominent role. With a larger acceleration, one easily increases the shear above the shear modulus of graphite. Slow motion, however, might reduce the shear force below the shear modulus of the graphite flakes and slip might not occur in that case. 

On the other hand, the spring applies the force $F_s$ that holds the body (C in Fig.\,\ref{FigureSTM}(a)) and the prism (A in Fig.\,\ref{FigureSTM}(a)) together ($F_s= kz$, with  $k=0.0521 N/mm$ and $z\approx 1$ mm). This force should be as large as possible, but small enough so that $\eta F_s$ (with $\eta\approx$ 0.1, the friction coefficient of graphite) does not exceed the force obtained when accelerating the two surfaces with respect to each other, $\eta F_s<F_m$.

Thus, the velocity of the fast motion that leads to slip determines the operation of the whole setup.

As the capacitance of shear piezos is usually large (of the order of a nF), the electrical response of the RC circuit build up by the piezos and the wiring used to operate them, provides the fastest timescale for motion of the piezos. We usually use low resistance copper wires to operate the coarse approach, obtaining time constants somewhat below 10$^{-7}$ s. This is also within the bandwidth given by our filtering system described above.

It is worth considering lateral motion by the spring B in Fig.\,\ref{FigureSTM}(a). Lateral motion in stick-slip motors has been considered in Ref.\,\onlinecite{Li_2015}. Here we note that the fast motion of the shear piezos during slip is on a time scale which is of the order of magnitude of the first lateral resonance frequency of the spring. The spring being very small (CuBe, manufactured by H\"aberli AG\cite{Haberli}, with wire diameter of 0.15 mm and outer diameter of 1.6 mm), the first lateral resonance frequency is of about 10 kHz, estimated using Ref.\,\onlinecite{Michalczyk2019}. The corresponding oscillation corresponds to bending the spring along its length. Exciting this resonance thus leads to lateral vibrations, which almost certainly reduce sticking in graphite. Lateral vibrations are indeed known to favor detachment of flakes in layered materials\cite{D1MA00118C}. This explains our observation that the coarse approach also moves with very small voltage amplitudes, for which the acceleration is so small that we obtain $F_m$ such that $F_m \approx F_s$.

The coarse approach system is firmly attached to the bottom part of the microscope (F in Fig.\,\ref{FigureSTM}(b)). There, we mount the sample on a slider which can be operated from room temperature as described in Ref.\onlinecite{Suderow2011}. 

We have built STM heads in grade 3 and grade 5 Titanium as well as in Shapal, obtaining similar performances. We have calculated using finite elements the resonance frequencies of the STM head. We obtain 12.4\,kHz and 14.6\,kHz for the smallest resonance frequencies. This is of order of the resonance frequency of the piezo tube, which is given above and is of about 10 kHz.  In  Fig.\,\ref{FigureSTM}(d,e) we show the displacement observed in a finite element calculation at the first and second resonances. The weakest part of the design, according to the finite element calculation, is the connection between the part holding the coarse approach and the bottom part  (C and F in Fig.\,\ref{FigureSTM}(a,b)). Increasing the diameter of the STM could lead to an improvement. However, the resonance frequency of the piezo tube remains of the same order, so that other factors influencing the piezo tube, such as the weight on the tip holder, are more relevant than the vibrations of the STM head.

We use tubes fabricated by EBL\cite{EBLTubes}, with 13 mm length, 3.2 mm diameter and 0.3 mm wall thickness. These piezo tubes are small and allow a relatively large scanning range because of the reduced wall thickness (about 2$\mu$m at low temperatures). However, they can easily get depolarized when applying a voltage of several hundred V. The depoling electric field of the piezo-tube is of 9.1 kV/cm at cryogenic temperatures, which is enough to apply without problems voltages up to 500 V. The depoling field decreases strongly at room temperature, so the piezo tubes have to be used with care at ambient conditions. In Fig.\,\ref{FigureSTM}(c) we show schematically the system we use to mount the piezo tube and the tip. We use a small Ti piece where we glue the piezo tube using epoxy (Stycast) (G in Fig.\,\ref{FigureSTM}(c)). Electrical contacts are made using fine copper wires of 0.05 mm to reduce forces acting on the tube. The tip is glued on a M1 screw (M in Fig.\,\ref{FigureSTM}(c)), which is then screwed firmly into a M1 nut (K in Fig.\,\ref{FigureSTM}(c)). We use a thin copper washer (L in Fig.\,\ref{FigureSTM}(c)) to make the electrical connection to the tip. We also position another washer (J in Fig.\,\ref{FigureSTM}(c)) which we connect to ground to screen the field from the high voltage signals coming from the piezo tube. The STM tip itself is made with a short fine wire, cut at room temperature to have a pointed end, and improved at low temperature in-situ by repeated indentation as described in Ref.\,\onlinecite{Rodrigo_2004}. This procedure allows to remove any trace of oxide that may be in the tip and the sample, giving the clean and reproducible spectroscopic data shown here. In total, the tip and its support system weights less than 40\,mg, so that the resonance frequency of the piezo tube is not much modified by the tip.

Thus, our STM head has a compact design and features a piezo tube with tip mounting system with a particularly large resonance frequency and a coarse approach system that is simple to build and is based on graphite enhanced stick-slip motion.

There are numerous other approaches for cryogenic coarse motors\cite{Pan1999,doi:10.1063/1.3694972,doi:10.1063/1.1150218,act6010007,doi:10.1063/1.4878624,doi:10.1063/1.3681444,doi:10.1063/1.5083994}. Some are commercial and include compact designs. All approaches require at some point two parts that slip with respect to each other. The uncertainties associated with slip (essentially due to surface termination issues, as we discuss above), are minimized by using, for example, piezos mounted on multiple legs. While some legs stick to the moving part, one or several other legs slip. Here we have preferred to address carefully the uncertainties related to slip. We show that the use of graphite covering large and flat surfaces, mounted carefully to ensure that these are parallel to each other, leads to successful results. The main advantage of our approach is that, during usual operation, the surfaces are firmly held to each other over a large area, making sure that the STM head is stable and compact. Furthermore, our design allows for a considerable simplification of the coarse approach.

Let us note that our coarse approach is inspired by the well known Pan-design\cite{Pan1999}. We use a very similar positioning of tip and sample. The Pan design has no spring pulling on the movable prism, but an additional plate with piezos pushing the movable prism from above\cite{Pan1999}. We have also used this design in the past, but we moved to the design described here because it allows us to have enough access to the sample holder by freeing the front end of the STM head. This is needed to make a sample arrangement on the sample holder that allows successful cleaving at very low temperatures and in-situ tip preparation, as described in Refs.\onlinecite{Suderow2011,Rodrigo_2004,Herrera2021,PhysRevB.97.134501,PhysRevB.97.014505}.

\begin{figure*}[htbp]
		\centering
		\begin{center}
			\includegraphics[width = 2\columnwidth]{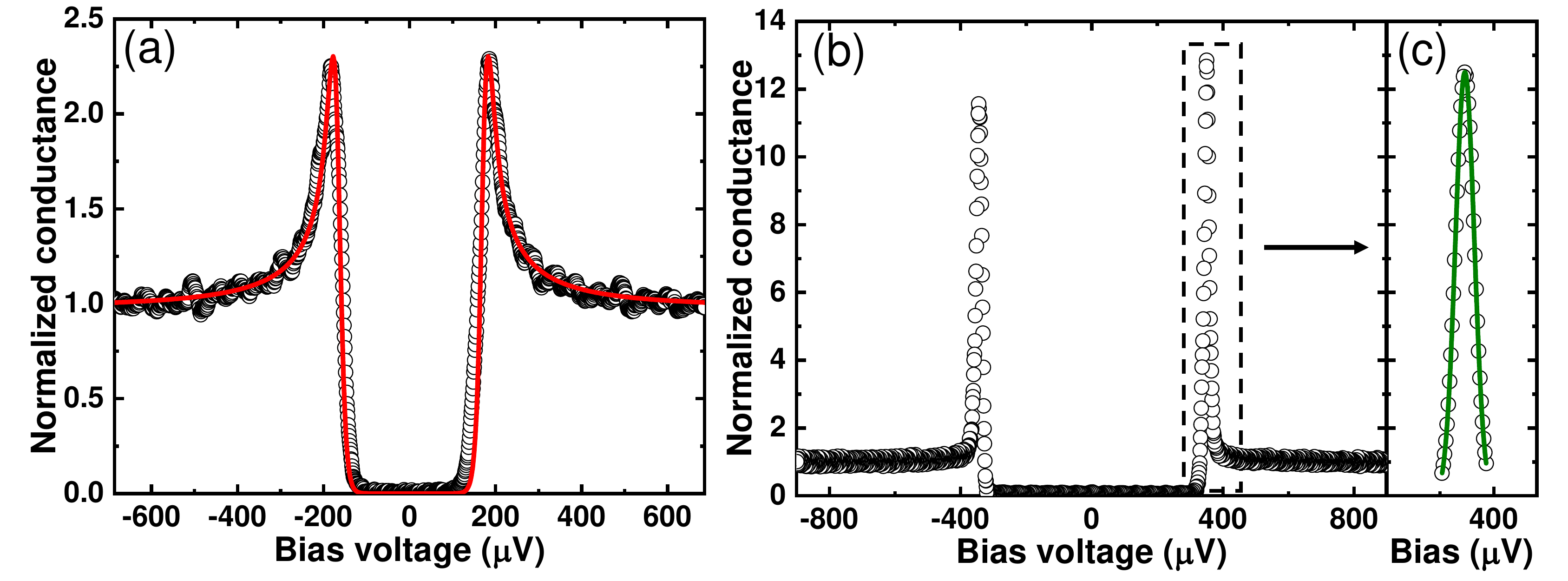}
		\end{center}
		\caption{(a) We show as points the normalized tunneling conductance vs bias voltage measured with an Al tip on a sample of Au. Red line is a fit using the BCS theory with $\Delta$ = 0.17 mV. From this fit, we obtain a temperature of 84 mK. (b) Normalized tunneling conductance curve vs bias voltage measured with an Al tip on a sample of Al. We zoom into the bias voltage range of the quasiparticle peaks in (c). Green line is a Gaussian peak, whose width at half height provides a measure of the energy resolution of our set-up, giving $\delta$V $\sim$ 8 $\mu$V.}\label{FigGap}
		\end{figure*}

\section{Results}

We now present results operating our STM head at high magnetic fields and very low temperatures. We first discuss the capability of making tunneling conductance studies at very low temperatures by making measurements of the superconducting gap in a simple superconductor, Al, and observing the voltage dependent features of the vortex lattice in 2H-NbSe$_2$. We then discuss results at high magnetic fields. We have made single atom point contacts of Al and Au at high magnetic fields, observing the conductance quantization in long term experiments made at a fixed magnetic field. We have build histograms out of more than $10^5$ conductance vs distance curves, each with 2048 points on a distance covering
a few nm distance. Furthermore, we have measured the atomic surface lattice of 2H-NbSe$_2$ and its charge density wave. The two latter experiments required long measurement periods at low temperatures and magnetic fields of 20 T.

\subsection{Resolution of $\mu$eV in the density of states from tunneling conductance vs bias voltage measurements}

\begin{figure*}[htbp]
		\centering
		\begin{center}
			\includegraphics[width = 2\columnwidth]{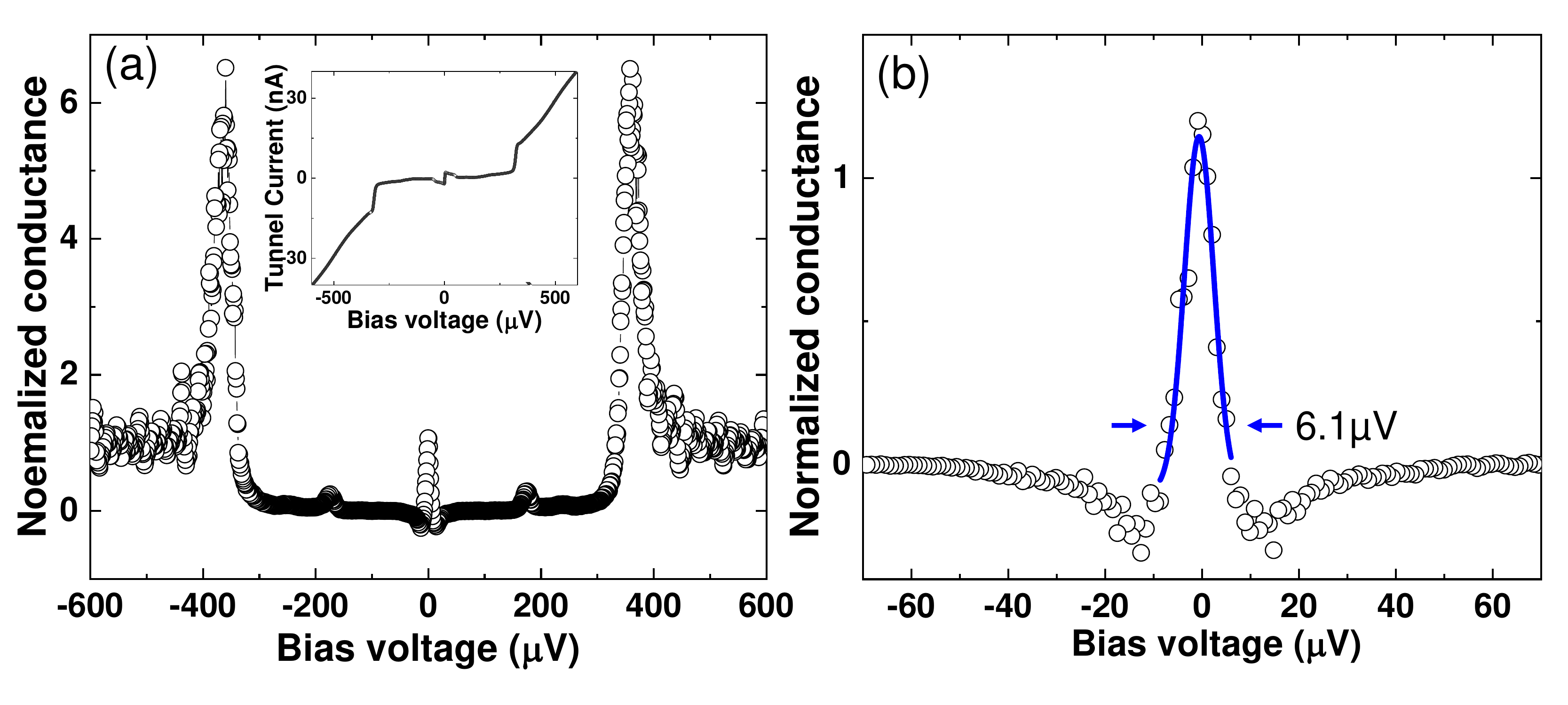}
		\end{center}
		\caption{(a) Normalized tunneling conductance (points, line is a guide to the eye) measured with an Al tip on a sample of Al (tunneling resistance of 250 k$\Omega$). In addition to the quasiparticle peaks (located at $\pm2\Delta$), we identify two peaks $\pm\Delta$ due to multiple Andreev reflections, and a pronounced peak at zero bias due to the Josephson effect. In the inset we show the tunneling current vs bias voltage. (b) Zoom around zero bias of (a). The  blue line is a Gaussian fit which provides an estimation of the energy resolution of our system of order of $\delta$V $\sim$ 6 $\mu$V.}\label{FigJosephson}
		\end{figure*}

We can write for the tunneling conductance of a STM as a function of the bias voltage

\begin{equation}
dI/dV \propto \int_{-\infty}^{+\infty} -\frac{\partial f(E+eV)}{\partial E}\,N_{s}(E)\,dE 
\label{equtunnel}\end{equation}

where $E$ is the energy, $N_{s}(E)$ is the density of states of the sample and $f(E)$ the Fermi function, provided that we neglect the energy dependence of the tunneling matrix elements and the density of states of the tip\cite{Suderow2014,RevModPhys.79.353,Voigtlander}. If we measure a superconductor with a Bardeen-Cooper-Schrieffer (BCS) density of states, $N_{s,tip}(E)\propto\frac{|E|}{\sqrt{E^2-\Delta ^2}}$, we see that the divergence at the superconducting gap $\Delta$ is smeared by the derivative of the Fermi function. Thus, we can use $dI/dV(V)$ to measure the temperature of the junction, provided that the superconductor is a single gap system with no magnetic impurities nor other sources of influence on the BCS $N(E)$\cite{GUILLAMON2008537,PhysRevLett.41.1509,PhysRevB.94.144508}.

The latter conditions are very difficult to obtain in a realistic superconductor. See for example the case of Pb, which shows two values of the superconducting gap\cite{PhysRevLett.114.157001}. We can take Al as the system which probably best approaches a single gap superconductor following BCS theory, although we note that even in Al, there are small but visible variations in the temperature dependence of the superconducting gap with respect to BCS theory\cite{doi:10.1139/p68-021}. We show in Fig.\,\ref{FigGap}(a) the curve obtained using a tip of Al and a sample of Au as points and a fit to Eq.\,\ref{equtunnel} using a BCS density of states. We take the superconducting gap as a fitting parameter, obtaining $\Delta=0.17$ mV and a temperature of 84 mK. The latter value is among the lowest obtained with STM so far\cite{Assig2013,Misra2013,Battisti2018,doi:10.1063/1.2400024,Moussy2001,GUILLAMON2008537,Rodrigo_2004,Schwenk2020}.

We have then analyzed the situation where both tip and sample are of Al. In that case, the integral of the current contains two $N_{s,tip}(E)$ which diverge at the same energy $E$, at $E=\Delta$ ($I(V)\propto\int dE(f(E-eV)-f(E))N_t(E-eV)N_s(E)$). Then, the Fermi function does not smear the divergence of the density of states and the $I(V)$ curves shows a sharp upturn at $\pm 2\Delta$\cite{doi:10.1139/p68-021,GUILLAMON2008537,Rodrigo_2004}. Accordingly, the tunneling conductance shows two peaks $\pm 2\Delta$. The height and width of these peaks can be due to multiple gaps, inelastic scattering or other uncertainties that influence $\Delta$\cite{PhysRevLett.41.1509,Rodrigo2004a,Rodrigo_2004}. In Fig.\,\ref{FigGap}(b,c) we show as points the tunneling conductance measured between a superconducting Al tip and a sample of Al, at our base temperature and at zero magnetic field. We fit the peak at $\Delta$ with a Gaussian distribution from which we obtain $\delta V \sim 8.7$ $\mu$V. This value is again among the lowest obtained with a STM measuring a superconductor-superconductor tunnel junction. In particular, it is considerably smaller than most of the values obtained using evaporated junctions between aluminum electrodes with an insulator in between\cite{doi:10.1139/p68-021,Court_2007}. The superconducting gap and critical temperature of Al increases considerably in evaporated samples. Thus, the obtained gap values $\Delta$ are up to two times the value we find here and show a scatter of several tens of $\mu$eV, which we do not observe in our experiment\cite{doi:10.1139/p68-021,Court_2007}.

We can also measure at a reduced tunneling conductance, approaching tip and sample\cite{GUILLAMON2008537,Rodrigo_2004}. In that case, multiple tunneling paths are visible due to Andreev reflections at the junction\cite{PhysRevB.54.7366,Agrait2003,PhysRevB.74.132501,Suderow_2000,PhysRevLett.78.3535}. Andreev reflection also smears the peaks in the tunneling conductance at $\pm 2\Delta$. We show in Fig.\,\ref{FigJosephson}(a) results at a tunneling conductance of about 5\% of the quantum of conductance ($G_0=2e^2/h=7.748 \cdot 10^{-5}\Omega^{-1}$). We see the peaks at $\pm 2\Delta$ and peaks at $\pm \Delta$ due to multiple Andreev reflections. There is an additional peak at zero bias. A close up view of the tunneling current vs bias voltage (inset of Fig.\,\ref{FigJosephson}(a)) shows that the zero bias peak in the tunneling conductance is due to the Josephson current between tip and sample. This has been observed previously in STM experiments using Pb or other superconductors as a tip\cite{Rodrigo_2004,PhysRevLett.87.097004,Ast2016,Huang2020,Senkpiel2020,PhysRevB.101.134507,PhysRevLett.119.147702,liu2021discovery,Hamidian2016,PhysRevB.97.174510,Cho2019}, and using Al in Ref.\,\cite{Rodrigo_2004}. The detailed features of the Josephson current are analyzed in those works. Here we analyze the width of the zero bias peak in the tunneling conductance due to the Josephson current. As shown in Ref.\cite{Rodrigo_2004,PhysRevLett.87.097004,Ast2016,Huang2020,Senkpiel2020,PhysRevB.101.134507,PhysRevLett.119.147702,liu2021discovery,Hamidian2016,PhysRevB.97.174510,Cho2019}, this depends on a number of parameters, the Josephson coupling energy, resistance and capacitance of the junction, etc. Nonetheless, as shown in Ref.\,\onlinecite{Schwenk2020}, it is also a good indicator of the performance of the experiment. We find here very similar curves as in previous work and obtained a peak smeared by $\delta V \sim 6$ $\mu$V, which is of the same order as the value for $\delta V$ provided above.

Thus, the capability of our wiring and improved filtering system to resolve fine features in the tunneling conductance is among the best reported so far.

\subsection{Vortex lattice observation}

\begin{figure*}[htbp]
		\centering
		\begin{center}
			\includegraphics[width = 1\textwidth]{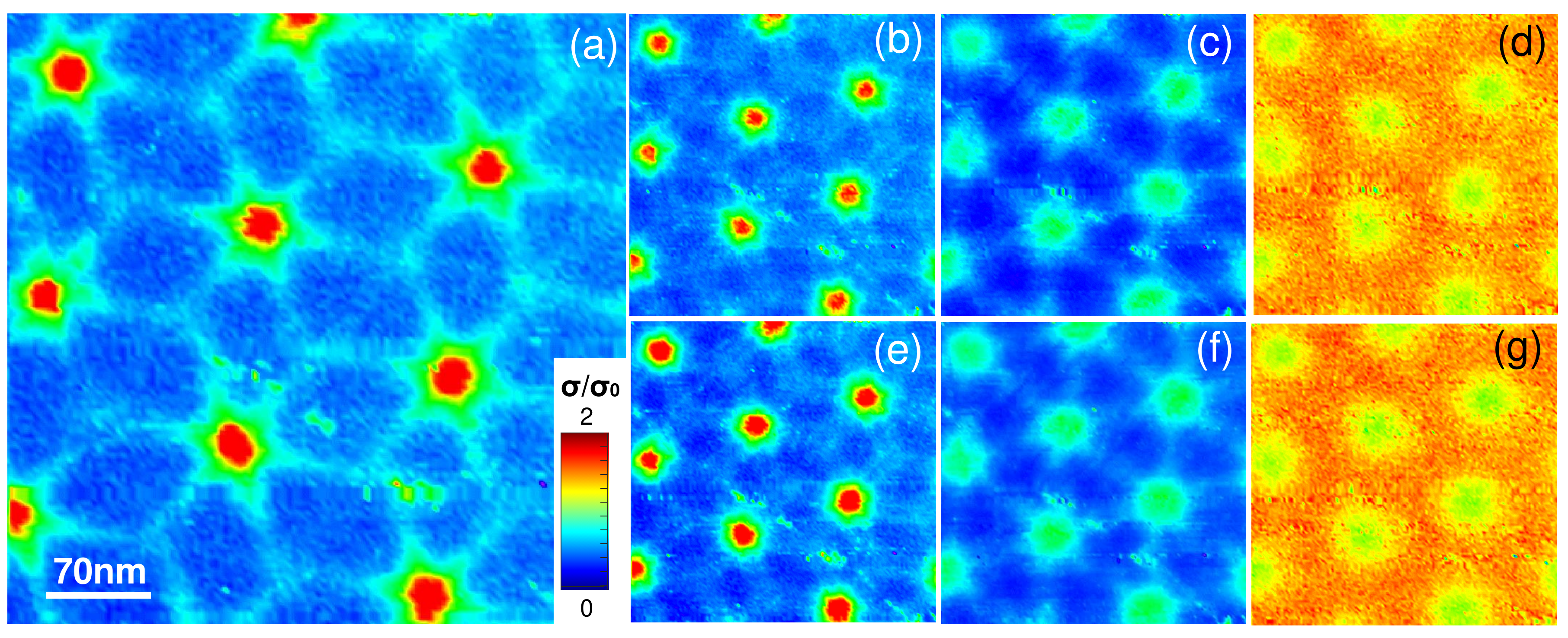}
		\end{center}
		\caption{Vortex lattice imaging in 2H-NbSe$_2$. Normalized tunneling conductance $\sigma/\sigma_0$ as a function of the position for different values of the bias voltage, 0\,mV (a), -0.3\,mV (b), -1\,mV (c), -1.5\,mV (d), 0.3\,mV (e), 1\,mV (f), 1.5\,mV (g). Magnetic field is of 0.5\,T. Maps follow the color scale given by the bar at the right in (a). The tunneling conductance $\sigma$ is normalized to its value $\sigma_0$ above 5 mV, well above the superconducting gap. All images are on the same location and the scale bar is shown in (a).
}\label{FigVortices}
		\end{figure*}
		
The observation of the superconducting vortex lattice was a hallmark in STM studies, allowing to visualize vortices in real space and at high magnetic fields\cite{Suderow2014}. This was achieved soon after the invention of the STM in the layered superconductor 2H-NbSe$_2$\cite{PhysRevLett.62.214,Hess1990}. This experiment is now considered as an excellent benchmark to show the capacities of a STM system. To visualize vortices, we measure the tunneling conductance as a function of the bias voltage at each position within a field of view. We normalize the tunneling conductance for voltages smaller than the superconducting gap to its value measured well above the gap. The features first observed in Refs.\,\onlinecite{PhysRevLett.62.214,Hess1990} can be summarized as follows. The tunneling conductance at the center of a vortex core shows a pronounced peak at zero bias. The peak splits as a function of the bias voltage when leaving the vortex core. The full gap is recovered in the space between vortices, at a length scale of order of the superconducting coherence length\cite{Hayashi1998,PhysRevB.94.014517}. Consequently, the normalized tunneling conductance as a function of the position shows a peak at the center of the vortex cores at zero bias (red points in Fig.\,\ref{FigVortices}(a)). The peak spreads with a sixfold symmetry around each vortex, as shown by the blue rays around red points in Fig.\,\ref{FigVortices}(a). Notice that these rays are not oriented towards neighboring vortices. This can be attributed to the superconducting gap anisotropy of 2H-NbSe$_2$\cite{Guillamon2008c}. Each one of the sixfold rays splits into two when increasing the bias voltage (Fig.\,\ref{FigVortices}(b,e)). At a higher bias voltage (Fig.\,\ref{FigVortices}(c,f)), the rays are now oriented towards neighboring vortices. Finally, when reaching the quasiparticle peaks (Fig.\,\ref{FigVortices}(d,g)), vortices become rounded. The observation of these features shows that the STM is stable during long experiments, and capable of taking reproducible measurements of the tunneling conductance as a function of the position over large fields of view.
	 	
\subsection{Single atom point contacts of Au and Al at high magnetic fields}

Another benchmark for stability is the reproducible creation and monitoring of single atom quantum point contacts\cite{Agrait2003}. To this end, we stay at one fixed position, and use tip and sample of the same material (here we have studied Au and Al). We first cut the feedback loop and repeatedly and strongly indent the sample to remove oxides and impurities\cite{Rodrigo_2004,GUILLAMON2008537,Suderow_2000,Agrait1993,Sheer1997,PhysRevLett.76.2302,Cuevas1998}. We then program long term conductance vs distance experiments in which we vary the distance between tip and sample by 1 nm at a fixed bias voltage of 100 mV. We use a current to voltage converter which allows us to measure tunneling as well as the contact regime (down to about 100$\Omega$). We record about 10$^4$ curves, each one with 2048 points. We do this repeatedly at each magnetic field, so as to gather about 10$^5$ at each field.

\begin{figure*}[htbp]
		\centering
		\begin{center}
			\includegraphics[width = 0.8\textwidth]{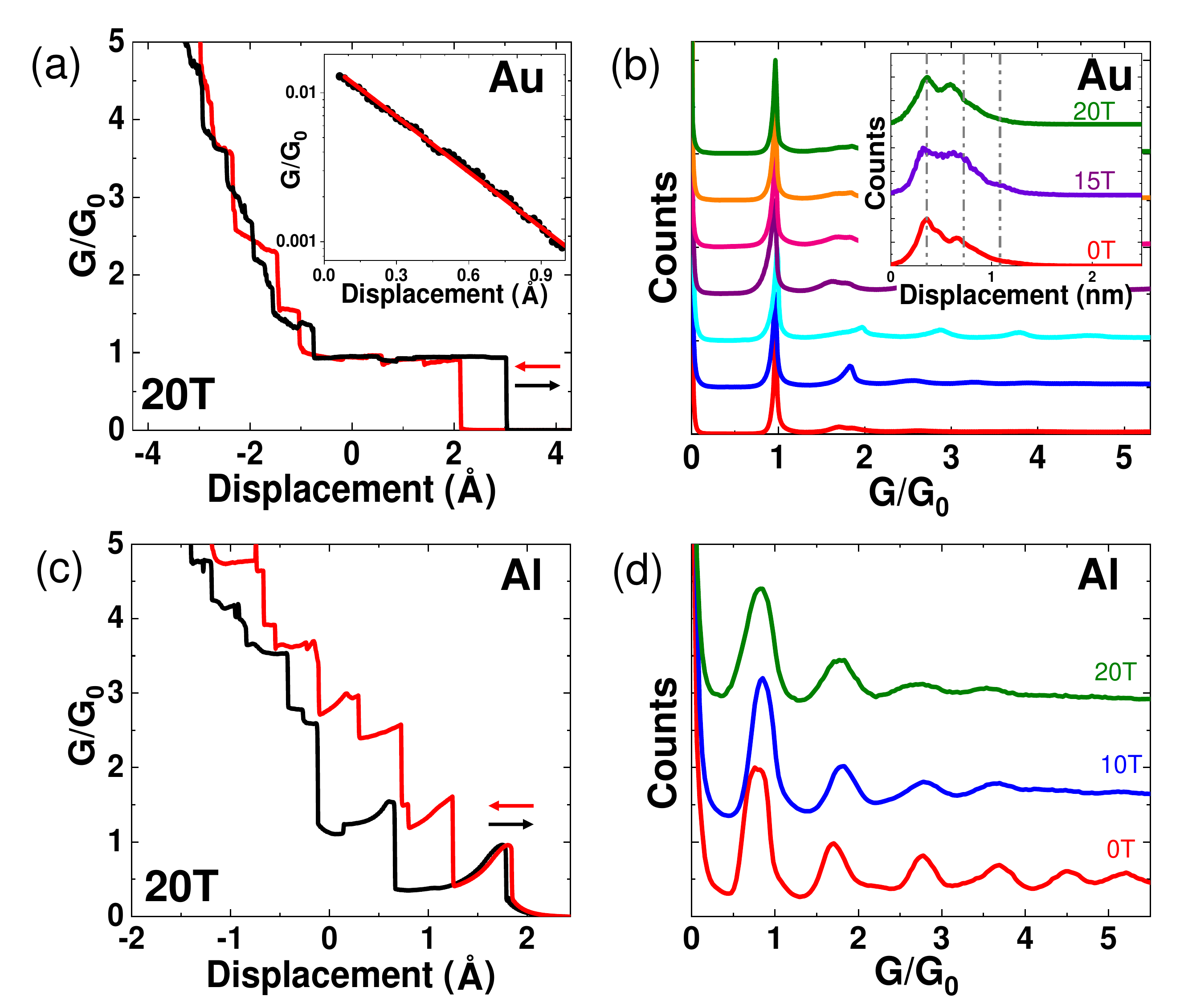}
		\end{center}
		\caption{(a) Conductance vs distance curves with 2048 points taken at a magnetic field of 20\,T are shown by black and red lines.  The conductance is normalized to $G_0$ ($G_0=\frac{2e^2}{h}$). Arrows indicate the direction of the tip, approaching (red) and separating (black). Bias voltage is of 100 mV. In the inset we show the conductance vs distance in the tunnelling regime. The expected dependence using a work function of $5.5 \pm 0.9$\,eV is shown as a red line. (b) Histogram over all the tunneling conductance values obtained in ramps as shown in (a) for different magnetic fields (from bottom to top, 0T, red, 10 T, blue, 14 T, cyan, 15 T, violet, 18 T, rose, 19 T, orange and 20 T in green). Histograms are shifted for clarity and have been normalized to the height of the 1$G_0$ peak without another normalization or averaging procedure. Curves are taken at slightly different locations. Each histogram comprises between $10^5$ and $10^6$ curves. We use the same bin width of 0.012 G/G$_0$ for all histograms. Bias voltage is of 100 mV. In the inset we show a histogram of the size of the last contact before breaking (the plateau at $G_0$ shown in (a)), at 0 T, red, 15 T, violet and 20 T in green. Vertical grey lines are integer multiples of 3.8 \AA. (c) Same as (a) but with Al, taken at 20 T. (d) same as (b) in Al, at magnetic fields of 0T, red, 10 T, blue and 20 T, green. 
}\label{FigIZ}
		\end{figure*}

In the tunneling range (inset of Fig.\,\ref{FigIZ}(a)) we find an exponential dependence of the conductance vs distance, which follows $I \propto e^{-\sqrt{\phi}z}$, $\phi$ being the work function of Au (we find approximately 6 eV) and  $z$ the distance between tip and sample. When tip and sample touch, we observe a characteristic staircase pattern. The first contact is very well defined and its conductance is close to the quantum of conductance $G_0$\cite{Agrait2003}. In particular, in Au, it is exactly $G_0$ and the value of the conductance just changes by a few percent when pulling or stretching. In Al, by contrast, it has a characteristic dependence showing that the conductance decreases when going away from the first touching point\cite{Agrait1993,Sheer1997,Scheer1998}. In both cases, when the joining point between both electrodes is no longer a single atom, but clusters of atoms, the conductance quantization is lost\cite{Agrait2003}.

The difference between the behavior in Au and Al can be discussed in terms of the valence orbitals and associated conduction channels through the atom making the contact between both electrodes. Calculations show that, in a Au ($4f^{14}5d^{10}6s^1$) single-atom contact, the conduction is realized mainly through the 6s orbital with an almost perfect transmission\cite{Cuevas1997,Cuevas1998}. Under elastic deformation, the transmission through this single channel changes very little, which explains the well defined step at $G_0$ in Au in Fig.\,\ref{FigIZ}(a)\cite{PhysRevLett.76.2302}. In an Al ($3s^23p^1$) single-atom contact, however, the conduction occurs through 3 channels. There is a hybridized $s$-$p_z$ orbital, and two degenerate $p_x-p_y$ orbitals\cite{Cuevas1998,Sheer1997}. The sum of the transmission coefficients of the three channels lies close to $G_0$, and varies strongly with elastic deformation of the junction, leading to a deformed first step around $G_0$ (Fig.\,\ref{FigIZ}(c)).

We show results of these experiments made at 20 T in Au Fig.\,\ref{FigIZ}(a) and in Al Fig.\,\ref{FigIZ}(c). Using all the 10$^5$ ramps made at each magnetic field, we construct histograms by counting the number of times we find a given conductance value. To this end, the 2048 points of the ramp, being 1 nm long, is separated into 100 bins. The histograms thus represent the number of counts for a given conductance value $G$. In Au Fig.\,\ref{FigIZ}(b), we find neat peaks at $G_0$ and less well defined peaks at higher conductance values. In Al Fig.\,\ref{FigIZ}(d) we find the same result, albeit with the differences characteristic of single atom point contacts of Al discussed above.

Furthermore, we have measured the length of the last single atom point contact in Au by tracing the length of the plateau at $G_0$ in the ramps. We
define the $G_0$ plateau by considering that changes above 20\% of $G_0$ lead to a change in the plateau from single atom point contact to complete rupture or
to a larger atomic point contact to complete rupture or to a larger atomic point contact\cite{Agrait1993,PhysRevLett.76.2302,Yanson1998,KOLESNYCHENKO20001257,PhysRevLett.83.2242}. We then count the appearance of a set of lengths and build with these a histogram of the $G_0$ plateau length. The resulting histogram is shown in the inset of Fig.\,\ref{FigIZ}(b). It was shown that the appearance of peaks at multiples of a certain distance (which is of 3.8 \AA, that is, of order of the Au atom size, modified by the presence of Helium exchange gas \cite{Yanson1998,KOLESNYCHENKO20001257,PhysRevLett.83.2242}) is due to the formation of single atom chains between tip and sample\cite{Yanson1998}. Here we can identify peaks in the distance histograms at the same distances as shown before\cite{Yanson1998}. These distances do not vary with the magnetic field, as shown in the inset of Fig.\,\ref{FigIZ}(b). 

To the best of our knowledge, these experiments have not been performed previously at such large magnetic fields, so that it is worth discussing briefly the implications. Our main result is that there is practically no change in the value of the conductance of a single atom when applying magnetic fields of 20 T. Of course, the flux through the contact is still negligible. Eventual magnetic field induced effects could then be attributed to Zeeman splitting of the bandstructure, influenced by spin orbit scattering, or magnetism. The Zeeman splitting at 20 T remains at the meV range, which is far below the Fermi energy of Al as well as of Au. None of these materials present features in this energy range in their bandstructure, nor spin orbit coupling sufficiently large to obtain a Zeeman splitting of the order of the main features in the band structure. The band structure of Au for instance is composed of holes, with the top of the band lying at about 2 eV. This explains thus the absence of magnetic field induced effects in the quantum point contact. It is worth noting, however, that Au features band inversion and that its surface states thus hold topological properties with a helical spin structure\cite{Yan2015}. Furthermore, the surface states are electron like and have the bottom of their band much closer to the Fermi level than the bulk states. Thus, there could be some magnetic field sensitivity which might require more detailed experiments.

\begin{figure}[htbp]
		\centering
		\begin{center}
			\includegraphics[width = 1\columnwidth]{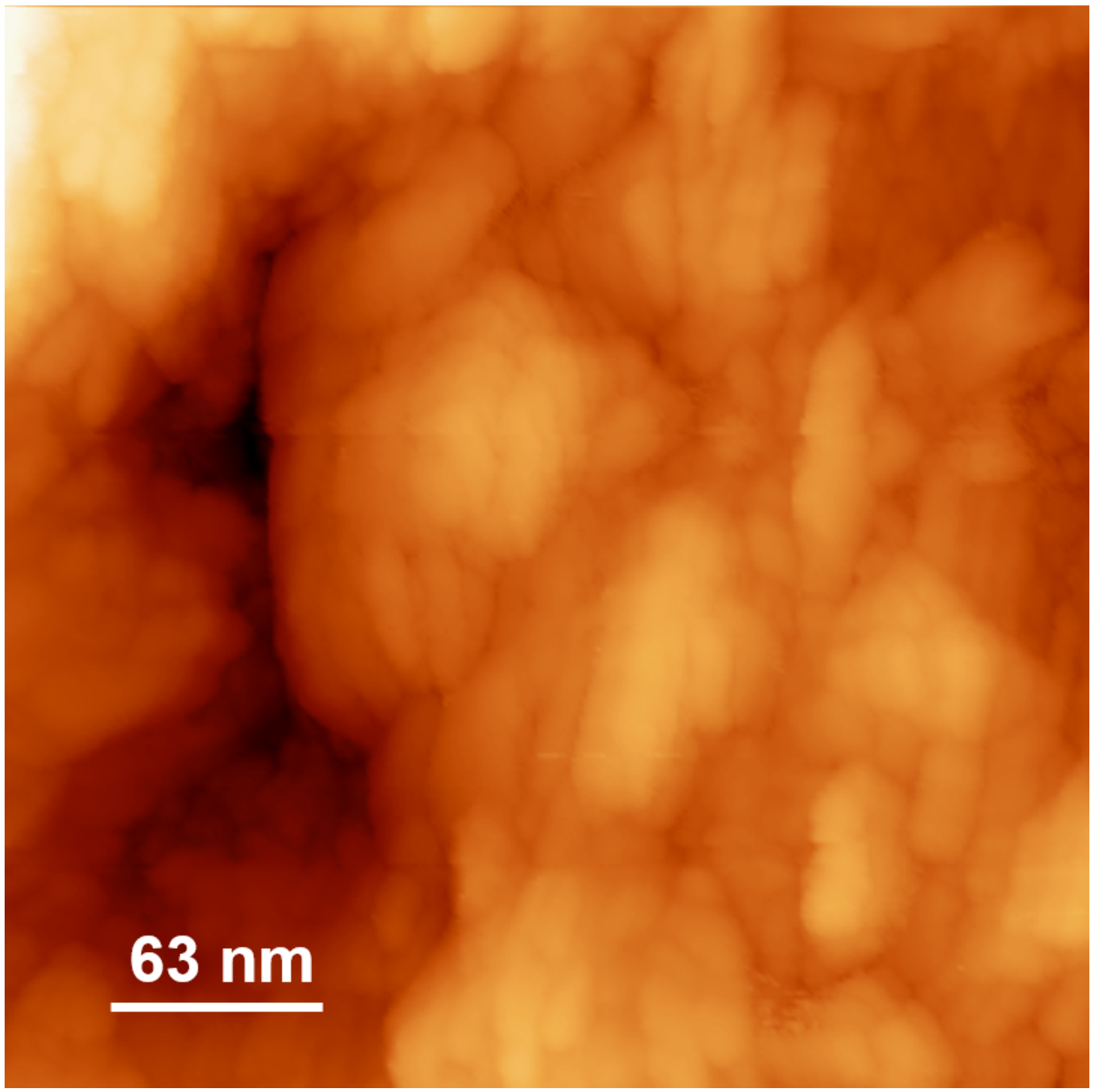}
		\end{center}
		\caption{We show a topography image taken at constant current (2nA, bias voltage of 100 mV), during a ramp from 20\,T to 22\,T at 100 mK. We use tip and sample of Au. We can identify the characteristic bubble like behavior of gold surfaces. The color scale from black to orange is of 55 nm. Scanning is made by making consecutive lines along the x-direction.}\label{Topo22T}
		\end{figure}

It is also worth noting that previous experiments at smaller magnetic fields have shown that single atom point contacts of Pt show a small magnetoresistance up to fields of 8 T, which is absent in Au or Al\cite{PhysRevB.72.224418,Strigl2015}. Contacts of Co or of other magnetic materials are also strongly influenced by the magnetic field\cite{Sokolov2007,Calvo2009}. Such studies are often made using a break junction, which has considerably less elements influencing the magnetic field behavior\cite{doi:10.1063/1.1146558}. In our STM we observe only a small magnetostriction when ramping the magnetic field scans that produces lateral and z motion of order of a nm for field scans of several T. For example, the topographic image in Fig.\,\ref{Topo22T} is made on Au when ramping between 20 T and 22 T and does not show any particular distortion.

\subsection{Atomic resolution and bias voltage dependent conductance maps in 2H-NbSe$_2$ at high magnetic fields}

\begin{figure*}[htbp]
		\centering
		\begin{center}
			\includegraphics[width = 0.9\textwidth]{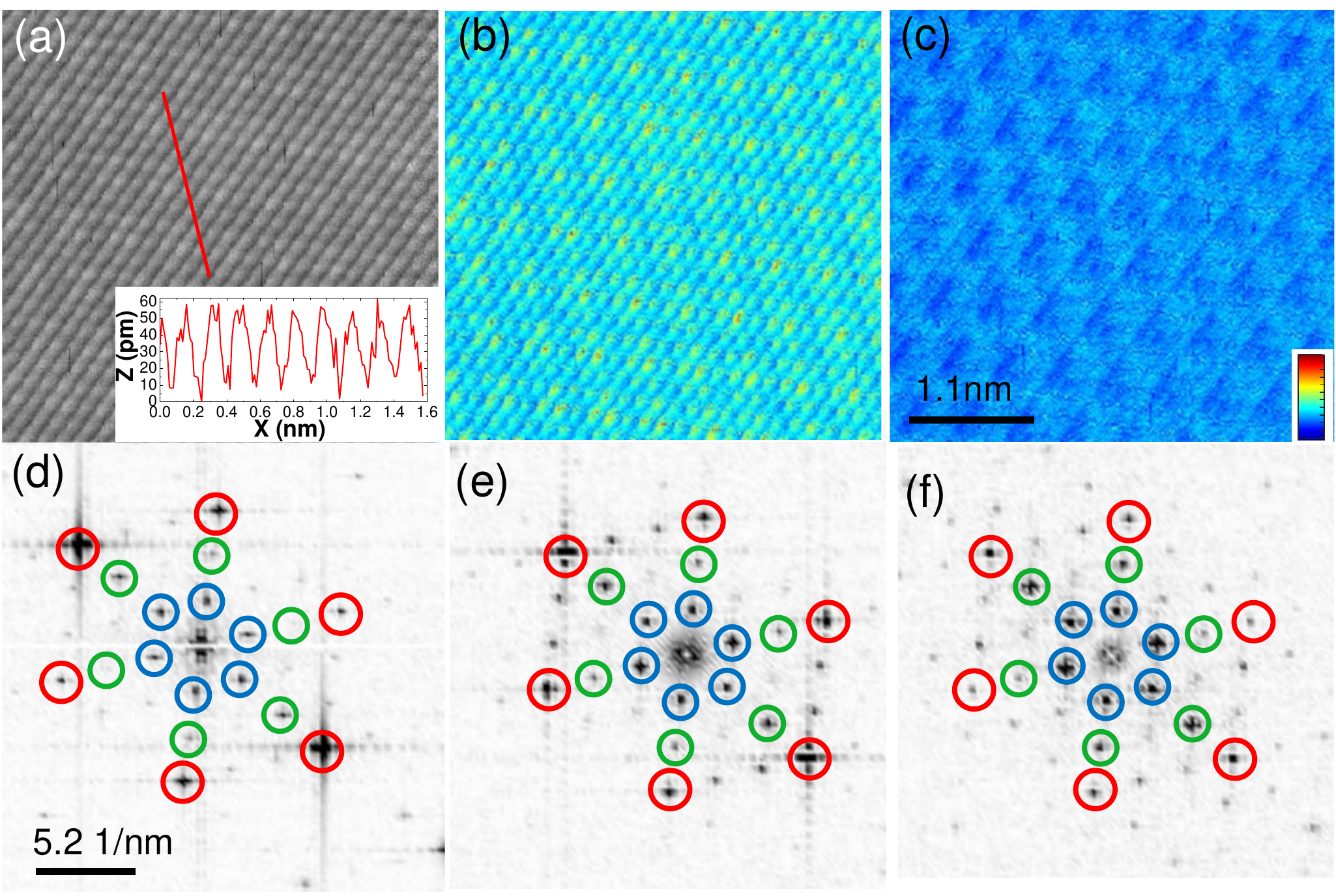}
		\end{center}
		\caption{(a) Atomic resolution topography images of 2H-NbSe$_2$ obtained at a magnetic field of 20 T and a temperature of 100 mK (bias voltage of 10 mV and tunneling current of 3.2 nA). Inset shows height vs distance profile along the red line. The tunneling conductance in the same field of view is shown as a function of the position in (b) for -2.5mV and in (c) for +2.5mV. Color scale at the bottom right of (c) provides the conductance in (b,c) and goes from zero to 0.66$\mu$S. Fourier transforms of (a,b,c) are shown in (d,e,f). Blue circles provide the CDW Bragg peaks, red circles the Bragg peaks due to the hexagonal surface Se atomic lattice and green circles the first harmonics of the CDW modulation located along the main crystal axis.
}\label{FigAtoms}
		\end{figure*}

Finally we present topography images with atomic resolution, obtained on the layered superconducting material 2H-NbSe$_2$ at 20 T and 100 mK in Fig.\,\ref{FigAtoms}(a). In the Fourier transform, we can identify (Fig.\,\ref{FigAtoms}(d)) the atomic and the charge density wave Bragg peaks, both located at the same positions as in zero field. These patterns are similar than those studied before\cite{guillamon2008intrinsic,guillamon2008superconducting,soumyanarayanan2013quantum}. In the same field of view, we have taken maps of the conductance vs bias voltage. We show results at $\pm$ 2.5 mV in Fig.\ref{FigAtoms}(b,c) and their respective Fourier transforms in Fig.\,\ref{FigAtoms}(e,f). We see that the charge density wave and atomic lattice also produce modulations in the density of states, with patterns that strongly change as a function of the bias voltage. The observation of precise Bragg peaks in topography and in spectroscopy shows that the STM is operating at 20 T with the same performance as at 0 T.

\section{Conclusion and outlook}

In summary, we have presented the design and the performance of a STM functioning at dilution refrigerator temperature and at magnetic fields up to 22 T. The set-up allows to perform STM measurements at millikelvin temperatures and with a resolution in energy which is of a few $\mu$eV. We have discussed an improved vibration isolation and important aspects of stick-slip motors. We have shown first results at very high magnetic fields on a comprehensive set of benchmarking systems. We provide in particular images of the charge density wave of 2H-NbSe$_2$ and discuss the behavior of Au and Al single atom point contacts at very high magnetic fields.

\begin{acknowledgments}
We thank discussions with Nicol\'as Agra\"it, Gabino Rubio, Sebasti\'an Vieira, Jose Gabriel Rodrigo, Alexey Ustinov, Christoph Strunk and Jenny Hoffman.This work was supported by the European Research Council PNICTEYES through grant agreement 679080, by the Spanish Research State Agency (FIS2017-84330-R, CEX2018-000805-M, RYC-2014-15093) and by the Comunidad de Madrid through program NANOFRONTMAG-CM (S2013/MIT-2850). We acknowledge collaborations through EU program Cost CA16218 (Nanocohybri). \end{acknowledgments}

\section*{Data Availability Statement}
The data that support the findings of this study are available from the corresponding author upon reasonable request.


\begin{thebibliography}{98}%
\makeatletter
\providecommand \@ifxundefined [1]{%
 \@ifx{#1\undefined}
}%
\providecommand \@ifnum [1]{%
 \ifnum #1\expandafter \@firstoftwo
 \else \expandafter \@secondoftwo
 \fi
}%
\providecommand \@ifx [1]{%
 \ifx #1\expandafter \@firstoftwo
 \else \expandafter \@secondoftwo
 \fi
}%
\providecommand \natexlab [1]{#1}%
\providecommand \enquote  [1]{``#1''}%
\providecommand \bibnamefont  [1]{#1}%
\providecommand \bibfnamefont [1]{#1}%
\providecommand \citenamefont [1]{#1}%
\providecommand \href@noop [0]{\@secondoftwo}%
\providecommand \href [0]{\begingroup \@sanitize@url \@href}%
\providecommand \@href[1]{\@@startlink{#1}\@@href}%
\providecommand \@@href[1]{\endgroup#1\@@endlink}%
\providecommand \@sanitize@url [0]{\catcode `\\12\catcode `\$12\catcode
  `\&12\catcode `\#12\catcode `\^12\catcode `\_12\catcode `\%12\relax}%
\providecommand \@@startlink[1]{}%
\providecommand \@@endlink[0]{}%
\providecommand \url  [0]{\begingroup\@sanitize@url \@url }%
\providecommand \@url [1]{\endgroup\@href {#1}{\urlprefix }}%
\providecommand \urlprefix  [0]{URL }%
\providecommand \Eprint [0]{\href }%
\providecommand \doibase [0]{http://dx.doi.org/}%
\providecommand \selectlanguage [0]{\@gobble}%
\providecommand \bibinfo  [0]{\@secondoftwo}%
\providecommand \bibfield  [0]{\@secondoftwo}%
\providecommand \translation [1]{[#1]}%
\providecommand \BibitemOpen [0]{}%
\providecommand \bibitemStop [0]{}%
\providecommand \bibitemNoStop [0]{.\EOS\space}%
\providecommand \EOS [0]{\spacefactor3000\relax}%
\providecommand \BibitemShut  [1]{\csname bibitem#1\endcsname}%
\let\auto@bib@innerbib\@empty
\bibitem [{\citenamefont {Binnig}\ \emph {et~al.}(1982)\citenamefont {Binnig},
  \citenamefont {Rohrer}, \citenamefont {Gerber},\ and\ \citenamefont
  {Weibel}}]{Binnig1982}%
  \BibitemOpen
  \bibfield  {author} {\bibinfo {author} {\bibfnamefont {G.}~\bibnamefont
  {Binnig}}, \bibinfo {author} {\bibfnamefont {H.}~\bibnamefont {Rohrer}},
  \bibinfo {author} {\bibfnamefont {C.}~\bibnamefont {Gerber}}, \ and\ \bibinfo
  {author} {\bibfnamefont {E.}~\bibnamefont {Weibel}},\ }\bibfield  {title}
  {\enquote {\bibinfo {title} {Tunneling through a controllable vacuum gap},}\
  }\href {\doibase 10.1063/1.92999} {\bibfield  {journal} {\bibinfo  {journal}
  {Applied Physics Letters}\ }\textbf {\bibinfo {volume} {40}},\ \bibinfo
  {pages} {178--180} (\bibinfo {year} {1982})}\BibitemShut {NoStop}%
\bibitem [{\citenamefont {Voigtlaender}(2015)}]{Voigtlander}%
  \BibitemOpen
  \bibfield  {author} {\bibinfo {author} {\bibfnamefont {B.}~\bibnamefont
  {Voigtlaender}},\ }\href {\doibase 10.1007/978-3-662-45240-0} {\emph
  {\bibinfo {title} {Scanning Probe Microscopy}}}\ (\bibinfo  {publisher}
  {Springer-Verlag Berlin Heidelberg},\ \bibinfo {year} {2015})\BibitemShut
  {NoStop}%
\bibitem [{\citenamefont {Wiesendanger}(1994)}]{wiesendanger_1994}%
  \BibitemOpen
  \bibfield  {author} {\bibinfo {author} {\bibfnamefont {R.}~\bibnamefont
  {Wiesendanger}},\ }\href {\doibase 10.1017/CBO9780511524356} {\emph {\bibinfo
  {title} {Scanning Probe Microscopy and Spectroscopy: Methods and
  Applications}}}\ (\bibinfo  {publisher} {Cambridge University Press},\
  \bibinfo {year} {1994})\BibitemShut {NoStop}%
\bibitem [{\citenamefont {Giambattista}\ \emph {et~al.}(1988)\citenamefont
  {Giambattista}, \citenamefont {Johnson}, \citenamefont {Coleman},
  \citenamefont {Drake},\ and\ \citenamefont {Hansma}}]{PhysRevB.37.2741}%
  \BibitemOpen
  \bibfield  {author} {\bibinfo {author} {\bibfnamefont {B.}~\bibnamefont
  {Giambattista}}, \bibinfo {author} {\bibfnamefont {A.}~\bibnamefont
  {Johnson}}, \bibinfo {author} {\bibfnamefont {R.~V.}\ \bibnamefont
  {Coleman}}, \bibinfo {author} {\bibfnamefont {B.}~\bibnamefont {Drake}}, \
  and\ \bibinfo {author} {\bibfnamefont {P.~K.}\ \bibnamefont {Hansma}},\
  }\bibfield  {title} {\enquote {\bibinfo {title} {Charge-density waves
  observed at 4.2 {K} by scanning-tunneling microscopy},}\ }\href {\doibase
  10.1103/PhysRevB.37.2741} {\bibfield  {journal} {\bibinfo  {journal} {Phys.
  Rev. B}\ }\textbf {\bibinfo {volume} {37}},\ \bibinfo {pages} {2741--2744}
  (\bibinfo {year} {1988})}\BibitemShut {NoStop}%
\bibitem [{\citenamefont {Hamidian}\ \emph {et~al.}(2016)\citenamefont
  {Hamidian}, \citenamefont {Edkins}, \citenamefont {Joo}, \citenamefont
  {Kostin}, \citenamefont {Eisaki}, \citenamefont {Uchida}, \citenamefont
  {Lawler}, \citenamefont {Kim}, \citenamefont {Mackenzie}, \citenamefont
  {Fujita}, \citenamefont {Lee},\ and\ \citenamefont {Davis}}]{Hamidian2016}%
  \BibitemOpen
  \bibfield  {author} {\bibinfo {author} {\bibfnamefont {M.~H.}\ \bibnamefont
  {Hamidian}}, \bibinfo {author} {\bibfnamefont {S.~D.}\ \bibnamefont
  {Edkins}}, \bibinfo {author} {\bibfnamefont {S.~H.}\ \bibnamefont {Joo}},
  \bibinfo {author} {\bibfnamefont {A.}~\bibnamefont {Kostin}}, \bibinfo
  {author} {\bibfnamefont {H.}~\bibnamefont {Eisaki}}, \bibinfo {author}
  {\bibfnamefont {S.}~\bibnamefont {Uchida}}, \bibinfo {author} {\bibfnamefont
  {M.~J.}\ \bibnamefont {Lawler}}, \bibinfo {author} {\bibfnamefont {E.-A.}\
  \bibnamefont {Kim}}, \bibinfo {author} {\bibfnamefont {A.~P.}\ \bibnamefont
  {Mackenzie}}, \bibinfo {author} {\bibfnamefont {K.}~\bibnamefont {Fujita}},
  \bibinfo {author} {\bibfnamefont {J.}~\bibnamefont {Lee}}, \ and\ \bibinfo
  {author} {\bibfnamefont {J.~C.~S.}\ \bibnamefont {Davis}},\ }\bibfield
  {title} {\enquote {\bibinfo {title} {Detection of a {Cooper}-pair density
  wave in {Bi$_2$Sr$_2$CaCu$_2$O$_{8+x}$}},}\ }\href {\doibase
  10.1038/nature17411} {\bibfield  {journal} {\bibinfo  {journal} {Nature}\
  }\textbf {\bibinfo {volume} {532}},\ \bibinfo {pages} {343--347} (\bibinfo
  {year} {2016})}\BibitemShut {NoStop}%
\bibitem [{\citenamefont {Hess}\ \emph {et~al.}(1989)\citenamefont {Hess},
  \citenamefont {Robinson}, \citenamefont {Dynes}, \citenamefont {Valles},\
  and\ \citenamefont {Waszczak}}]{PhysRevLett.62.214}%
  \BibitemOpen
  \bibfield  {author} {\bibinfo {author} {\bibfnamefont {H.~F.}\ \bibnamefont
  {Hess}}, \bibinfo {author} {\bibfnamefont {R.~B.}\ \bibnamefont {Robinson}},
  \bibinfo {author} {\bibfnamefont {R.~C.}\ \bibnamefont {Dynes}}, \bibinfo
  {author} {\bibfnamefont {J.~M.}\ \bibnamefont {Valles}}, \ and\ \bibinfo
  {author} {\bibfnamefont {J.~V.}\ \bibnamefont {Waszczak}},\ }\bibfield
  {title} {\enquote {\bibinfo {title} {Scanning-tunneling-microscope
  observation of the {Abrikosov} flux lattice and the density of states near
  and inside a fluxoid},}\ }\href {\doibase 10.1103/PhysRevLett.62.214}
  {\bibfield  {journal} {\bibinfo  {journal} {Phys. Rev. Lett.}\ }\textbf
  {\bibinfo {volume} {62}},\ \bibinfo {pages} {214--216} (\bibinfo {year}
  {1989})}\BibitemShut {NoStop}%
\bibitem [{\citenamefont {Fischer}\ \emph {et~al.}(2007)\citenamefont
  {Fischer}, \citenamefont {Kugler}, \citenamefont {Maggio-Aprile},
  \citenamefont {Berthod},\ and\ \citenamefont {Renner}}]{RevModPhys.79.353}%
  \BibitemOpen
  \bibfield  {author} {\bibinfo {author} {\bibfnamefont {O.}~\bibnamefont
  {Fischer}}, \bibinfo {author} {\bibfnamefont {M.}~\bibnamefont {Kugler}},
  \bibinfo {author} {\bibfnamefont {I.}~\bibnamefont {Maggio-Aprile}}, \bibinfo
  {author} {\bibfnamefont {C.}~\bibnamefont {Berthod}}, \ and\ \bibinfo
  {author} {\bibfnamefont {C.}~\bibnamefont {Renner}},\ }\bibfield  {title}
  {\enquote {\bibinfo {title} {Scanning tunneling spectroscopy of
  high-temperature superconductors},}\ }\href {\doibase
  10.1103/RevModPhys.79.353} {\bibfield  {journal} {\bibinfo  {journal} {Rev.
  Mod. Phys.}\ }\textbf {\bibinfo {volume} {79}},\ \bibinfo {pages} {353--419}
  (\bibinfo {year} {2007})}\BibitemShut {NoStop}%
\bibitem [{\citenamefont {Suderow}\ \emph {et~al.}(2014)\citenamefont
  {Suderow}, \citenamefont {Guillam{\'{o}}n}, \citenamefont {Rodrigo},\ and\
  \citenamefont {Vieira}}]{Suderow2014}%
  \BibitemOpen
  \bibfield  {author} {\bibinfo {author} {\bibfnamefont {H.}~\bibnamefont
  {Suderow}}, \bibinfo {author} {\bibfnamefont {I.}~\bibnamefont
  {Guillam{\'{o}}n}}, \bibinfo {author} {\bibfnamefont {J.~G.}\ \bibnamefont
  {Rodrigo}}, \ and\ \bibinfo {author} {\bibfnamefont {S.}~\bibnamefont
  {Vieira}},\ }\bibfield  {title} {\enquote {\bibinfo {title} {Imaging
  superconducting vortex cores and lattices with a scanning tunneling
  microscope},}\ }\href {\doibase 10.1088/0953-2048/27/6/063001} {\bibfield
  {journal} {\bibinfo  {journal} {Superconductor Science and Technology}\
  }\textbf {\bibinfo {volume} {27}},\ \bibinfo {pages} {063001} (\bibinfo
  {year} {2014})}\BibitemShut {NoStop}%
\bibitem [{\citenamefont {Matsui}\ \emph {et~al.}(2005)\citenamefont {Matsui},
  \citenamefont {Kambara}, \citenamefont {Niimi}, \citenamefont {Tagami},
  \citenamefont {Tsukada},\ and\ \citenamefont
  {Fukuyama}}]{PhysRevLett.94.226403}%
  \BibitemOpen
  \bibfield  {author} {\bibinfo {author} {\bibfnamefont {T.}~\bibnamefont
  {Matsui}}, \bibinfo {author} {\bibfnamefont {H.}~\bibnamefont {Kambara}},
  \bibinfo {author} {\bibfnamefont {Y.}~\bibnamefont {Niimi}}, \bibinfo
  {author} {\bibfnamefont {K.}~\bibnamefont {Tagami}}, \bibinfo {author}
  {\bibfnamefont {M.}~\bibnamefont {Tsukada}}, \ and\ \bibinfo {author}
  {\bibfnamefont {H.}~\bibnamefont {Fukuyama}},\ }\bibfield  {title} {\enquote
  {\bibinfo {title} {{STS} observations of {Landau} levels at graphite
  surfaces},}\ }\href {\doibase 10.1103/PhysRevLett.94.226403} {\bibfield
  {journal} {\bibinfo  {journal} {Phys. Rev. Lett.}\ }\textbf {\bibinfo
  {volume} {94}},\ \bibinfo {pages} {226403} (\bibinfo {year}
  {2005})}\BibitemShut {NoStop}%
\bibitem [{\citenamefont {Jeon}\ \emph {et~al.}(2014)\citenamefont {Jeon},
  \citenamefont {Zhou}, \citenamefont {Gyenis}, \citenamefont {Feldman},
  \citenamefont {Kimchi}, \citenamefont {Potter}, \citenamefont {Gibson},
  \citenamefont {Cava}, \citenamefont {Vishwanath},\ and\ \citenamefont
  {Yazdani}}]{Jeon2014}%
  \BibitemOpen
  \bibfield  {author} {\bibinfo {author} {\bibfnamefont {S.}~\bibnamefont
  {Jeon}}, \bibinfo {author} {\bibfnamefont {B.~B.}\ \bibnamefont {Zhou}},
  \bibinfo {author} {\bibfnamefont {A.}~\bibnamefont {Gyenis}}, \bibinfo
  {author} {\bibfnamefont {B.~E.}\ \bibnamefont {Feldman}}, \bibinfo {author}
  {\bibfnamefont {I.}~\bibnamefont {Kimchi}}, \bibinfo {author} {\bibfnamefont
  {A.~C.}\ \bibnamefont {Potter}}, \bibinfo {author} {\bibfnamefont {Q.~D.}\
  \bibnamefont {Gibson}}, \bibinfo {author} {\bibfnamefont {R.~J.}\
  \bibnamefont {Cava}}, \bibinfo {author} {\bibfnamefont {A.}~\bibnamefont
  {Vishwanath}}, \ and\ \bibinfo {author} {\bibfnamefont {A.}~\bibnamefont
  {Yazdani}},\ }\bibfield  {title} {\enquote {\bibinfo {title} {Landau
  quantization and quasiparticle interference in the three-dimensional {Dirac}
  semimetal {Cd$_3$As$_2$}},}\ }\href {\doibase 10.1038/nmat4023} {\bibfield
  {journal} {\bibinfo  {journal} {Nature Materials}\ }\textbf {\bibinfo
  {volume} {13}},\ \bibinfo {pages} {851--856} (\bibinfo {year}
  {2014})}\BibitemShut {NoStop}%
\bibitem [{\citenamefont {Aoki}, \citenamefont {Ishida},\ and\ \citenamefont
  {Flouquet}(2019)}]{doi:10.7566/JPSJ.88.022001}%
  \BibitemOpen
  \bibfield  {author} {\bibinfo {author} {\bibfnamefont {D.}~\bibnamefont
  {Aoki}}, \bibinfo {author} {\bibfnamefont {K.}~\bibnamefont {Ishida}}, \ and\
  \bibinfo {author} {\bibfnamefont {J.}~\bibnamefont {Flouquet}},\ }\bibfield
  {title} {\enquote {\bibinfo {title} {Review of {U-based} ferromagnetic
  superconductors: Comparison between {UGe$_2$, URhGe, and UCoGe}},}\ }\href
  {\doibase 10.7566/JPSJ.88.022001} {\bibfield  {journal} {\bibinfo  {journal}
  {Journal of the Physical Society of Japan}\ }\textbf {\bibinfo {volume}
  {88}},\ \bibinfo {pages} {022001} (\bibinfo {year} {2019})}\BibitemShut
  {NoStop}%
\bibitem [{\citenamefont {Lu}\ \emph {et~al.}(2015)\citenamefont {Lu},
  \citenamefont {Zheliuk}, \citenamefont {Leermakers}, \citenamefont {Yuan},
  \citenamefont {Zeitler}, \citenamefont {Law},\ and\ \citenamefont
  {Ye}}]{Lu1353}%
  \BibitemOpen
  \bibfield  {author} {\bibinfo {author} {\bibfnamefont {J.~M.}\ \bibnamefont
  {Lu}}, \bibinfo {author} {\bibfnamefont {O.}~\bibnamefont {Zheliuk}},
  \bibinfo {author} {\bibfnamefont {I.}~\bibnamefont {Leermakers}}, \bibinfo
  {author} {\bibfnamefont {N.~F.~Q.}\ \bibnamefont {Yuan}}, \bibinfo {author}
  {\bibfnamefont {U.}~\bibnamefont {Zeitler}}, \bibinfo {author} {\bibfnamefont
  {K.~T.}\ \bibnamefont {Law}}, \ and\ \bibinfo {author} {\bibfnamefont
  {J.~T.}\ \bibnamefont {Ye}},\ }\bibfield  {title} {\enquote {\bibinfo {title}
  {Evidence for two-dimensional {Ising} superconductivity in gated
  {MoS$_2$}},}\ }\href {\doibase 10.1126/science.aab2277} {\bibfield  {journal}
  {\bibinfo  {journal} {Science}\ }\textbf {\bibinfo {volume} {350}},\ \bibinfo
  {pages} {1353--1357} (\bibinfo {year} {2015})}\BibitemShut {NoStop}%
\bibitem [{\citenamefont {Saito}\ \emph {et~al.}(2016)\citenamefont {Saito},
  \citenamefont {Nakamura}, \citenamefont {Bahramy}, \citenamefont {Kohama},
  \citenamefont {Ye}, \citenamefont {Kasahara}, \citenamefont {Nakagawa},
  \citenamefont {Onga}, \citenamefont {Tokunaga}, \citenamefont {Nojima},
  \citenamefont {Yanase},\ and\ \citenamefont {Iwasa}}]{Saito2016}%
  \BibitemOpen
  \bibfield  {author} {\bibinfo {author} {\bibfnamefont {Y.}~\bibnamefont
  {Saito}}, \bibinfo {author} {\bibfnamefont {Y.}~\bibnamefont {Nakamura}},
  \bibinfo {author} {\bibfnamefont {M.~S.}\ \bibnamefont {Bahramy}}, \bibinfo
  {author} {\bibfnamefont {Y.}~\bibnamefont {Kohama}}, \bibinfo {author}
  {\bibfnamefont {J.}~\bibnamefont {Ye}}, \bibinfo {author} {\bibfnamefont
  {Y.}~\bibnamefont {Kasahara}}, \bibinfo {author} {\bibfnamefont
  {Y.}~\bibnamefont {Nakagawa}}, \bibinfo {author} {\bibfnamefont
  {M.}~\bibnamefont {Onga}}, \bibinfo {author} {\bibfnamefont {M.}~\bibnamefont
  {Tokunaga}}, \bibinfo {author} {\bibfnamefont {T.}~\bibnamefont {Nojima}},
  \bibinfo {author} {\bibfnamefont {Y.}~\bibnamefont {Yanase}}, \ and\ \bibinfo
  {author} {\bibfnamefont {Y.}~\bibnamefont {Iwasa}},\ }\bibfield  {title}
  {\enquote {\bibinfo {title} {Superconductivity protected by spin--valley
  locking in ion-gated {MoS$_2$}},}\ }\href {\doibase 10.1038/nphys3580}
  {\bibfield  {journal} {\bibinfo  {journal} {Nature Physics}\ }\textbf
  {\bibinfo {volume} {12}},\ \bibinfo {pages} {144--149} (\bibinfo {year}
  {2016})}\BibitemShut {NoStop}%
\bibitem [{\citenamefont {Paschen}\ and\ \citenamefont
  {Si}(2021)}]{Paschen2021}%
  \BibitemOpen
  \bibfield  {author} {\bibinfo {author} {\bibfnamefont {S.}~\bibnamefont
  {Paschen}}\ and\ \bibinfo {author} {\bibfnamefont {Q.}~\bibnamefont {Si}},\
  }\bibfield  {title} {\enquote {\bibinfo {title} {Quantum phases driven by
  strong correlations},}\ }\href {\doibase 10.1038/s42254-020-00262-6}
  {\bibfield  {journal} {\bibinfo  {journal} {Nature Reviews Physics}\ }\textbf
  {\bibinfo {volume} {3}},\ \bibinfo {pages} {9--26} (\bibinfo {year}
  {2021})}\BibitemShut {NoStop}%
\bibitem [{\citenamefont {Ast}\ \emph {et~al.}(2016)\citenamefont {Ast},
  \citenamefont {Jäck}, \citenamefont {Senkpiel}, \citenamefont {Eltschka},
  \citenamefont {Etzkorn}, \citenamefont {Ankerhold},\ and\ \citenamefont
  {Kern}}]{Ast2016}%
  \BibitemOpen
  \bibfield  {author} {\bibinfo {author} {\bibfnamefont {C.~R.}\ \bibnamefont
  {Ast}}, \bibinfo {author} {\bibfnamefont {B.}~\bibnamefont {Jäck}}, \bibinfo
  {author} {\bibfnamefont {J.}~\bibnamefont {Senkpiel}}, \bibinfo {author}
  {\bibfnamefont {M.}~\bibnamefont {Eltschka}}, \bibinfo {author}
  {\bibfnamefont {M.}~\bibnamefont {Etzkorn}}, \bibinfo {author} {\bibfnamefont
  {J.}~\bibnamefont {Ankerhold}}, \ and\ \bibinfo {author} {\bibfnamefont
  {K.}~\bibnamefont {Kern}},\ }\bibfield  {title} {\enquote {\bibinfo {title}
  {Sensing the quantum limit in scanning tunnelling spectroscopy},}\ }\href
  {https://doi.org/10.1038/ncomms13009} {\bibfield  {journal} {\bibinfo
  {journal} {Nature Communications}\ }\textbf {\bibinfo {volume} {7}},\
  \bibinfo {pages} {13009} (\bibinfo {year} {2016})}\BibitemShut {NoStop}%
\bibitem [{\citenamefont {Rodrigo}, \citenamefont {Suderow},\ and\
  \citenamefont {Vieira}(2004)}]{Rodrigo2004a}%
  \BibitemOpen
  \bibfield  {author} {\bibinfo {author} {\bibfnamefont {J.~G.}\ \bibnamefont
  {Rodrigo}}, \bibinfo {author} {\bibfnamefont {H.}~\bibnamefont {Suderow}}, \
  and\ \bibinfo {author} {\bibfnamefont {S.}~\bibnamefont {Vieira}},\
  }\bibfield  {title} {\enquote {\bibinfo {title} {On the use of {STM}
  superconducting tips at very low temperatures},}\ }\href
  {https://doi.org/10.1140/epjb/e2004-00273-y} {\bibfield  {journal} {\bibinfo
  {journal} {The European Physical Journal B - Condensed Matter and Complex
  Systems}\ }\textbf {\bibinfo {volume} {40}},\ \bibinfo {pages} {483--488}
  (\bibinfo {year} {2004})}\BibitemShut {NoStop}%
\bibitem [{\citenamefont {Naaman}, \citenamefont {Teizer},\ and\ \citenamefont
  {Dynes}(2001)}]{PhysRevLett.87.097004}%
  \BibitemOpen
  \bibfield  {author} {\bibinfo {author} {\bibfnamefont {O.}~\bibnamefont
  {Naaman}}, \bibinfo {author} {\bibfnamefont {W.}~\bibnamefont {Teizer}}, \
  and\ \bibinfo {author} {\bibfnamefont {R.~C.}\ \bibnamefont {Dynes}},\
  }\bibfield  {title} {\enquote {\bibinfo {title} {Fluctuation dominated
  {Josephson} tunneling with a scanning tunneling microscope},}\ }\href
  {\doibase 10.1103/PhysRevLett.87.097004} {\bibfield  {journal} {\bibinfo
  {journal} {Phys. Rev. Lett.}\ }\textbf {\bibinfo {volume} {87}},\ \bibinfo
  {pages} {097004} (\bibinfo {year} {2001})}\BibitemShut {NoStop}%
\bibitem [{\citenamefont {Tao}\ \emph {et~al.}(2017)\citenamefont {Tao},
  \citenamefont {Singh}, \citenamefont {Rossi}, \citenamefont {Gerritsen},
  \citenamefont {Hendriksen}, \citenamefont {Khajetoorians}, \citenamefont
  {Christianen}, \citenamefont {Maan}, \citenamefont {Zeitler},\ and\
  \citenamefont {Bryant}}]{Tao2017}%
  \BibitemOpen
  \bibfield  {author} {\bibinfo {author} {\bibfnamefont {W.}~\bibnamefont
  {Tao}}, \bibinfo {author} {\bibfnamefont {S.}~\bibnamefont {Singh}}, \bibinfo
  {author} {\bibfnamefont {L.}~\bibnamefont {Rossi}}, \bibinfo {author}
  {\bibfnamefont {J.~W.}\ \bibnamefont {Gerritsen}}, \bibinfo {author}
  {\bibfnamefont {B.~L.~M.}\ \bibnamefont {Hendriksen}}, \bibinfo {author}
  {\bibfnamefont {A.~A.}\ \bibnamefont {Khajetoorians}}, \bibinfo {author}
  {\bibfnamefont {P.~C.~M.}\ \bibnamefont {Christianen}}, \bibinfo {author}
  {\bibfnamefont {J.~C.}\ \bibnamefont {Maan}}, \bibinfo {author}
  {\bibfnamefont {U.}~\bibnamefont {Zeitler}}, \ and\ \bibinfo {author}
  {\bibfnamefont {B.}~\bibnamefont {Bryant}},\ }\bibfield  {title} {\enquote
  {\bibinfo {title} {A low-temperature scanning tunneling microscope capable of
  microscopy and spectroscopy in a {Bitter} magnet at up to 34 {T}},}\ }\href
  {\doibase 10.1063/1.4995372} {\bibfield  {journal} {\bibinfo  {journal}
  {Review of Scientific Instruments}\ }\textbf {\bibinfo {volume} {88}},\
  \bibinfo {pages} {093706} (\bibinfo {year} {2017})}\BibitemShut {NoStop}%
\bibitem [{\citenamefont {Song}\ \emph {et~al.}(2010)\citenamefont {Song},
  \citenamefont {Otte}, \citenamefont {Shvarts}, \citenamefont {Zhao},
  \citenamefont {Kuk}, \citenamefont {Blankenship}, \citenamefont {Band},
  \citenamefont {Hess},\ and\ \citenamefont {Stroscio}}]{Song2010}%
  \BibitemOpen
  \bibfield  {author} {\bibinfo {author} {\bibfnamefont {Y.~J.}\ \bibnamefont
  {Song}}, \bibinfo {author} {\bibfnamefont {A.~F.}\ \bibnamefont {Otte}},
  \bibinfo {author} {\bibfnamefont {V.}~\bibnamefont {Shvarts}}, \bibinfo
  {author} {\bibfnamefont {Z.}~\bibnamefont {Zhao}}, \bibinfo {author}
  {\bibfnamefont {Y.}~\bibnamefont {Kuk}}, \bibinfo {author} {\bibfnamefont
  {S.~R.}\ \bibnamefont {Blankenship}}, \bibinfo {author} {\bibfnamefont
  {A.}~\bibnamefont {Band}}, \bibinfo {author} {\bibfnamefont {F.~M.}\
  \bibnamefont {Hess}}, \ and\ \bibinfo {author} {\bibfnamefont {J.~A.}\
  \bibnamefont {Stroscio}},\ }\bibfield  {title} {\enquote {\bibinfo {title}
  {Invited review article: A 10 {mK} scanning probe microscopy facility},}\
  }\href {\doibase 10.1063/1.3520482} {\bibfield  {journal} {\bibinfo
  {journal} {Review of Scientific Instruments}\ }\textbf {\bibinfo {volume}
  {81}},\ \bibinfo {pages} {121101} (\bibinfo {year} {2010})}\BibitemShut
  {NoStop}%
\bibitem [{\citenamefont {Misra}\ \emph {et~al.}(2013)\citenamefont {Misra},
  \citenamefont {Zhou}, \citenamefont {Drozdov}, \citenamefont {Seo},
  \citenamefont {Urban}, \citenamefont {Gyenis}, \citenamefont {Kingsley},
  \citenamefont {Jones},\ and\ \citenamefont {Yazdani}}]{Misra2013}%
  \BibitemOpen
  \bibfield  {author} {\bibinfo {author} {\bibfnamefont {S.}~\bibnamefont
  {Misra}}, \bibinfo {author} {\bibfnamefont {B.~B.}\ \bibnamefont {Zhou}},
  \bibinfo {author} {\bibfnamefont {I.~K.}\ \bibnamefont {Drozdov}}, \bibinfo
  {author} {\bibfnamefont {J.}~\bibnamefont {Seo}}, \bibinfo {author}
  {\bibfnamefont {L.}~\bibnamefont {Urban}}, \bibinfo {author} {\bibfnamefont
  {A.}~\bibnamefont {Gyenis}}, \bibinfo {author} {\bibfnamefont {S.~C.~J.}\
  \bibnamefont {Kingsley}}, \bibinfo {author} {\bibfnamefont {H.}~\bibnamefont
  {Jones}}, \ and\ \bibinfo {author} {\bibfnamefont {A.}~\bibnamefont
  {Yazdani}},\ }\bibfield  {title} {\enquote {\bibinfo {title} {Design and
  performance of an ultra-high vacuum scanning tunneling microscope operating
  at dilution refrigerator temperatures and high magnetic fields},}\ }\href
  {\doibase 10.1063/1.4822271} {\bibfield  {journal} {\bibinfo  {journal}
  {Review of Scientific Instruments}\ }\textbf {\bibinfo {volume} {84}},\
  \bibinfo {pages} {103903} (\bibinfo {year} {2013})}\BibitemShut {NoStop}%
\bibitem [{\citenamefont {Singh}\ \emph {et~al.}(2013)\citenamefont {Singh},
  \citenamefont {Enayat}, \citenamefont {White},\ and\ \citenamefont
  {Wahl}}]{Singh2013}%
  \BibitemOpen
  \bibfield  {author} {\bibinfo {author} {\bibfnamefont {U.~R.}\ \bibnamefont
  {Singh}}, \bibinfo {author} {\bibfnamefont {M.}~\bibnamefont {Enayat}},
  \bibinfo {author} {\bibfnamefont {S.~C.}\ \bibnamefont {White}}, \ and\
  \bibinfo {author} {\bibfnamefont {P.}~\bibnamefont {Wahl}},\ }\bibfield
  {title} {\enquote {\bibinfo {title} {Construction and performance of a
  dilution-refrigerator based spectroscopic-imaging scanning tunneling
  microscope},}\ }\href {\doibase 10.1063/1.4788941} {\bibfield  {journal}
  {\bibinfo  {journal} {Review of Scientific Instruments}\ }\textbf {\bibinfo
  {volume} {84}},\ \bibinfo {pages} {013708} (\bibinfo {year}
  {2013})}\BibitemShut {NoStop}%
\bibitem [{\citenamefont {Assig}\ \emph {et~al.}(2013)\citenamefont {Assig},
  \citenamefont {Etzkorn}, \citenamefont {Enders}, \citenamefont {Stiepany},
  \citenamefont {Ast},\ and\ \citenamefont {Kern}}]{Assig2013}%
  \BibitemOpen
  \bibfield  {author} {\bibinfo {author} {\bibfnamefont {M.}~\bibnamefont
  {Assig}}, \bibinfo {author} {\bibfnamefont {M.}~\bibnamefont {Etzkorn}},
  \bibinfo {author} {\bibfnamefont {A.}~\bibnamefont {Enders}}, \bibinfo
  {author} {\bibfnamefont {W.}~\bibnamefont {Stiepany}}, \bibinfo {author}
  {\bibfnamefont {C.~R.}\ \bibnamefont {Ast}}, \ and\ \bibinfo {author}
  {\bibfnamefont {K.}~\bibnamefont {Kern}},\ }\bibfield  {title} {\enquote
  {\bibinfo {title} {A 10 {mK} scanning tunneling microscope operating in ultra
  high vacuum and high magnetic fields},}\ }\href {\doibase 10.1063/1.4793793}
  {\bibfield  {journal} {\bibinfo  {journal} {Review of Scientific
  Instruments}\ }\textbf {\bibinfo {volume} {84}},\ \bibinfo {pages} {033903}
  (\bibinfo {year} {2013})}\BibitemShut {NoStop}%
\bibitem [{\citenamefont {Roychowdhury}\ \emph {et~al.}(2014)\citenamefont
  {Roychowdhury}, \citenamefont {Gubrud}, \citenamefont {Dana}, \citenamefont
  {Anderson}, \citenamefont {Lobb}, \citenamefont {Wellstood},\ and\
  \citenamefont {Dreyer}}]{Roychowdhury2014}%
  \BibitemOpen
  \bibfield  {author} {\bibinfo {author} {\bibfnamefont {A.}~\bibnamefont
  {Roychowdhury}}, \bibinfo {author} {\bibfnamefont {M.~A.}\ \bibnamefont
  {Gubrud}}, \bibinfo {author} {\bibfnamefont {R.}~\bibnamefont {Dana}},
  \bibinfo {author} {\bibfnamefont {J.~R.}\ \bibnamefont {Anderson}}, \bibinfo
  {author} {\bibfnamefont {C.~J.}\ \bibnamefont {Lobb}}, \bibinfo {author}
  {\bibfnamefont {F.~C.}\ \bibnamefont {Wellstood}}, \ and\ \bibinfo {author}
  {\bibfnamefont {M.}~\bibnamefont {Dreyer}},\ }\bibfield  {title} {\enquote
  {\bibinfo {title} {A 30 {mK}, 13.5 {T} scanning tunneling microscope with two
  independent tips},}\ }\href {\doibase 10.1063/1.4871056} {\bibfield
  {journal} {\bibinfo  {journal} {Review of Scientific Instruments}\ }\textbf
  {\bibinfo {volume} {85}},\ \bibinfo {pages} {043706} (\bibinfo {year}
  {2014})}\BibitemShut {NoStop}%
\bibitem [{\citenamefont {Machida}, \citenamefont {Kohsaka},\ and\
  \citenamefont {Hanaguri}(2018)}]{Machida2018}%
  \BibitemOpen
  \bibfield  {author} {\bibinfo {author} {\bibfnamefont {T.}~\bibnamefont
  {Machida}}, \bibinfo {author} {\bibfnamefont {Y.}~\bibnamefont {Kohsaka}}, \
  and\ \bibinfo {author} {\bibfnamefont {T.}~\bibnamefont {Hanaguri}},\
  }\bibfield  {title} {\enquote {\bibinfo {title} {A scanning tunneling
  microscope for spectroscopic imaging below 90 {mK} in magnetic fields up to
  17.5 {T}},}\ }\href {\doibase 10.1063/1.5049619} {\bibfield  {journal}
  {\bibinfo  {journal} {Review of Scientific Instruments}\ }\textbf {\bibinfo
  {volume} {89}},\ \bibinfo {pages} {093707} (\bibinfo {year}
  {2018})}\BibitemShut {NoStop}%
\bibitem [{\citenamefont {Park}\ and\ \citenamefont
  {Barrett}(1993)}]{ParkStroscio}%
  \BibitemOpen
  \bibfield  {author} {\bibinfo {author} {\bibfnamefont {S.-I.}\ \bibnamefont
  {Park}}\ and\ \bibinfo {author} {\bibfnamefont {R.~C.}\ \bibnamefont
  {Barrett}},\ }\href@noop {} {\emph {\bibinfo {title} {Scanning Tunneling
  Microscopy}}},\ edited by\ \bibinfo {editor} {\bibfnamefont {J.~A.}\
  \bibnamefont {Stroscio}}\ and\ \bibinfo {editor} {\bibfnamefont {W.~J.}\
  \bibnamefont {Kaiser}},\ Methods in Experimental Physics\ (\bibinfo
  {publisher} {Springer},\ \bibinfo {year} {1993})\ \bibinfo {note} {{ISBN}:
  978-0-12-475972-5}\BibitemShut {NoStop}%
\bibitem [{\citenamefont {Ge}, \citenamefont {Ovadia},\ and\ \citenamefont
  {Hoffman}(2019)}]{doi:10.1063/1.5111989}%
  \BibitemOpen
  \bibfield  {author} {\bibinfo {author} {\bibfnamefont {J.-F.}\ \bibnamefont
  {Ge}}, \bibinfo {author} {\bibfnamefont {M.}~\bibnamefont {Ovadia}}, \ and\
  \bibinfo {author} {\bibfnamefont {J.~E.}\ \bibnamefont {Hoffman}},\
  }\bibfield  {title} {\enquote {\bibinfo {title} {Achieving low noise in
  scanning tunneling spectroscopy},}\ }\href {\doibase 10.1063/1.5111989}
  {\bibfield  {journal} {\bibinfo  {journal} {Review of Scientific
  Instruments}\ }\textbf {\bibinfo {volume} {90}},\ \bibinfo {pages} {101401}
  (\bibinfo {year} {2019})}\BibitemShut {NoStop}%
\bibitem [{\citenamefont {Hoffman}(2003)}]{hoffmanphd}%
  \BibitemOpen
  \bibfield  {author} {\bibinfo {author} {\bibfnamefont {J.~E.}\ \bibnamefont
  {Hoffman}},\ }\href@noop {} {\enquote {\bibinfo {title} {A search for
  alternative electronic order in the high temperature superconductor
  {Bi$_2$Sr$_2$CaCu$_2$O$_{8+\delta}$} by scanning tunneling microscopy},}\ }
  (\bibinfo {year} {2003}),\ \Eprint {http://arxiv.org/abs/PhD Thesis, Cornell}
  {PhD Thesis, Cornell} \BibitemShut {NoStop}%
\bibitem [{\citenamefont {Martin-Vega}\ and\ \citenamefont
  {et~al}(2021)}]{WinSPM}%
  \BibitemOpen
  \bibfield  {author} {\bibinfo {author} {\bibfnamefont {F.}~\bibnamefont
  {Martin-Vega}}\ and\ \bibinfo {author} {\bibnamefont {et~al}},\ }\bibfield
  {title} {\enquote {\bibinfo {title} {Soft real-time {USB} controlled feedback
  for scanning tunneling microscopy and spectroscopy: {Application} to
  {2H-NbSe$_2$}, {WTe$_2$}, {FeSe} and {Co$_3$Sn$_3$S$_2$} at very low
  temperatures},}\ }\href@noop {} {\bibfield  {journal} {\bibinfo  {journal}
  {Submitted}\ } (\bibinfo {year} {2021})}\BibitemShut {NoStop}%
\bibitem [{\citenamefont {Galvis}\ \emph {et~al.}(2015)\citenamefont {Galvis},
  \citenamefont {Herrera}, \citenamefont {Guillamón}, \citenamefont
  {Azpeitia}, \citenamefont {Luccas}, \citenamefont {Munuera}, \citenamefont
  {Cuenca}, \citenamefont {Higuera}, \citenamefont {Díaz}, \citenamefont
  {Pazos}, \citenamefont {García-Hernandez}, \citenamefont {Buendía},
  \citenamefont {Vieira},\ and\ \citenamefont
  {Suderow}}]{doi:10.1063/1.4905531}%
  \BibitemOpen
  \bibfield  {author} {\bibinfo {author} {\bibfnamefont {J.~A.}\ \bibnamefont
  {Galvis}}, \bibinfo {author} {\bibfnamefont {E.}~\bibnamefont {Herrera}},
  \bibinfo {author} {\bibfnamefont {I.}~\bibnamefont {Guillamón}}, \bibinfo
  {author} {\bibfnamefont {J.}~\bibnamefont {Azpeitia}}, \bibinfo {author}
  {\bibfnamefont {R.~F.}\ \bibnamefont {Luccas}}, \bibinfo {author}
  {\bibfnamefont {C.}~\bibnamefont {Munuera}}, \bibinfo {author} {\bibfnamefont
  {M.}~\bibnamefont {Cuenca}}, \bibinfo {author} {\bibfnamefont {J.~A.}\
  \bibnamefont {Higuera}}, \bibinfo {author} {\bibfnamefont {N.}~\bibnamefont
  {Díaz}}, \bibinfo {author} {\bibfnamefont {M.}~\bibnamefont {Pazos}},
  \bibinfo {author} {\bibfnamefont {M.}~\bibnamefont {García-Hernandez}},
  \bibinfo {author} {\bibfnamefont {A.}~\bibnamefont {Buendía}}, \bibinfo
  {author} {\bibfnamefont {S.}~\bibnamefont {Vieira}}, \ and\ \bibinfo {author}
  {\bibfnamefont {H.}~\bibnamefont {Suderow}},\ }\bibfield  {title} {\enquote
  {\bibinfo {title} {Three axis vector magnet set-up for cryogenic scanning
  probe microscopy},}\ }\href {\doibase 10.1063/1.4905531} {\bibfield
  {journal} {\bibinfo  {journal} {Review of Scientific Instruments}\ }\textbf
  {\bibinfo {volume} {86}},\ \bibinfo {pages} {013706} (\bibinfo {year}
  {2015})}\BibitemShut {NoStop}%
\bibitem [{\citenamefont {Suderow}, \citenamefont {Guillam{\'{o}}n},\ and\
  \citenamefont {Vieira}(2011)}]{Suderow2011}%
  \BibitemOpen
  \bibfield  {author} {\bibinfo {author} {\bibfnamefont {H.}~\bibnamefont
  {Suderow}}, \bibinfo {author} {\bibfnamefont {I.}~\bibnamefont
  {Guillam{\'{o}}n}}, \ and\ \bibinfo {author} {\bibfnamefont {S.}~\bibnamefont
  {Vieira}},\ }\bibfield  {title} {\enquote {\bibinfo {title} {Compact very low
  temperature scanning tunneling microscope with mechanically driven horizontal
  linear positioning stage},}\ }\href {\doibase 10.1063/1.3567008} {\bibfield
  {journal} {\bibinfo  {journal} {Review of Scientific Instruments}\ }\textbf
  {\bibinfo {volume} {82}},\ \bibinfo {pages} {033711} (\bibinfo {year}
  {2011})}\BibitemShut {NoStop}%
\bibitem [{\citenamefont {Rodrigo}\ \emph {et~al.}(2004)\citenamefont
  {Rodrigo}, \citenamefont {Suderow}, \citenamefont {Vieira}, \citenamefont
  {Bascones},\ and\ \citenamefont {Guinea}}]{Rodrigo_2004}%
  \BibitemOpen
  \bibfield  {author} {\bibinfo {author} {\bibfnamefont {J.~G.}\ \bibnamefont
  {Rodrigo}}, \bibinfo {author} {\bibfnamefont {H.}~\bibnamefont {Suderow}},
  \bibinfo {author} {\bibfnamefont {S.}~\bibnamefont {Vieira}}, \bibinfo
  {author} {\bibfnamefont {E.}~\bibnamefont {Bascones}}, \ and\ \bibinfo
  {author} {\bibfnamefont {F.}~\bibnamefont {Guinea}},\ }\bibfield  {title}
  {\enquote {\bibinfo {title} {Superconducting nanostructures fabricated with
  the scanning tunnelling microscope},}\ }\href {\doibase
  10.1088/0953-8984/16/34/r01} {\bibfield  {journal} {\bibinfo  {journal}
  {Journal of Physics: Condensed Matter}\ }\textbf {\bibinfo {volume} {16}},\
  \bibinfo {pages} {R1151--R1182} (\bibinfo {year} {2004})}\BibitemShut
  {NoStop}%
\bibitem [{\citenamefont {Herrera}\ \emph {et~al.}(2021)\citenamefont
  {Herrera}, \citenamefont {Barrena}, \citenamefont {Guillam{\'o}n},
  \citenamefont {Galvis}, \citenamefont {Herrera}, \citenamefont {Castilla},
  \citenamefont {Aoki}, \citenamefont {Flouquet},\ and\ \citenamefont
  {Suderow}}]{Herrera2021}%
  \BibitemOpen
  \bibfield  {author} {\bibinfo {author} {\bibfnamefont {E.}~\bibnamefont
  {Herrera}}, \bibinfo {author} {\bibfnamefont {V.}~\bibnamefont {Barrena}},
  \bibinfo {author} {\bibfnamefont {I.}~\bibnamefont {Guillam{\'o}n}}, \bibinfo
  {author} {\bibfnamefont {J.~A.}\ \bibnamefont {Galvis}}, \bibinfo {author}
  {\bibfnamefont {W.~J.}\ \bibnamefont {Herrera}}, \bibinfo {author}
  {\bibfnamefont {J.}~\bibnamefont {Castilla}}, \bibinfo {author}
  {\bibfnamefont {D.}~\bibnamefont {Aoki}}, \bibinfo {author} {\bibfnamefont
  {J.}~\bibnamefont {Flouquet}}, \ and\ \bibinfo {author} {\bibfnamefont
  {H.}~\bibnamefont {Suderow}},\ }\bibfield  {title} {\enquote {\bibinfo
  {title} {{1D} charge density wave in the hidden order state of
  {URu$_2$Si$_2$}},}\ }\href {\doibase 10.1038/s42005-021-00598-0} {\bibfield
  {journal} {\bibinfo  {journal} {Communications Physics}\ }\textbf {\bibinfo
  {volume} {4}},\ \bibinfo {pages} {98} (\bibinfo {year} {2021})}\BibitemShut
  {NoStop}%
\bibitem [{ecc()}]{eccosorb}%
  \BibitemOpen
  \href@noop {} {}\bibinfo {note}
  {Https://www.eccosorb.eu/Eccosorb.html}\BibitemShut {NoStop}%
\bibitem [{Mag()}]{Magnet}%
  \BibitemOpen
  \href
  {https://www.oxinst.com/news/oxford-instruments-commissions-22-tesla-superconducting-magnet-system/}
  {\enquote {\bibinfo {title} {Commissioning note by {Oxford Instruments}.}}\
  }\BibitemShut {NoStop}%
\bibitem [{\citenamefont {Guillam{\'{o}}n}\ \emph
  {et~al.}(2008{\natexlab{a}})\citenamefont {Guillam{\'{o}}n}, \citenamefont
  {Suderow}, \citenamefont {Vieira},\ and\ \citenamefont
  {Rodiere}}]{GUILLAMON2008537}%
  \BibitemOpen
  \bibfield  {author} {\bibinfo {author} {\bibfnamefont {I.}~\bibnamefont
  {Guillam{\'{o}}n}}, \bibinfo {author} {\bibfnamefont {H.}~\bibnamefont
  {Suderow}}, \bibinfo {author} {\bibfnamefont {S.}~\bibnamefont {Vieira}}, \
  and\ \bibinfo {author} {\bibfnamefont {P.}~\bibnamefont {Rodiere}},\
  }\bibfield  {title} {\enquote {\bibinfo {title} {Scanning tunneling
  spectroscopy with superconducting tips of {Al}},}\ }\href {\doibase
  https://doi.org/10.1016/j.physc.2007.11.066} {\bibfield  {journal} {\bibinfo
  {journal} {Physica C: Superconductivity and its Applications}\ }\textbf
  {\bibinfo {volume} {468}},\ \bibinfo {pages} {537--542} (\bibinfo {year}
  {2008}{\natexlab{a}})},\ \bibinfo {note} {proceedings of the Fifth
  International Conference on Vortex Matter in Nanostructured
  Superconductors}\BibitemShut {NoStop}%
\bibitem [{\citenamefont {Schwenk}\ \emph {et~al.}(2020)\citenamefont
  {Schwenk}, \citenamefont {Kim}, \citenamefont {Berwanger}, \citenamefont
  {Ghahari}, \citenamefont {Walkup}, \citenamefont {Slot}, \citenamefont {Le},
  \citenamefont {Cullen}, \citenamefont {Blankenship}, \citenamefont
  {Vranjkovic}, \citenamefont {Hug}, \citenamefont {Kuk}, \citenamefont
  {Giessibl},\ and\ \citenamefont {Stroscio}}]{Schwenk2020}%
  \BibitemOpen
  \bibfield  {author} {\bibinfo {author} {\bibfnamefont {J.}~\bibnamefont
  {Schwenk}}, \bibinfo {author} {\bibfnamefont {S.}~\bibnamefont {Kim}},
  \bibinfo {author} {\bibfnamefont {J.}~\bibnamefont {Berwanger}}, \bibinfo
  {author} {\bibfnamefont {F.}~\bibnamefont {Ghahari}}, \bibinfo {author}
  {\bibfnamefont {D.}~\bibnamefont {Walkup}}, \bibinfo {author} {\bibfnamefont
  {M.~R.}\ \bibnamefont {Slot}}, \bibinfo {author} {\bibfnamefont {S.~T.}\
  \bibnamefont {Le}}, \bibinfo {author} {\bibfnamefont {W.~G.}\ \bibnamefont
  {Cullen}}, \bibinfo {author} {\bibfnamefont {S.~R.}\ \bibnamefont
  {Blankenship}}, \bibinfo {author} {\bibfnamefont {S.}~\bibnamefont
  {Vranjkovic}}, \bibinfo {author} {\bibfnamefont {H.~J.}\ \bibnamefont {Hug}},
  \bibinfo {author} {\bibfnamefont {Y.}~\bibnamefont {Kuk}}, \bibinfo {author}
  {\bibfnamefont {F.~J.}\ \bibnamefont {Giessibl}}, \ and\ \bibinfo {author}
  {\bibfnamefont {J.~A.}\ \bibnamefont {Stroscio}},\ }\bibfield  {title}
  {\enquote {\bibinfo {title} {Achieving $\mu$ev tunneling resolution in an
  in-operando scanning tunneling microscopy, atomic force microscopy, and
  magnetotransport system for quantum materials research},}\ }\href {\doibase
  10.1063/5.0005320} {\bibfield  {journal} {\bibinfo  {journal} {Review of
  Scientific Instruments}\ }\textbf {\bibinfo {volume} {91}},\ \bibinfo {pages}
  {071101} (\bibinfo {year} {2020})}\BibitemShut {NoStop}%
\bibitem [{\citenamefont {Lukashenko}\ and\ \citenamefont
  {Ustinov}(2008)}]{Lukashenko2008}%
  \BibitemOpen
  \bibfield  {author} {\bibinfo {author} {\bibfnamefont {A.}~\bibnamefont
  {Lukashenko}}\ and\ \bibinfo {author} {\bibfnamefont {A.~V.}\ \bibnamefont
  {Ustinov}},\ }\bibfield  {title} {\enquote {\bibinfo {title} {Improved powder
  filters for qubit measurements},}\ }\href {\doibase 10.1063/1.2827515}
  {\bibfield  {journal} {\bibinfo  {journal} {Review of Scientific
  Instruments}\ }\textbf {\bibinfo {volume} {79}},\ \bibinfo {pages} {014701}
  (\bibinfo {year} {2008})}\BibitemShut {NoStop}%
\bibitem [{\citenamefont {Bladh}\ \emph {et~al.}(2003)\citenamefont {Bladh},
  \citenamefont {Gunnarsson}, \citenamefont {H\"{u}rfeld}, \citenamefont
  {Devi}, \citenamefont {Kristoffersson}, \citenamefont {Sm\u{a}lander},
  \citenamefont {Pehrson}, \citenamefont {Claeson}, \citenamefont {Delsing},\
  and\ \citenamefont {Taslakov}}]{Bladh2003}%
  \BibitemOpen
  \bibfield  {author} {\bibinfo {author} {\bibfnamefont {K.}~\bibnamefont
  {Bladh}}, \bibinfo {author} {\bibfnamefont {D.}~\bibnamefont {Gunnarsson}},
  \bibinfo {author} {\bibfnamefont {E.}~\bibnamefont {H\"{u}rfeld}}, \bibinfo
  {author} {\bibfnamefont {S.}~\bibnamefont {Devi}}, \bibinfo {author}
  {\bibfnamefont {C.}~\bibnamefont {Kristoffersson}}, \bibinfo {author}
  {\bibfnamefont {B.}~\bibnamefont {Sm\u{a}lander}}, \bibinfo {author}
  {\bibfnamefont {S.}~\bibnamefont {Pehrson}}, \bibinfo {author} {\bibfnamefont
  {T.}~\bibnamefont {Claeson}}, \bibinfo {author} {\bibfnamefont
  {P.}~\bibnamefont {Delsing}}, \ and\ \bibinfo {author} {\bibfnamefont
  {M.}~\bibnamefont {Taslakov}},\ }\bibfield  {title} {\enquote {\bibinfo
  {title} {Comparison of cryogenic filters for use in single electronics
  experiments},}\ }\href {\doibase 10.1063/1.1540721} {\bibfield  {journal}
  {\bibinfo  {journal} {Review of Scientific Instruments}\ }\textbf {\bibinfo
  {volume} {74}},\ \bibinfo {pages} {1323--1327} (\bibinfo {year}
  {2003})}\BibitemShut {NoStop}%
\bibitem [{\citenamefont {Milliken}\ \emph {et~al.}(2007)\citenamefont
  {Milliken}, \citenamefont {Rozen}, \citenamefont {Keefe},\ and\ \citenamefont
  {Koch}}]{Milliken2007}%
  \BibitemOpen
  \bibfield  {author} {\bibinfo {author} {\bibfnamefont {F.~P.}\ \bibnamefont
  {Milliken}}, \bibinfo {author} {\bibfnamefont {J.~R.}\ \bibnamefont {Rozen}},
  \bibinfo {author} {\bibfnamefont {G.~A.}\ \bibnamefont {Keefe}}, \ and\
  \bibinfo {author} {\bibfnamefont {R.~H.}\ \bibnamefont {Koch}},\ }\bibfield
  {title} {\enquote {\bibinfo {title} {50 {$\Omega$} characteristiv impedance
  low-pass metal powder filters},}\ }\href {\doibase 10.1063/1.2431770}
  {\bibfield  {journal} {\bibinfo  {journal} {Review of Scientific
  Instruments}\ }\textbf {\bibinfo {volume} {78}},\ \bibinfo {pages} {024701}
  (\bibinfo {year} {2007})}\BibitemShut {NoStop}%
\bibitem [{\citenamefont {Thalmann}\ \emph {et~al.}(2017)\citenamefont
  {Thalmann}, \citenamefont {Pernau}, \citenamefont {Strunk}, \citenamefont
  {Scheer},\ and\ \citenamefont {Pietsch}}]{Thalmann2017}%
  \BibitemOpen
  \bibfield  {author} {\bibinfo {author} {\bibfnamefont {M.}~\bibnamefont
  {Thalmann}}, \bibinfo {author} {\bibfnamefont {H.-F.}\ \bibnamefont
  {Pernau}}, \bibinfo {author} {\bibfnamefont {C.}~\bibnamefont {Strunk}},
  \bibinfo {author} {\bibfnamefont {E.}~\bibnamefont {Scheer}}, \ and\ \bibinfo
  {author} {\bibfnamefont {T.}~\bibnamefont {Pietsch}},\ }\bibfield  {title}
  {\enquote {\bibinfo {title} {Comparison of cryogenic low-pass filters},}\
  }\href {\doibase 10.1063/1.4995076} {\bibfield  {journal} {\bibinfo
  {journal} {Review of Scientific Instruments}\ }\textbf {\bibinfo {volume}
  {88}},\ \bibinfo {pages} {114703} (\bibinfo {year} {2017})}\BibitemShut
  {NoStop}%
\bibitem [{\citenamefont {Hunstig}(2017)}]{act6010007}%
  \BibitemOpen
  \bibfield  {author} {\bibinfo {author} {\bibfnamefont {M.}~\bibnamefont
  {Hunstig}},\ }\bibfield  {title} {\enquote {\bibinfo {title} {Piezoelectric
  inertia motors—a critical review of history, concepts, design,
  applications, and perspectives},}\ }\href {\doibase 10.3390/act6010007}
  {\bibfield  {journal} {\bibinfo  {journal} {Actuators}\ }\textbf {\bibinfo
  {volume} {6}} (\bibinfo {year} {2017}),\ 10.3390/act6010007}\BibitemShut
  {NoStop}%
\bibitem [{\citenamefont {Roch}\ \emph {et~al.}(2021)\citenamefont {Roch},
  \citenamefont {Brener}, \citenamefont {Molinari},\ and\ \citenamefont
  {Bouchbinder}}]{roch2021velocitydriven}%
  \BibitemOpen
  \bibfield  {author} {\bibinfo {author} {\bibfnamefont {T.}~\bibnamefont
  {Roch}}, \bibinfo {author} {\bibfnamefont {E.~A.}\ \bibnamefont {Brener}},
  \bibinfo {author} {\bibfnamefont {J.-F.}\ \bibnamefont {Molinari}}, \ and\
  \bibinfo {author} {\bibfnamefont {E.}~\bibnamefont {Bouchbinder}},\
  }\href@noop {} {\enquote {\bibinfo {title} {Velocity-driven frictional
  sliding: Coarsening and steady-state pulse trains},}\ } (\bibinfo {year}
  {2021}),\ \Eprint {http://arxiv.org/abs/2104.13110} {arXiv:2104.13110
  [cond-mat.soft]} \BibitemShut {NoStop}%
\bibitem [{\citenamefont {Liu}\ \emph {et~al.}(2012)\citenamefont {Liu},
  \citenamefont {Zhang}, \citenamefont {Yang}, \citenamefont {Liu},
  \citenamefont {Yang},\ and\ \citenamefont {Zheng}}]{Liu2012}%
  \BibitemOpen
  \bibfield  {author} {\bibinfo {author} {\bibfnamefont {Z.}~\bibnamefont
  {Liu}}, \bibinfo {author} {\bibfnamefont {S.-M.}\ \bibnamefont {Zhang}},
  \bibinfo {author} {\bibfnamefont {J.-R.}\ \bibnamefont {Yang}}, \bibinfo
  {author} {\bibfnamefont {J.~Z.}\ \bibnamefont {Liu}}, \bibinfo {author}
  {\bibfnamefont {Y.-L.}\ \bibnamefont {Yang}}, \ and\ \bibinfo {author}
  {\bibfnamefont {Q.-S.}\ \bibnamefont {Zheng}},\ }\bibfield  {title} {\enquote
  {\bibinfo {title} {Interlayer shear strength of single crystalline
  graphite},}\ }\href {\doibase 10.1007/s10409-012-0137-0} {\bibfield
  {journal} {\bibinfo  {journal} {Acta Mechanica Sinica}\ }\textbf {\bibinfo
  {volume} {28}},\ \bibinfo {pages} {978--982} (\bibinfo {year}
  {2012})}\BibitemShut {NoStop}%
\bibitem [{\citenamefont {Li}\ \emph {et~al.}(2015)\citenamefont {Li},
  \citenamefont {Zhou}, \citenamefont {Zhao}, \citenamefont {Shao},
  \citenamefont {Hou},\ and\ \citenamefont {Xu}}]{Li_2015}%
  \BibitemOpen
  \bibfield  {author} {\bibinfo {author} {\bibfnamefont {J.}~\bibnamefont
  {Li}}, \bibinfo {author} {\bibfnamefont {X.}~\bibnamefont {Zhou}}, \bibinfo
  {author} {\bibfnamefont {H.}~\bibnamefont {Zhao}}, \bibinfo {author}
  {\bibfnamefont {M.}~\bibnamefont {Shao}}, \bibinfo {author} {\bibfnamefont
  {P.}~\bibnamefont {Hou}}, \ and\ \bibinfo {author} {\bibfnamefont
  {X.}~\bibnamefont {Xu}},\ }\bibfield  {title} {\enquote {\bibinfo {title}
  {Design and experimental performances of a piezoelectric linear actuator by
  means of lateral motion},}\ }\href {\doibase 10.1088/0964-1726/24/6/065007}
  {\bibfield  {journal} {\bibinfo  {journal} {Smart Materials and Structures}\
  }\textbf {\bibinfo {volume} {24}},\ \bibinfo {pages} {065007} (\bibinfo
  {year} {2015})}\BibitemShut {NoStop}%
\bibitem [{Hab()}]{Haberli}%
  \BibitemOpen
  \href@noop {} {}\bibinfo {note} {Https://www.haeberli-ag.ch/de/}\BibitemShut
  {NoStop}%
\bibitem [{\citenamefont {Michalczyk}\ and\ \citenamefont
  {Bera}(2019)}]{Michalczyk2019}%
  \BibitemOpen
  \bibfield  {author} {\bibinfo {author} {\bibfnamefont {K.}~\bibnamefont
  {Michalczyk}}\ and\ \bibinfo {author} {\bibfnamefont {P.}~\bibnamefont
  {Bera}},\ }\bibfield  {title} {\enquote {\bibinfo {title} {A simple formula
  for predicting the first natural frequency of transverse vibrations of
  axially loaded helical springs},}\ }\href {\doibase 10.15632/jtam-pl/110243}
  {\bibfield  {journal} {\bibinfo  {journal} {Journal of Theoretical and
  Applied Mechanics}\ }\textbf {\bibinfo {volume} {57}},\ \bibinfo {pages}
  {779--790} (\bibinfo {year} {2019})}\BibitemShut {NoStop}%
\bibitem [{\citenamefont {Azpeitia}\ \emph {et~al.}(2021)\citenamefont
  {Azpeitia}, \citenamefont {Frisenda}, \citenamefont {Lee}, \citenamefont
  {Bouwmeester}, \citenamefont {Zhang}, \citenamefont {Mompean}, \citenamefont
  {van~der Zant}, \citenamefont {García-Hernández},\ and\ \citenamefont
  {Castellanos-Gomez}}]{D1MA00118C}%
  \BibitemOpen
  \bibfield  {author} {\bibinfo {author} {\bibfnamefont {J.}~\bibnamefont
  {Azpeitia}}, \bibinfo {author} {\bibfnamefont {R.}~\bibnamefont {Frisenda}},
  \bibinfo {author} {\bibfnamefont {M.}~\bibnamefont {Lee}}, \bibinfo {author}
  {\bibfnamefont {D.}~\bibnamefont {Bouwmeester}}, \bibinfo {author}
  {\bibfnamefont {W.}~\bibnamefont {Zhang}}, \bibinfo {author} {\bibfnamefont
  {F.}~\bibnamefont {Mompean}}, \bibinfo {author} {\bibfnamefont {H.~S.~J.}\
  \bibnamefont {van~der Zant}}, \bibinfo {author} {\bibfnamefont
  {M.}~\bibnamefont {García-Hernández}}, \ and\ \bibinfo {author}
  {\bibfnamefont {A.}~\bibnamefont {Castellanos-Gomez}},\ }\bibfield  {title}
  {\enquote {\bibinfo {title} {Integrating superconducting {Van der Waals}
  materials on paper substrates},}\ }\href {\doibase 10.1039/D1MA00118C}
  {\bibfield  {journal} {\bibinfo  {journal} {Mater. Adv.}\ }\textbf {\bibinfo
  {volume} {2}},\ \bibinfo {pages} {3274--3281} (\bibinfo {year}
  {2021})}\BibitemShut {NoStop}%
\bibitem [{EBL()}]{EBLTubes}%
  \BibitemOpen
  \href@noop {} {}\bibinfo {note} {Https://eblproducts.com/}\BibitemShut
  {NoStop}%
\bibitem [{\citenamefont {Pan}, \citenamefont {Hudson},\ and\ \citenamefont
  {Davis}(1999)}]{Pan1999}%
  \BibitemOpen
  \bibfield  {author} {\bibinfo {author} {\bibfnamefont {S.~H.}\ \bibnamefont
  {Pan}}, \bibinfo {author} {\bibfnamefont {E.~W.}\ \bibnamefont {Hudson}}, \
  and\ \bibinfo {author} {\bibfnamefont {J.~C.}\ \bibnamefont {Davis}},\
  }\bibfield  {title} {\enquote {\bibinfo {title} {$^3${He} refrigerator based
  very low temperature scanning tunneling microscope},}\ }\href {\doibase
  10.1063/1.1149605} {\bibfield  {journal} {\bibinfo  {journal} {Review of
  Scientific Instruments}\ }\textbf {\bibinfo {volume} {70}},\ \bibinfo {pages}
  {1459--1463} (\bibinfo {year} {1999})}\BibitemShut {NoStop}%
\bibitem [{\citenamefont {Drevniok}\ \emph {et~al.}(2012)\citenamefont
  {Drevniok}, \citenamefont {Paul}, \citenamefont {Hairsine},\ and\
  \citenamefont {McLean}}]{doi:10.1063/1.3694972}%
  \BibitemOpen
  \bibfield  {author} {\bibinfo {author} {\bibfnamefont {B.}~\bibnamefont
  {Drevniok}}, \bibinfo {author} {\bibfnamefont {W.~M.~P.}\ \bibnamefont
  {Paul}}, \bibinfo {author} {\bibfnamefont {K.~R.}\ \bibnamefont {Hairsine}},
  \ and\ \bibinfo {author} {\bibfnamefont {A.~B.}\ \bibnamefont {McLean}},\
  }\bibfield  {title} {\enquote {\bibinfo {title} {Methods and instrumentation
  for piezoelectric motors},}\ }\href {\doibase 10.1063/1.3694972} {\bibfield
  {journal} {\bibinfo  {journal} {Review of Scientific Instruments}\ }\textbf
  {\bibinfo {volume} {83}},\ \bibinfo {pages} {033706} (\bibinfo {year}
  {2012})}\BibitemShut {NoStop}%
\bibitem [{\citenamefont {Pietzsch}\ \emph {et~al.}(2000)\citenamefont
  {Pietzsch}, \citenamefont {Kubetzka}, \citenamefont {Haude}, \citenamefont
  {Bode},\ and\ \citenamefont {Wiesendanger}}]{doi:10.1063/1.1150218}%
  \BibitemOpen
  \bibfield  {author} {\bibinfo {author} {\bibfnamefont {O.}~\bibnamefont
  {Pietzsch}}, \bibinfo {author} {\bibfnamefont {A.}~\bibnamefont {Kubetzka}},
  \bibinfo {author} {\bibfnamefont {D.}~\bibnamefont {Haude}}, \bibinfo
  {author} {\bibfnamefont {M.}~\bibnamefont {Bode}}, \ and\ \bibinfo {author}
  {\bibfnamefont {R.}~\bibnamefont {Wiesendanger}},\ }\bibfield  {title}
  {\enquote {\bibinfo {title} {A low-temperature ultrahigh vacuum scanning
  tunneling microscope with a split-coil magnet and a rotary motion stepper
  motor for high spatial resolution studies of surface magnetism},}\ }\href
  {\doibase 10.1063/1.1150218} {\bibfield  {journal} {\bibinfo  {journal}
  {Review of Scientific Instruments}\ }\textbf {\bibinfo {volume} {71}},\
  \bibinfo {pages} {424--430} (\bibinfo {year} {2000})}\BibitemShut {NoStop}%
\bibitem [{\citenamefont {den Heijer}\ \emph {et~al.}(2014)\citenamefont {den
  Heijer}, \citenamefont {Fokkema}, \citenamefont {Saedi}, \citenamefont
  {Schakel},\ and\ \citenamefont {Rost}}]{doi:10.1063/1.4878624}%
  \BibitemOpen
  \bibfield  {author} {\bibinfo {author} {\bibfnamefont {M.}~\bibnamefont {den
  Heijer}}, \bibinfo {author} {\bibfnamefont {V.}~\bibnamefont {Fokkema}},
  \bibinfo {author} {\bibfnamefont {A.}~\bibnamefont {Saedi}}, \bibinfo
  {author} {\bibfnamefont {P.}~\bibnamefont {Schakel}}, \ and\ \bibinfo
  {author} {\bibfnamefont {M.~J.}\ \bibnamefont {Rost}},\ }\bibfield  {title}
  {\enquote {\bibinfo {title} {Improving the accuracy of walking piezo
  motors},}\ }\href {\doibase 10.1063/1.4878624} {\bibfield  {journal}
  {\bibinfo  {journal} {Review of Scientific Instruments}\ }\textbf {\bibinfo
  {volume} {85}},\ \bibinfo {pages} {055007} (\bibinfo {year}
  {2014})}\BibitemShut {NoStop}%
\bibitem [{\citenamefont {Cherepanov}, \citenamefont {Coenen},\ and\
  \citenamefont {Voigtländer}(2012)}]{doi:10.1063/1.3681444}%
  \BibitemOpen
  \bibfield  {author} {\bibinfo {author} {\bibfnamefont {V.}~\bibnamefont
  {Cherepanov}}, \bibinfo {author} {\bibfnamefont {P.}~\bibnamefont {Coenen}},
  \ and\ \bibinfo {author} {\bibfnamefont {B.}~\bibnamefont {Voigtländer}},\
  }\bibfield  {title} {\enquote {\bibinfo {title} {A nanopositioner for
  scanning probe microscopy: The {KoalaDrive}},}\ }\href {\doibase
  10.1063/1.3681444} {\bibfield  {journal} {\bibinfo  {journal} {Review of
  Scientific Instruments}\ }\textbf {\bibinfo {volume} {83}},\ \bibinfo {pages}
  {023703} (\bibinfo {year} {2012})}\BibitemShut {NoStop}%
\bibitem [{\citenamefont {Wang}\ \emph {et~al.}(2019)\citenamefont {Wang},
  \citenamefont {Huang}, \citenamefont {Sun}, \citenamefont {Hu}, \citenamefont
  {Fu},\ and\ \citenamefont {Lin}}]{doi:10.1063/1.5083994}%
  \BibitemOpen
  \bibfield  {author} {\bibinfo {author} {\bibfnamefont {P.}~\bibnamefont
  {Wang}}, \bibinfo {author} {\bibfnamefont {K.}~\bibnamefont {Huang}},
  \bibinfo {author} {\bibfnamefont {J.}~\bibnamefont {Sun}}, \bibinfo {author}
  {\bibfnamefont {J.}~\bibnamefont {Hu}}, \bibinfo {author} {\bibfnamefont
  {H.}~\bibnamefont {Fu}}, \ and\ \bibinfo {author} {\bibfnamefont
  {X.}~\bibnamefont {Lin}},\ }\bibfield  {title} {\enquote {\bibinfo {title}
  {Piezo-driven sample rotation system with ultra-low electron temperature},}\
  }\href {\doibase 10.1063/1.5083994} {\bibfield  {journal} {\bibinfo
  {journal} {Review of Scientific Instruments}\ }\textbf {\bibinfo {volume}
  {90}},\ \bibinfo {pages} {023905} (\bibinfo {year} {2019})}\BibitemShut
  {NoStop}%
\bibitem [{\citenamefont {Fente}\ \emph
  {et~al.}(2018{\natexlab{a}})\citenamefont {Fente}, \citenamefont {Meier},
  \citenamefont {Kong}, \citenamefont {Kogan}, \citenamefont {Bud'ko},
  \citenamefont {Canfield}, \citenamefont {Guillam\'on},\ and\ \citenamefont
  {Suderow}}]{PhysRevB.97.134501}%
  \BibitemOpen
  \bibfield  {author} {\bibinfo {author} {\bibfnamefont {A.}~\bibnamefont
  {Fente}}, \bibinfo {author} {\bibfnamefont {W.~R.}\ \bibnamefont {Meier}},
  \bibinfo {author} {\bibfnamefont {T.}~\bibnamefont {Kong}}, \bibinfo {author}
  {\bibfnamefont {V.~G.}\ \bibnamefont {Kogan}}, \bibinfo {author}
  {\bibfnamefont {S.~L.}\ \bibnamefont {Bud'ko}}, \bibinfo {author}
  {\bibfnamefont {P.~C.}\ \bibnamefont {Canfield}}, \bibinfo {author}
  {\bibfnamefont {I.}~\bibnamefont {Guillam\'on}}, \ and\ \bibinfo {author}
  {\bibfnamefont {H.}~\bibnamefont {Suderow}},\ }\bibfield  {title} {\enquote
  {\bibinfo {title} {Influence of multiband sign-changing superconductivity on
  vortex cores and vortex pinning in stoichiometric high-{$T_c$}
  {CaKFe$_4$As$_4$}},}\ }\href {\doibase 10.1103/PhysRevB.97.134501} {\bibfield
   {journal} {\bibinfo  {journal} {Phys. Rev. B}\ }\textbf {\bibinfo {volume}
  {97}},\ \bibinfo {pages} {134501} (\bibinfo {year}
  {2018}{\natexlab{a}})}\BibitemShut {NoStop}%
\bibitem [{\citenamefont {Fente}\ \emph
  {et~al.}(2018{\natexlab{b}})\citenamefont {Fente}, \citenamefont
  {Correa-Orellana}, \citenamefont {B\"ohmer}, \citenamefont {Kreyssig},
  \citenamefont {Ran}, \citenamefont {Bud'ko}, \citenamefont {Canfield},
  \citenamefont {Mompean}, \citenamefont {Garc\'{\i}a-Hern\'andez},
  \citenamefont {Munuera}, \citenamefont {Guillam\'on},\ and\ \citenamefont
  {Suderow}}]{PhysRevB.97.014505}%
  \BibitemOpen
  \bibfield  {author} {\bibinfo {author} {\bibfnamefont {A.}~\bibnamefont
  {Fente}}, \bibinfo {author} {\bibfnamefont {A.}~\bibnamefont
  {Correa-Orellana}}, \bibinfo {author} {\bibfnamefont {A.~E.}\ \bibnamefont
  {B\"ohmer}}, \bibinfo {author} {\bibfnamefont {A.}~\bibnamefont {Kreyssig}},
  \bibinfo {author} {\bibfnamefont {S.}~\bibnamefont {Ran}}, \bibinfo {author}
  {\bibfnamefont {S.~L.}\ \bibnamefont {Bud'ko}}, \bibinfo {author}
  {\bibfnamefont {P.~C.}\ \bibnamefont {Canfield}}, \bibinfo {author}
  {\bibfnamefont {F.~J.}\ \bibnamefont {Mompean}}, \bibinfo {author}
  {\bibfnamefont {M.}~\bibnamefont {Garc\'{\i}a-Hern\'andez}}, \bibinfo
  {author} {\bibfnamefont {C.}~\bibnamefont {Munuera}}, \bibinfo {author}
  {\bibfnamefont {I.}~\bibnamefont {Guillam\'on}}, \ and\ \bibinfo {author}
  {\bibfnamefont {H.}~\bibnamefont {Suderow}},\ }\bibfield  {title} {\enquote
  {\bibinfo {title} {Direct visualization of phase separation between
  superconducting and nematic domains in {Co}-doped {CaFe$_2$As$_2$} close to a
  first-order phase transition},}\ }\href {\doibase 10.1103/PhysRevB.97.014505}
  {\bibfield  {journal} {\bibinfo  {journal} {Phys. Rev. B}\ }\textbf {\bibinfo
  {volume} {97}},\ \bibinfo {pages} {014505} (\bibinfo {year}
  {2018}{\natexlab{b}})}\BibitemShut {NoStop}%
\bibitem [{\citenamefont {Dynes}, \citenamefont {Narayanamurti},\ and\
  \citenamefont {Garno}(1978)}]{PhysRevLett.41.1509}%
  \BibitemOpen
  \bibfield  {author} {\bibinfo {author} {\bibfnamefont {R.~C.}\ \bibnamefont
  {Dynes}}, \bibinfo {author} {\bibfnamefont {V.}~\bibnamefont
  {Narayanamurti}}, \ and\ \bibinfo {author} {\bibfnamefont {J.~P.}\
  \bibnamefont {Garno}},\ }\bibfield  {title} {\enquote {\bibinfo {title}
  {Direct measurement of quasiparticle-lifetime broadening in a strong-coupled
  superconductor},}\ }\href {\doibase 10.1103/PhysRevLett.41.1509} {\bibfield
  {journal} {\bibinfo  {journal} {Phys. Rev. Lett.}\ }\textbf {\bibinfo
  {volume} {41}},\ \bibinfo {pages} {1509--1512} (\bibinfo {year}
  {1978})}\BibitemShut {NoStop}%
\bibitem [{\citenamefont {Herman}\ and\ \citenamefont
  {Hlubina}(2016)}]{PhysRevB.94.144508}%
  \BibitemOpen
  \bibfield  {author} {\bibinfo {author} {\bibfnamefont {F.~c.~v.}\
  \bibnamefont {Herman}}\ and\ \bibinfo {author} {\bibfnamefont
  {R.}~\bibnamefont {Hlubina}},\ }\bibfield  {title} {\enquote {\bibinfo
  {title} {Microscopic interpretation of the dynes formula for the tunneling
  density of states},}\ }\href {\doibase 10.1103/PhysRevB.94.144508} {\bibfield
   {journal} {\bibinfo  {journal} {Phys. Rev. B}\ }\textbf {\bibinfo {volume}
  {94}},\ \bibinfo {pages} {144508} (\bibinfo {year} {2016})}\BibitemShut
  {NoStop}%
\bibitem [{\citenamefont {Ruby}\ \emph {et~al.}(2015)\citenamefont {Ruby},
  \citenamefont {Heinrich}, \citenamefont {Pascual},\ and\ \citenamefont
  {Franke}}]{PhysRevLett.114.157001}%
  \BibitemOpen
  \bibfield  {author} {\bibinfo {author} {\bibfnamefont {M.}~\bibnamefont
  {Ruby}}, \bibinfo {author} {\bibfnamefont {B.~W.}\ \bibnamefont {Heinrich}},
  \bibinfo {author} {\bibfnamefont {J.~I.}\ \bibnamefont {Pascual}}, \ and\
  \bibinfo {author} {\bibfnamefont {K.~J.}\ \bibnamefont {Franke}},\ }\bibfield
   {title} {\enquote {\bibinfo {title} {Experimental demonstration of a
  two-band superconducting state for lead using scanning tunneling
  spectroscopy},}\ }\href {\doibase 10.1103/PhysRevLett.114.157001} {\bibfield
  {journal} {\bibinfo  {journal} {Phys. Rev. Lett.}\ }\textbf {\bibinfo
  {volume} {114}},\ \bibinfo {pages} {157001} (\bibinfo {year}
  {2015})}\BibitemShut {NoStop}%
\bibitem [{\citenamefont {Blackford}\ and\ \citenamefont
  {March}(1968)}]{doi:10.1139/p68-021}%
  \BibitemOpen
  \bibfield  {author} {\bibinfo {author} {\bibfnamefont {B.~L.}\ \bibnamefont
  {Blackford}}\ and\ \bibinfo {author} {\bibfnamefont {R.~H.}\ \bibnamefont
  {March}},\ }\bibfield  {title} {\enquote {\bibinfo {title} {Temperature
  dependence of the energy gap in superconducting {Al–Al$_2$O$_3$–Al}
  tunnel junctions},}\ }\href {\doibase 10.1139/p68-021} {\bibfield  {journal}
  {\bibinfo  {journal} {Canadian Journal of Physics}\ }\textbf {\bibinfo
  {volume} {46}},\ \bibinfo {pages} {141--145} (\bibinfo {year}
  {1968})}\BibitemShut {NoStop}%
\bibitem [{\citenamefont {Battisti}\ \emph {et~al.}(2018)\citenamefont
  {Battisti}, \citenamefont {Verdoes}, \citenamefont {van Oosten},
  \citenamefont {Bastiaans},\ and\ \citenamefont {Allan}}]{Battisti2018}%
  \BibitemOpen
  \bibfield  {author} {\bibinfo {author} {\bibfnamefont {I.}~\bibnamefont
  {Battisti}}, \bibinfo {author} {\bibfnamefont {G.}~\bibnamefont {Verdoes}},
  \bibinfo {author} {\bibfnamefont {K.}~\bibnamefont {van Oosten}}, \bibinfo
  {author} {\bibfnamefont {K.~M.}\ \bibnamefont {Bastiaans}}, \ and\ \bibinfo
  {author} {\bibfnamefont {M.~P.}\ \bibnamefont {Allan}},\ }\bibfield  {title}
  {\enquote {\bibinfo {title} {Definition of design guidelines, construction,
  and performance of an ultra-stable scanning tunneling microscope for
  spectroscopic imaging},}\ }\href {\doibase 10.1063/1.5064442} {\bibfield
  {journal} {\bibinfo  {journal} {Review of Scientific Instruments}\ }\textbf
  {\bibinfo {volume} {89}},\ \bibinfo {pages} {123705} (\bibinfo {year}
  {2018})}\BibitemShut {NoStop}%
\bibitem [{\citenamefont {le~Sueur}\ and\ \citenamefont
  {Joyez}(2006)}]{doi:10.1063/1.2400024}%
  \BibitemOpen
  \bibfield  {author} {\bibinfo {author} {\bibfnamefont {H.}~\bibnamefont
  {le~Sueur}}\ and\ \bibinfo {author} {\bibfnamefont {P.}~\bibnamefont
  {Joyez}},\ }\bibfield  {title} {\enquote {\bibinfo {title} {Room-temperature
  tunnel current amplifier and experimental setup for high resolution
  electronic spectroscopy in millikelvin scanning tunneling microscope
  experiments},}\ }\href {\doibase 10.1063/1.2400024} {\bibfield  {journal}
  {\bibinfo  {journal} {Review of Scientific Instruments}\ }\textbf {\bibinfo
  {volume} {77}},\ \bibinfo {pages} {123701} (\bibinfo {year}
  {2006})}\BibitemShut {NoStop}%
\bibitem [{\citenamefont {Moussy}, \citenamefont {Courtois},\ and\
  \citenamefont {Pannetier}(2001)}]{Moussy2001}%
  \BibitemOpen
  \bibfield  {author} {\bibinfo {author} {\bibfnamefont {N.}~\bibnamefont
  {Moussy}}, \bibinfo {author} {\bibfnamefont {H.}~\bibnamefont {Courtois}}, \
  and\ \bibinfo {author} {\bibfnamefont {B.}~\bibnamefont {Pannetier}},\
  }\bibfield  {title} {\enquote {\bibinfo {title} {A very low temperature
  scanning tunneling microscope for the local spectroscopy of mesoscopic
  structures},}\ }\href {\doibase 10.1063/1.1331328} {\bibfield  {journal}
  {\bibinfo  {journal} {Review of Scientific Instruments}\ }\textbf {\bibinfo
  {volume} {72}},\ \bibinfo {pages} {128--131} (\bibinfo {year}
  {2001})}\BibitemShut {NoStop}%
\bibitem [{\citenamefont {Court}, \citenamefont {Ferguson},\ and\ \citenamefont
  {Clark}(2007)}]{Court_2007}%
  \BibitemOpen
  \bibfield  {author} {\bibinfo {author} {\bibfnamefont {N.~A.}\ \bibnamefont
  {Court}}, \bibinfo {author} {\bibfnamefont {A.~J.}\ \bibnamefont {Ferguson}},
  \ and\ \bibinfo {author} {\bibfnamefont {R.~G.}\ \bibnamefont {Clark}},\
  }\bibfield  {title} {\enquote {\bibinfo {title} {Energy gap measurement of
  nanostructured aluminium thin films for single {Cooper}-pair devices},}\
  }\href {\doibase 10.1088/0953-2048/21/01/015013} {\bibfield  {journal}
  {\bibinfo  {journal} {Superconductor Science and Technology}\ }\textbf
  {\bibinfo {volume} {21}},\ \bibinfo {pages} {015013} (\bibinfo {year}
  {2007})}\BibitemShut {NoStop}%
\bibitem [{\citenamefont {Cuevas}, \citenamefont {Mart\'{\i}n-Rodero},\ and\
  \citenamefont {Yeyati}(1996)}]{PhysRevB.54.7366}%
  \BibitemOpen
  \bibfield  {author} {\bibinfo {author} {\bibfnamefont {J.~C.}\ \bibnamefont
  {Cuevas}}, \bibinfo {author} {\bibfnamefont {A.}~\bibnamefont
  {Mart\'{\i}n-Rodero}}, \ and\ \bibinfo {author} {\bibfnamefont {A.~L.}\
  \bibnamefont {Yeyati}},\ }\bibfield  {title} {\enquote {\bibinfo {title}
  {Hamiltonian approach to the transport properties of superconducting quantum
  point contacts},}\ }\href {\doibase 10.1103/PhysRevB.54.7366} {\bibfield
  {journal} {\bibinfo  {journal} {Phys. Rev. B}\ }\textbf {\bibinfo {volume}
  {54}},\ \bibinfo {pages} {7366--7379} (\bibinfo {year} {1996})}\BibitemShut
  {NoStop}%
\bibitem [{\citenamefont {Agrait}, \citenamefont {Yeyati},\ and\ \citenamefont
  {{van Ruitenbeek}}(2003)}]{Agrait2003}%
  \BibitemOpen
  \bibfield  {author} {\bibinfo {author} {\bibfnamefont {N.}~\bibnamefont
  {Agrait}}, \bibinfo {author} {\bibfnamefont {A.~L.}\ \bibnamefont {Yeyati}},
  \ and\ \bibinfo {author} {\bibfnamefont {J.~M.}\ \bibnamefont {{van
  Ruitenbeek}}},\ }\bibfield  {title} {\enquote {\bibinfo {title} {Quantum
  properties of atomic-sized conductors},}\ }\href {\doibase
  https://doi.org/10.1016/S0370-1573(02)00633-6} {\bibfield  {journal}
  {\bibinfo  {journal} {Physics Reports}\ }\textbf {\bibinfo {volume} {377}},\
  \bibinfo {pages} {81--279} (\bibinfo {year} {2003})}\BibitemShut {NoStop}%
\bibitem [{\citenamefont {Ternes}\ \emph {et~al.}(2006)\citenamefont {Ternes},
  \citenamefont {Schneider}, \citenamefont {Cuevas}, \citenamefont {Lutz},
  \citenamefont {Hirjibehedin},\ and\ \citenamefont
  {Heinrich}}]{PhysRevB.74.132501}%
  \BibitemOpen
  \bibfield  {author} {\bibinfo {author} {\bibfnamefont {M.}~\bibnamefont
  {Ternes}}, \bibinfo {author} {\bibfnamefont {W.-D.}\ \bibnamefont
  {Schneider}}, \bibinfo {author} {\bibfnamefont {J.-C.}\ \bibnamefont
  {Cuevas}}, \bibinfo {author} {\bibfnamefont {C.~P.}\ \bibnamefont {Lutz}},
  \bibinfo {author} {\bibfnamefont {C.~F.}\ \bibnamefont {Hirjibehedin}}, \
  and\ \bibinfo {author} {\bibfnamefont {A.~J.}\ \bibnamefont {Heinrich}},\
  }\bibfield  {title} {\enquote {\bibinfo {title} {Subgap structure in
  asymmetric superconducting tunnel junctions},}\ }\href {\doibase
  10.1103/PhysRevB.74.132501} {\bibfield  {journal} {\bibinfo  {journal} {Phys.
  Rev. B}\ }\textbf {\bibinfo {volume} {74}},\ \bibinfo {pages} {132501}
  (\bibinfo {year} {2006})}\BibitemShut {NoStop}%
\bibitem [{\citenamefont {Suderow}\ \emph {et~al.}(2000)\citenamefont
  {Suderow}, \citenamefont {Bascones}, \citenamefont {Belzig}, \citenamefont
  {Guinea},\ and\ \citenamefont {Vieira}}]{Suderow_2000}%
  \BibitemOpen
  \bibfield  {author} {\bibinfo {author} {\bibfnamefont {H.}~\bibnamefont
  {Suderow}}, \bibinfo {author} {\bibfnamefont {E.}~\bibnamefont {Bascones}},
  \bibinfo {author} {\bibfnamefont {W.}~\bibnamefont {Belzig}}, \bibinfo
  {author} {\bibfnamefont {F.}~\bibnamefont {Guinea}}, \ and\ \bibinfo {author}
  {\bibfnamefont {S.}~\bibnamefont {Vieira}},\ }\bibfield  {title} {\enquote
  {\bibinfo {title} {Andreev scattering in nanoscopic junctions in a magnetic
  field},}\ }\href {\doibase 10.1209/epl/i2000-00544-3} {\bibfield  {journal}
  {\bibinfo  {journal} {Europhysics Letters ({EPL})}\ }\textbf {\bibinfo
  {volume} {50}},\ \bibinfo {pages} {749--755} (\bibinfo {year}
  {2000})}\BibitemShut {NoStop}%
\bibitem [{\citenamefont {Scheer}\ \emph
  {et~al.}(1997{\natexlab{a}})\citenamefont {Scheer}, \citenamefont {Joyez},
  \citenamefont {Esteve}, \citenamefont {Urbina},\ and\ \citenamefont
  {Devoret}}]{PhysRevLett.78.3535}%
  \BibitemOpen
  \bibfield  {author} {\bibinfo {author} {\bibfnamefont {E.}~\bibnamefont
  {Scheer}}, \bibinfo {author} {\bibfnamefont {P.}~\bibnamefont {Joyez}},
  \bibinfo {author} {\bibfnamefont {D.}~\bibnamefont {Esteve}}, \bibinfo
  {author} {\bibfnamefont {C.}~\bibnamefont {Urbina}}, \ and\ \bibinfo {author}
  {\bibfnamefont {M.~H.}\ \bibnamefont {Devoret}},\ }\bibfield  {title}
  {\enquote {\bibinfo {title} {Conduction channel transmissions of atomic-size
  aluminum contacts},}\ }\href {\doibase 10.1103/PhysRevLett.78.3535}
  {\bibfield  {journal} {\bibinfo  {journal} {Phys. Rev. Lett.}\ }\textbf
  {\bibinfo {volume} {78}},\ \bibinfo {pages} {3535--3538} (\bibinfo {year}
  {1997}{\natexlab{a}})}\BibitemShut {NoStop}%
\bibitem [{\citenamefont {Huang}\ \emph {et~al.}(2020)\citenamefont {Huang},
  \citenamefont {Padurariu}, \citenamefont {Senkpiel}, \citenamefont {Drost},
  \citenamefont {Yeyati}, \citenamefont {Cuevas}, \citenamefont {Kubala},
  \citenamefont {Ankerhold}, \citenamefont {Kern},\ and\ \citenamefont
  {Ast}}]{Huang2020}%
  \BibitemOpen
  \bibfield  {author} {\bibinfo {author} {\bibfnamefont {H.}~\bibnamefont
  {Huang}}, \bibinfo {author} {\bibfnamefont {C.}~\bibnamefont {Padurariu}},
  \bibinfo {author} {\bibfnamefont {J.}~\bibnamefont {Senkpiel}}, \bibinfo
  {author} {\bibfnamefont {R.}~\bibnamefont {Drost}}, \bibinfo {author}
  {\bibfnamefont {A.~L.}\ \bibnamefont {Yeyati}}, \bibinfo {author}
  {\bibfnamefont {J.~C.}\ \bibnamefont {Cuevas}}, \bibinfo {author}
  {\bibfnamefont {B.}~\bibnamefont {Kubala}}, \bibinfo {author} {\bibfnamefont
  {J.}~\bibnamefont {Ankerhold}}, \bibinfo {author} {\bibfnamefont
  {K.}~\bibnamefont {Kern}}, \ and\ \bibinfo {author} {\bibfnamefont {C.~R.}\
  \bibnamefont {Ast}},\ }\bibfield  {title} {\enquote {\bibinfo {title}
  {Tunnelling dynamics between superconducting bound states at the atomic
  limit},}\ }\href {\doibase 10.1038/s41567-020-0971-0} {\bibfield  {journal}
  {\bibinfo  {journal} {Nature Physics}\ }\textbf {\bibinfo {volume} {16}},\
  \bibinfo {pages} {1227--1231} (\bibinfo {year} {2020})}\BibitemShut {NoStop}%
\bibitem [{\citenamefont {Senkpiel}\ \emph {et~al.}(2020)\citenamefont
  {Senkpiel}, \citenamefont {Dambach}, \citenamefont {Etzkorn}, \citenamefont
  {Drost}, \citenamefont {Padurariu}, \citenamefont {Kubala}, \citenamefont
  {Belzig}, \citenamefont {Yeyati}, \citenamefont {Cuevas}, \citenamefont
  {Ankerhold}, \citenamefont {Ast},\ and\ \citenamefont {Kern}}]{Senkpiel2020}%
  \BibitemOpen
  \bibfield  {author} {\bibinfo {author} {\bibfnamefont {J.}~\bibnamefont
  {Senkpiel}}, \bibinfo {author} {\bibfnamefont {S.}~\bibnamefont {Dambach}},
  \bibinfo {author} {\bibfnamefont {M.}~\bibnamefont {Etzkorn}}, \bibinfo
  {author} {\bibfnamefont {R.}~\bibnamefont {Drost}}, \bibinfo {author}
  {\bibfnamefont {C.}~\bibnamefont {Padurariu}}, \bibinfo {author}
  {\bibfnamefont {B.}~\bibnamefont {Kubala}}, \bibinfo {author} {\bibfnamefont
  {W.}~\bibnamefont {Belzig}}, \bibinfo {author} {\bibfnamefont {A.~L.}\
  \bibnamefont {Yeyati}}, \bibinfo {author} {\bibfnamefont {J.~C.}\
  \bibnamefont {Cuevas}}, \bibinfo {author} {\bibfnamefont {J.}~\bibnamefont
  {Ankerhold}}, \bibinfo {author} {\bibfnamefont {C.~R.}\ \bibnamefont {Ast}},
  \ and\ \bibinfo {author} {\bibfnamefont {K.}~\bibnamefont {Kern}},\
  }\bibfield  {title} {\enquote {\bibinfo {title} {Single channel {Josephson}
  effect in a high transmission atomic contact},}\ }\href {\doibase
  10.1038/s42005-020-00397-z} {\bibfield  {journal} {\bibinfo  {journal}
  {Communications Physics}\ }\textbf {\bibinfo {volume} {3}},\ \bibinfo {pages}
  {131} (\bibinfo {year} {2020})}\BibitemShut {NoStop}%
\bibitem [{\citenamefont {Kot}\ \emph {et~al.}(2020)\citenamefont {Kot},
  \citenamefont {Drost}, \citenamefont {Uhl}, \citenamefont {Ankerhold},
  \citenamefont {Cuevas},\ and\ \citenamefont {Ast}}]{PhysRevB.101.134507}%
  \BibitemOpen
  \bibfield  {author} {\bibinfo {author} {\bibfnamefont {P.}~\bibnamefont
  {Kot}}, \bibinfo {author} {\bibfnamefont {R.}~\bibnamefont {Drost}}, \bibinfo
  {author} {\bibfnamefont {M.}~\bibnamefont {Uhl}}, \bibinfo {author}
  {\bibfnamefont {J.}~\bibnamefont {Ankerhold}}, \bibinfo {author}
  {\bibfnamefont {J.~C.}\ \bibnamefont {Cuevas}}, \ and\ \bibinfo {author}
  {\bibfnamefont {C.~R.}\ \bibnamefont {Ast}},\ }\bibfield  {title} {\enquote
  {\bibinfo {title} {Microwave-assisted tunneling and interference effects in
  superconducting junctions under fast driving signals},}\ }\href {\doibase
  10.1103/PhysRevB.101.134507} {\bibfield  {journal} {\bibinfo  {journal}
  {Phys. Rev. B}\ }\textbf {\bibinfo {volume} {101}},\ \bibinfo {pages}
  {134507} (\bibinfo {year} {2020})}\BibitemShut {NoStop}%
\bibitem [{\citenamefont {J\"ack}\ \emph {et~al.}(2017)\citenamefont {J\"ack},
  \citenamefont {Senkpiel}, \citenamefont {Etzkorn}, \citenamefont {Ankerhold},
  \citenamefont {Ast},\ and\ \citenamefont {Kern}}]{PhysRevLett.119.147702}%
  \BibitemOpen
  \bibfield  {author} {\bibinfo {author} {\bibfnamefont {B.}~\bibnamefont
  {J\"ack}}, \bibinfo {author} {\bibfnamefont {J.}~\bibnamefont {Senkpiel}},
  \bibinfo {author} {\bibfnamefont {M.}~\bibnamefont {Etzkorn}}, \bibinfo
  {author} {\bibfnamefont {J.}~\bibnamefont {Ankerhold}}, \bibinfo {author}
  {\bibfnamefont {C.~R.}\ \bibnamefont {Ast}}, \ and\ \bibinfo {author}
  {\bibfnamefont {K.}~\bibnamefont {Kern}},\ }\bibfield  {title} {\enquote
  {\bibinfo {title} {Quantum brownian motion at strong dissipation probed by
  superconducting tunnel junctions},}\ }\href {\doibase
  10.1103/PhysRevLett.119.147702} {\bibfield  {journal} {\bibinfo  {journal}
  {Phys. Rev. Lett.}\ }\textbf {\bibinfo {volume} {119}},\ \bibinfo {pages}
  {147702} (\bibinfo {year} {2017})}\BibitemShut {NoStop}%
\bibitem [{\citenamefont {Liu}\ \emph {et~al.}(2021)\citenamefont {Liu},
  \citenamefont {Chong}, \citenamefont {Sharma},\ and\ \citenamefont
  {Davis}}]{liu2021discovery}%
  \BibitemOpen
  \bibfield  {author} {\bibinfo {author} {\bibfnamefont {X.}~\bibnamefont
  {Liu}}, \bibinfo {author} {\bibfnamefont {Y.~X.}\ \bibnamefont {Chong}},
  \bibinfo {author} {\bibfnamefont {R.}~\bibnamefont {Sharma}}, \ and\ \bibinfo
  {author} {\bibfnamefont {J.~C.~S.}\ \bibnamefont {Davis}},\ }\href@noop {}
  {\enquote {\bibinfo {title} {Discovery of a {Cooper}-pair density wave state
  in a transition-metal dichalcogenide},}\ } (\bibinfo {year} {2021}),\ \Eprint
  {http://arxiv.org/abs/2007.15228} {arXiv:2007.15228 [cond-mat.supr-con]}
  \BibitemShut {NoStop}%
\bibitem [{\citenamefont {Wang}\ \emph {et~al.}(2018)\citenamefont {Wang},
  \citenamefont {Edkins}, \citenamefont {Hamidian}, \citenamefont {Davis},
  \citenamefont {Fradkin},\ and\ \citenamefont
  {Kivelson}}]{PhysRevB.97.174510}%
  \BibitemOpen
  \bibfield  {author} {\bibinfo {author} {\bibfnamefont {Y.}~\bibnamefont
  {Wang}}, \bibinfo {author} {\bibfnamefont {S.~D.}\ \bibnamefont {Edkins}},
  \bibinfo {author} {\bibfnamefont {M.~H.}\ \bibnamefont {Hamidian}}, \bibinfo
  {author} {\bibfnamefont {J.~C.~S.}\ \bibnamefont {Davis}}, \bibinfo {author}
  {\bibfnamefont {E.}~\bibnamefont {Fradkin}}, \ and\ \bibinfo {author}
  {\bibfnamefont {S.~A.}\ \bibnamefont {Kivelson}},\ }\bibfield  {title}
  {\enquote {\bibinfo {title} {Pair density waves in superconducting vortex
  halos},}\ }\href {\doibase 10.1103/PhysRevB.97.174510} {\bibfield  {journal}
  {\bibinfo  {journal} {Phys. Rev. B}\ }\textbf {\bibinfo {volume} {97}},\
  \bibinfo {pages} {174510} (\bibinfo {year} {2018})}\BibitemShut {NoStop}%
\bibitem [{\citenamefont {Cho}\ \emph {et~al.}(2019)\citenamefont {Cho},
  \citenamefont {Bastiaans}, \citenamefont {Chatzopoulos}, \citenamefont {Gu},\
  and\ \citenamefont {Allan}}]{Cho2019}%
  \BibitemOpen
  \bibfield  {author} {\bibinfo {author} {\bibfnamefont {D.}~\bibnamefont
  {Cho}}, \bibinfo {author} {\bibfnamefont {K.~M.}\ \bibnamefont {Bastiaans}},
  \bibinfo {author} {\bibfnamefont {D.}~\bibnamefont {Chatzopoulos}}, \bibinfo
  {author} {\bibfnamefont {G.~D.}\ \bibnamefont {Gu}}, \ and\ \bibinfo {author}
  {\bibfnamefont {M.~P.}\ \bibnamefont {Allan}},\ }\bibfield  {title} {\enquote
  {\bibinfo {title} {A strongly inhomogeneous superfluid in an iron-based
  superconductor},}\ }\href {\doibase 10.1038/s41586-019-1408-8} {\bibfield
  {journal} {\bibinfo  {journal} {Nature}\ }\textbf {\bibinfo {volume} {571}},\
  \bibinfo {pages} {541--545} (\bibinfo {year} {2019})}\BibitemShut {NoStop}%
\bibitem [{\citenamefont {Hess}, \citenamefont {Robinson},\ and\ \citenamefont
  {Waszczak}(1990)}]{Hess1990}%
  \BibitemOpen
  \bibfield  {author} {\bibinfo {author} {\bibfnamefont {H.~F.}\ \bibnamefont
  {Hess}}, \bibinfo {author} {\bibfnamefont {R.~B.}\ \bibnamefont {Robinson}},
  \ and\ \bibinfo {author} {\bibfnamefont {J.~V.}\ \bibnamefont {Waszczak}},\
  }\bibfield  {title} {\enquote {\bibinfo {title} {Vortex-core structure
  observed with a scanning tunneling microscope},}\ }\href {\doibase
  10.1103/PhysRevLett.64.2711} {\bibfield  {journal} {\bibinfo  {journal}
  {Phys. Rev. Lett.}\ }\textbf {\bibinfo {volume} {64}},\ \bibinfo {pages}
  {2711--2714} (\bibinfo {year} {1990})}\BibitemShut {NoStop}%
\bibitem [{\citenamefont {Hayashi}\ \emph {et~al.}(1998)\citenamefont
  {Hayashi}, \citenamefont {Isoshima}, \citenamefont {Ichioka},\ and\
  \citenamefont {Machida}}]{Hayashi1998}%
  \BibitemOpen
  \bibfield  {author} {\bibinfo {author} {\bibfnamefont {N.}~\bibnamefont
  {Hayashi}}, \bibinfo {author} {\bibfnamefont {T.}~\bibnamefont {Isoshima}},
  \bibinfo {author} {\bibfnamefont {M.}~\bibnamefont {Ichioka}}, \ and\
  \bibinfo {author} {\bibfnamefont {K.}~\bibnamefont {Machida}},\ }\bibfield
  {title} {\enquote {\bibinfo {title} {Low-lying quasiparticle excitations
  around a vortex core in quantum limit},}\ }\href {\doibase
  10.1103/PhysRevLett.80.2921} {\bibfield  {journal} {\bibinfo  {journal}
  {Phys. Rev. Lett.}\ }\textbf {\bibinfo {volume} {80}},\ \bibinfo {pages}
  {2921--2924} (\bibinfo {year} {1998})}\BibitemShut {NoStop}%
\bibitem [{\citenamefont {Fente}\ \emph {et~al.}(2016)\citenamefont {Fente},
  \citenamefont {Herrera}, \citenamefont {Guillam\'on}, \citenamefont
  {Suderow}, \citenamefont {Ma\~nas Valero}, \citenamefont {Galbiati},
  \citenamefont {Coronado},\ and\ \citenamefont {Kogan}}]{PhysRevB.94.014517}%
  \BibitemOpen
  \bibfield  {author} {\bibinfo {author} {\bibfnamefont {A.}~\bibnamefont
  {Fente}}, \bibinfo {author} {\bibfnamefont {E.}~\bibnamefont {Herrera}},
  \bibinfo {author} {\bibfnamefont {I.}~\bibnamefont {Guillam\'on}}, \bibinfo
  {author} {\bibfnamefont {H.}~\bibnamefont {Suderow}}, \bibinfo {author}
  {\bibfnamefont {S.}~\bibnamefont {Ma\~nas Valero}}, \bibinfo {author}
  {\bibfnamefont {M.}~\bibnamefont {Galbiati}}, \bibinfo {author}
  {\bibfnamefont {E.}~\bibnamefont {Coronado}}, \ and\ \bibinfo {author}
  {\bibfnamefont {V.~G.}\ \bibnamefont {Kogan}},\ }\bibfield  {title} {\enquote
  {\bibinfo {title} {Field dependence of the vortex core size probed by
  scanning tunneling microscopy},}\ }\href {\doibase
  10.1103/PhysRevB.94.014517} {\bibfield  {journal} {\bibinfo  {journal} {Phys.
  Rev. B}\ }\textbf {\bibinfo {volume} {94}},\ \bibinfo {pages} {014517}
  (\bibinfo {year} {2016})}\BibitemShut {NoStop}%
\bibitem [{\citenamefont {Guillam{\'{o}}n}\ \emph
  {et~al.}(2008{\natexlab{b}})\citenamefont {Guillam{\'{o}}n}, \citenamefont
  {Suderow}, \citenamefont {Guinea},\ and\ \citenamefont
  {Vieira}}]{Guillamon2008c}%
  \BibitemOpen
  \bibfield  {author} {\bibinfo {author} {\bibfnamefont {I.}~\bibnamefont
  {Guillam{\'{o}}n}}, \bibinfo {author} {\bibfnamefont {H.}~\bibnamefont
  {Suderow}}, \bibinfo {author} {\bibfnamefont {F.}~\bibnamefont {Guinea}}, \
  and\ \bibinfo {author} {\bibfnamefont {S.}~\bibnamefont {Vieira}},\
  }\bibfield  {title} {\enquote {\bibinfo {title} {Intrinsic atomic-scale
  modulations of the superconducting gap of 2{H}-{N}b{S}e$_{2}$},}\ }\href
  {\doibase 10.1103/PhysRevB.77.134505} {\bibfield  {journal} {\bibinfo
  {journal} {Phys. Rev. B}\ }\textbf {\bibinfo {volume} {77}},\ \bibinfo
  {pages} {134505} (\bibinfo {year} {2008}{\natexlab{b}})}\BibitemShut
  {NoStop}%
\bibitem [{\citenamefont {Agra\"{\i}t}, \citenamefont {Rodrigo},\ and\
  \citenamefont {Vieira}(1993)}]{Agrait1993}%
  \BibitemOpen
  \bibfield  {author} {\bibinfo {author} {\bibfnamefont {N.}~\bibnamefont
  {Agra\"{\i}t}}, \bibinfo {author} {\bibfnamefont {J.~G.}\ \bibnamefont
  {Rodrigo}}, \ and\ \bibinfo {author} {\bibfnamefont {S.}~\bibnamefont
  {Vieira}},\ }\bibfield  {title} {\enquote {\bibinfo {title} {Conductance
  steps and quantization in atomic-size contacts},}\ }\href {\doibase
  10.1103/PhysRevB.47.12345} {\bibfield  {journal} {\bibinfo  {journal} {Phys.
  Rev. B}\ }\textbf {\bibinfo {volume} {47}},\ \bibinfo {pages} {12345--12348}
  (\bibinfo {year} {1993})}\BibitemShut {NoStop}%
\bibitem [{\citenamefont {Scheer}\ \emph
  {et~al.}(1997{\natexlab{b}})\citenamefont {Scheer}, \citenamefont {Joyez},
  \citenamefont {Esteve}, \citenamefont {Urbina},\ and\ \citenamefont
  {Devoret}}]{Sheer1997}%
  \BibitemOpen
  \bibfield  {author} {\bibinfo {author} {\bibfnamefont {E.}~\bibnamefont
  {Scheer}}, \bibinfo {author} {\bibfnamefont {P.}~\bibnamefont {Joyez}},
  \bibinfo {author} {\bibfnamefont {D.}~\bibnamefont {Esteve}}, \bibinfo
  {author} {\bibfnamefont {C.}~\bibnamefont {Urbina}}, \ and\ \bibinfo {author}
  {\bibfnamefont {M.~H.}\ \bibnamefont {Devoret}},\ }\bibfield  {title}
  {\enquote {\bibinfo {title} {Conduction channel transmissions of atomic-size
  aluminum contacts},}\ }\href {\doibase 10.1103/PhysRevLett.78.3535}
  {\bibfield  {journal} {\bibinfo  {journal} {Phys. Rev. Lett.}\ }\textbf
  {\bibinfo {volume} {78}},\ \bibinfo {pages} {3535--3538} (\bibinfo {year}
  {1997}{\natexlab{b}})}\BibitemShut {NoStop}%
\bibitem [{\citenamefont {Rubio}, \citenamefont {Agra\"{\i}t},\ and\
  \citenamefont {Vieira}(1996)}]{PhysRevLett.76.2302}%
  \BibitemOpen
  \bibfield  {author} {\bibinfo {author} {\bibfnamefont {G.}~\bibnamefont
  {Rubio}}, \bibinfo {author} {\bibfnamefont {N.}~\bibnamefont {Agra\"{\i}t}},
  \ and\ \bibinfo {author} {\bibfnamefont {S.}~\bibnamefont {Vieira}},\
  }\bibfield  {title} {\enquote {\bibinfo {title} {Atomic-sized metallic
  contacts: Mechanical properties and electronic transport},}\ }\href {\doibase
  10.1103/PhysRevLett.76.2302} {\bibfield  {journal} {\bibinfo  {journal}
  {Phys. Rev. Lett.}\ }\textbf {\bibinfo {volume} {76}},\ \bibinfo {pages}
  {2302--2305} (\bibinfo {year} {1996})}\BibitemShut {NoStop}%
\bibitem [{\citenamefont {Cuevas}\ \emph {et~al.}(1998)\citenamefont {Cuevas},
  \citenamefont {Levy~Yeyati}, \citenamefont {Mart\'{\i}n-Rodero},
  \citenamefont {Rubio~Bollinger}, \citenamefont {Untiedt},\ and\ \citenamefont
  {Agra\"{\i}t}}]{Cuevas1998}%
  \BibitemOpen
  \bibfield  {author} {\bibinfo {author} {\bibfnamefont {J.~C.}\ \bibnamefont
  {Cuevas}}, \bibinfo {author} {\bibfnamefont {A.}~\bibnamefont {Levy~Yeyati}},
  \bibinfo {author} {\bibfnamefont {A.}~\bibnamefont {Mart\'{\i}n-Rodero}},
  \bibinfo {author} {\bibfnamefont {G.}~\bibnamefont {Rubio~Bollinger}},
  \bibinfo {author} {\bibfnamefont {C.}~\bibnamefont {Untiedt}}, \ and\
  \bibinfo {author} {\bibfnamefont {N.}~\bibnamefont {Agra\"{\i}t}},\
  }\bibfield  {title} {\enquote {\bibinfo {title} {Evolution of conducting
  channels in metallic atomic contacts under elastic deformation},}\ }\href
  {\doibase 10.1103/PhysRevLett.81.2990} {\bibfield  {journal} {\bibinfo
  {journal} {Phys. Rev. Lett.}\ }\textbf {\bibinfo {volume} {81}},\ \bibinfo
  {pages} {2990--2993} (\bibinfo {year} {1998})}\BibitemShut {NoStop}%
\bibitem [{\citenamefont {Scheer}\ \emph {et~al.}(1998)\citenamefont {Scheer},
  \citenamefont {Agra{\"i}t}, \citenamefont {Cuevas}, \citenamefont {Yeyati},
  \citenamefont {Ludoph}, \citenamefont {Mart{\'i}n-Rodero}, \citenamefont
  {Bollinger}, \citenamefont {van Ruitenbeek},\ and\ \citenamefont
  {Urbina}}]{Scheer1998}%
  \BibitemOpen
  \bibfield  {author} {\bibinfo {author} {\bibfnamefont {E.}~\bibnamefont
  {Scheer}}, \bibinfo {author} {\bibfnamefont {N.}~\bibnamefont {Agra{\"i}t}},
  \bibinfo {author} {\bibfnamefont {J.~C.}\ \bibnamefont {Cuevas}}, \bibinfo
  {author} {\bibfnamefont {A.~L.}\ \bibnamefont {Yeyati}}, \bibinfo {author}
  {\bibfnamefont {B.}~\bibnamefont {Ludoph}}, \bibinfo {author} {\bibfnamefont
  {A.}~\bibnamefont {Mart{\'i}n-Rodero}}, \bibinfo {author} {\bibfnamefont
  {G.~R.}\ \bibnamefont {Bollinger}}, \bibinfo {author} {\bibfnamefont {J.~M.}\
  \bibnamefont {van Ruitenbeek}}, \ and\ \bibinfo {author} {\bibfnamefont
  {C.}~\bibnamefont {Urbina}},\ }\bibfield  {title} {\enquote {\bibinfo {title}
  {The signature of chemical valence in the electrical conduction through a
  single-atom contact},}\ }\href {\doibase 10.1038/28112} {\bibfield  {journal}
  {\bibinfo  {journal} {Nature}\ }\textbf {\bibinfo {volume} {394}},\ \bibinfo
  {pages} {154--157} (\bibinfo {year} {1998})}\BibitemShut {NoStop}%
\bibitem [{\citenamefont {Cuevas}, \citenamefont {Yeyati},\ and\ \citenamefont
  {Mart\'{\i}n-Rodero}(1998)}]{Cuevas1997}%
  \BibitemOpen
  \bibfield  {author} {\bibinfo {author} {\bibfnamefont {J.~C.}\ \bibnamefont
  {Cuevas}}, \bibinfo {author} {\bibfnamefont {A.~L.}\ \bibnamefont {Yeyati}},
  \ and\ \bibinfo {author} {\bibfnamefont {A.}~\bibnamefont
  {Mart\'{\i}n-Rodero}},\ }\bibfield  {title} {\enquote {\bibinfo {title}
  {Microscopic origin of conducting channels in metallic atomic-size
  contacts},}\ }\href {\doibase 10.1103/PhysRevLett.80.1066} {\bibfield
  {journal} {\bibinfo  {journal} {Phys. Rev. Lett.}\ }\textbf {\bibinfo
  {volume} {80}},\ \bibinfo {pages} {1066--1069} (\bibinfo {year}
  {1998})}\BibitemShut {NoStop}%
\bibitem [{\citenamefont {Yanson}\ \emph {et~al.}(1998)\citenamefont {Yanson},
  \citenamefont {Bollinger}, \citenamefont {van~den Brom}, \citenamefont
  {Agra{\"i}t},\ and\ \citenamefont {van Ruitenbeek}}]{Yanson1998}%
  \BibitemOpen
  \bibfield  {author} {\bibinfo {author} {\bibfnamefont {A.~I.}\ \bibnamefont
  {Yanson}}, \bibinfo {author} {\bibfnamefont {G.~R.}\ \bibnamefont
  {Bollinger}}, \bibinfo {author} {\bibfnamefont {H.~E.}\ \bibnamefont {van~den
  Brom}}, \bibinfo {author} {\bibfnamefont {N.}~\bibnamefont {Agra{\"i}t}}, \
  and\ \bibinfo {author} {\bibfnamefont {J.~M.}\ \bibnamefont {van
  Ruitenbeek}},\ }\bibfield  {title} {\enquote {\bibinfo {title} {Formation and
  manipulation of a metallic wire of single gold atoms},}\ }\href {\doibase
  10.1038/27405} {\bibfield  {journal} {\bibinfo  {journal} {Nature}\ }\textbf
  {\bibinfo {volume} {395}},\ \bibinfo {pages} {783--785} (\bibinfo {year}
  {1998})}\BibitemShut {NoStop}%
\bibitem [{\citenamefont {Kolesnychenko}, \citenamefont {Shklyarevskii},\ and\
  \citenamefont {{van Kempen}}(2000)}]{KOLESNYCHENKO20001257}%
  \BibitemOpen
  \bibfield  {author} {\bibinfo {author} {\bibfnamefont {O.}~\bibnamefont
  {Kolesnychenko}}, \bibinfo {author} {\bibfnamefont {O.}~\bibnamefont
  {Shklyarevskii}}, \ and\ \bibinfo {author} {\bibfnamefont {H.}~\bibnamefont
  {{van Kempen}}},\ }\bibfield  {title} {\enquote {\bibinfo {title} {Anomalous
  increase of the work function in metals due to adsorbed helium},}\ }\href
  {\doibase https://doi.org/10.1016/S0921-4526(99)02519-3} {\bibfield
  {journal} {\bibinfo  {journal} {Physica B: Condensed Matter}\ }\textbf
  {\bibinfo {volume} {284-288}},\ \bibinfo {pages} {1257--1258} (\bibinfo
  {year} {2000})}\BibitemShut {NoStop}%
\bibitem [{\citenamefont {Kolesnychenko}, \citenamefont {Shklyarevskii},\ and\
  \citenamefont {van Kempen}(1999)}]{PhysRevLett.83.2242}%
  \BibitemOpen
  \bibfield  {author} {\bibinfo {author} {\bibfnamefont {O.~Y.}\ \bibnamefont
  {Kolesnychenko}}, \bibinfo {author} {\bibfnamefont {O.~I.}\ \bibnamefont
  {Shklyarevskii}}, \ and\ \bibinfo {author} {\bibfnamefont {H.}~\bibnamefont
  {van Kempen}},\ }\bibfield  {title} {\enquote {\bibinfo {title} {Giant
  influence of adsorbed helium on field emission resonance measurements},}\
  }\href {\doibase 10.1103/PhysRevLett.83.2242} {\bibfield  {journal} {\bibinfo
   {journal} {Phys. Rev. Lett.}\ }\textbf {\bibinfo {volume} {83}},\ \bibinfo
  {pages} {2242--2245} (\bibinfo {year} {1999})}\BibitemShut {NoStop}%
\bibitem [{\citenamefont {Yan}\ \emph {et~al.}(2015)\citenamefont {Yan},
  \citenamefont {Stadtm{\"u}ller}, \citenamefont {Haag}, \citenamefont
  {Jakobs}, \citenamefont {Seidel}, \citenamefont {Jungkenn}, \citenamefont
  {Mathias}, \citenamefont {Cinchetti}, \citenamefont {Aeschlimann},\ and\
  \citenamefont {Felser}}]{Yan2015}%
  \BibitemOpen
  \bibfield  {author} {\bibinfo {author} {\bibfnamefont {B.}~\bibnamefont
  {Yan}}, \bibinfo {author} {\bibfnamefont {B.}~\bibnamefont
  {Stadtm{\"u}ller}}, \bibinfo {author} {\bibfnamefont {N.}~\bibnamefont
  {Haag}}, \bibinfo {author} {\bibfnamefont {S.}~\bibnamefont {Jakobs}},
  \bibinfo {author} {\bibfnamefont {J.}~\bibnamefont {Seidel}}, \bibinfo
  {author} {\bibfnamefont {D.}~\bibnamefont {Jungkenn}}, \bibinfo {author}
  {\bibfnamefont {S.}~\bibnamefont {Mathias}}, \bibinfo {author} {\bibfnamefont
  {M.}~\bibnamefont {Cinchetti}}, \bibinfo {author} {\bibfnamefont
  {M.}~\bibnamefont {Aeschlimann}}, \ and\ \bibinfo {author} {\bibfnamefont
  {C.}~\bibnamefont {Felser}},\ }\bibfield  {title} {\enquote {\bibinfo {title}
  {Topological states on the gold surface},}\ }\href {\doibase
  10.1038/ncomms10167} {\bibfield  {journal} {\bibinfo  {journal} {Nature
  Communications}\ }\textbf {\bibinfo {volume} {6}},\ \bibinfo {pages} {10167}
  (\bibinfo {year} {2015})}\BibitemShut {NoStop}%
\bibitem [{\citenamefont {Fern\'andez-Rossier}\ \emph
  {et~al.}(2005)\citenamefont {Fern\'andez-Rossier}, \citenamefont {Jacob},
  \citenamefont {Untiedt},\ and\ \citenamefont
  {Palacios}}]{PhysRevB.72.224418}%
  \BibitemOpen
  \bibfield  {author} {\bibinfo {author} {\bibfnamefont {J.}~\bibnamefont
  {Fern\'andez-Rossier}}, \bibinfo {author} {\bibfnamefont {D.}~\bibnamefont
  {Jacob}}, \bibinfo {author} {\bibfnamefont {C.}~\bibnamefont {Untiedt}}, \
  and\ \bibinfo {author} {\bibfnamefont {J.~J.}\ \bibnamefont {Palacios}},\
  }\bibfield  {title} {\enquote {\bibinfo {title} {Transport in magnetically
  ordered {Pt} nanocontacts},}\ }\href {\doibase 10.1103/PhysRevB.72.224418}
  {\bibfield  {journal} {\bibinfo  {journal} {Phys. Rev. B}\ }\textbf {\bibinfo
  {volume} {72}},\ \bibinfo {pages} {224418} (\bibinfo {year}
  {2005})}\BibitemShut {NoStop}%
\bibitem [{\citenamefont {Strigl}\ \emph {et~al.}(2015)\citenamefont {Strigl},
  \citenamefont {Espy}, \citenamefont {B{\"u}ckle}, \citenamefont {Scheer},\
  and\ \citenamefont {Pietsch}}]{Strigl2015}%
  \BibitemOpen
  \bibfield  {author} {\bibinfo {author} {\bibfnamefont {F.}~\bibnamefont
  {Strigl}}, \bibinfo {author} {\bibfnamefont {C.}~\bibnamefont {Espy}},
  \bibinfo {author} {\bibfnamefont {M.}~\bibnamefont {B{\"u}ckle}}, \bibinfo
  {author} {\bibfnamefont {E.}~\bibnamefont {Scheer}}, \ and\ \bibinfo {author}
  {\bibfnamefont {T.}~\bibnamefont {Pietsch}},\ }\bibfield  {title} {\enquote
  {\bibinfo {title} {Emerging magnetic order in platinum atomic contacts and
  chains},}\ }\href {\doibase 10.1038/ncomms7172} {\bibfield  {journal}
  {\bibinfo  {journal} {Nature Communications}\ }\textbf {\bibinfo {volume}
  {6}},\ \bibinfo {pages} {6172} (\bibinfo {year} {2015})}\BibitemShut
  {NoStop}%
\bibitem [{\citenamefont {Sokolov}\ \emph {et~al.}(2007)\citenamefont
  {Sokolov}, \citenamefont {Zhang}, \citenamefont {Tsymbal}, \citenamefont
  {Redepenning},\ and\ \citenamefont {Doudin}}]{Sokolov2007}%
  \BibitemOpen
  \bibfield  {author} {\bibinfo {author} {\bibfnamefont {A.}~\bibnamefont
  {Sokolov}}, \bibinfo {author} {\bibfnamefont {C.}~\bibnamefont {Zhang}},
  \bibinfo {author} {\bibfnamefont {E.~Y.}\ \bibnamefont {Tsymbal}}, \bibinfo
  {author} {\bibfnamefont {J.}~\bibnamefont {Redepenning}}, \ and\ \bibinfo
  {author} {\bibfnamefont {B.}~\bibnamefont {Doudin}},\ }\bibfield  {title}
  {\enquote {\bibinfo {title} {Quantized magnetoresistance in atomic-size
  contacts},}\ }\href {\doibase 10.1038/nnano.2007.36} {\bibfield  {journal}
  {\bibinfo  {journal} {Nature Nanotechnology}\ }\textbf {\bibinfo {volume}
  {2}},\ \bibinfo {pages} {171--175} (\bibinfo {year} {2007})}\BibitemShut
  {NoStop}%
\bibitem [{\citenamefont {Calvo}\ \emph {et~al.}(2009)\citenamefont {Calvo},
  \citenamefont {Fern{\'a}ndez-Rossier}, \citenamefont {Palacios},
  \citenamefont {Jacob}, \citenamefont {Natelson},\ and\ \citenamefont
  {Untiedt}}]{Calvo2009}%
  \BibitemOpen
  \bibfield  {author} {\bibinfo {author} {\bibfnamefont {M.~R.}\ \bibnamefont
  {Calvo}}, \bibinfo {author} {\bibfnamefont {J.}~\bibnamefont
  {Fern{\'a}ndez-Rossier}}, \bibinfo {author} {\bibfnamefont {J.~J.}\
  \bibnamefont {Palacios}}, \bibinfo {author} {\bibfnamefont {D.}~\bibnamefont
  {Jacob}}, \bibinfo {author} {\bibfnamefont {D.}~\bibnamefont {Natelson}}, \
  and\ \bibinfo {author} {\bibfnamefont {C.}~\bibnamefont {Untiedt}},\
  }\bibfield  {title} {\enquote {\bibinfo {title} {The {Kondo} effect in
  ferromagnetic atomic contacts},}\ }\href {\doibase 10.1038/nature07878}
  {\bibfield  {journal} {\bibinfo  {journal} {Nature}\ }\textbf {\bibinfo
  {volume} {458}},\ \bibinfo {pages} {1150--1153} (\bibinfo {year}
  {2009})}\BibitemShut {NoStop}%
\bibitem [{\citenamefont {van Ruitenbeek}\ \emph {et~al.}(1996)\citenamefont
  {van Ruitenbeek}, \citenamefont {Alvarez}, \citenamefont {Piñeyro},
  \citenamefont {Grahmann}, \citenamefont {Joyez}, \citenamefont {Devoret},
  \citenamefont {Esteve},\ and\ \citenamefont
  {Urbina}}]{doi:10.1063/1.1146558}%
  \BibitemOpen
  \bibfield  {author} {\bibinfo {author} {\bibfnamefont {J.~M.}\ \bibnamefont
  {van Ruitenbeek}}, \bibinfo {author} {\bibfnamefont {A.}~\bibnamefont
  {Alvarez}}, \bibinfo {author} {\bibfnamefont {I.}~\bibnamefont {Piñeyro}},
  \bibinfo {author} {\bibfnamefont {C.}~\bibnamefont {Grahmann}}, \bibinfo
  {author} {\bibfnamefont {P.}~\bibnamefont {Joyez}}, \bibinfo {author}
  {\bibfnamefont {M.~H.}\ \bibnamefont {Devoret}}, \bibinfo {author}
  {\bibfnamefont {D.}~\bibnamefont {Esteve}}, \ and\ \bibinfo {author}
  {\bibfnamefont {C.}~\bibnamefont {Urbina}},\ }\bibfield  {title} {\enquote
  {\bibinfo {title} {Adjustable nanofabricated atomic size contacts},}\ }\href
  {\doibase 10.1063/1.1146558} {\bibfield  {journal} {\bibinfo  {journal}
  {Review of Scientific Instruments}\ }\textbf {\bibinfo {volume} {67}},\
  \bibinfo {pages} {108--111} (\bibinfo {year} {1996})}\BibitemShut {NoStop}%
\bibitem [{\citenamefont {Guillamon}\ \emph {et~al.}(2008)\citenamefont
  {Guillamon}, \citenamefont {Suderow}, \citenamefont {Guinea},\ and\
  \citenamefont {Vieira}}]{guillamon2008intrinsic}%
  \BibitemOpen
  \bibfield  {author} {\bibinfo {author} {\bibfnamefont {I.}~\bibnamefont
  {Guillamon}}, \bibinfo {author} {\bibfnamefont {H.}~\bibnamefont {Suderow}},
  \bibinfo {author} {\bibfnamefont {F.}~\bibnamefont {Guinea}}, \ and\ \bibinfo
  {author} {\bibfnamefont {S.}~\bibnamefont {Vieira}},\ }\bibfield  {title}
  {\enquote {\bibinfo {title} {Intrinsic atomic-scale modulations of the
  superconducting gap of {2H-NbSe}$_2$},}\ }\href {\doibase
  10.1103/PhysRevB.77.134505} {\bibfield  {journal} {\bibinfo  {journal}
  {Physical Review B}\ }\textbf {\bibinfo {volume} {77}},\ \bibinfo {pages}
  {134505} (\bibinfo {year} {2008})}\BibitemShut {NoStop}%
\bibitem [{\citenamefont {Guillam{\'o}n}\ \emph {et~al.}(2008)\citenamefont
  {Guillam{\'o}n}, \citenamefont {Suderow}, \citenamefont {Vieira},
  \citenamefont {Cario}, \citenamefont {Diener},\ and\ \citenamefont
  {Rodiere}}]{guillamon2008superconducting}%
  \BibitemOpen
  \bibfield  {author} {\bibinfo {author} {\bibfnamefont {I.}~\bibnamefont
  {Guillam{\'o}n}}, \bibinfo {author} {\bibfnamefont {H.}~\bibnamefont
  {Suderow}}, \bibinfo {author} {\bibfnamefont {S.}~\bibnamefont {Vieira}},
  \bibinfo {author} {\bibfnamefont {L.}~\bibnamefont {Cario}}, \bibinfo
  {author} {\bibfnamefont {P.}~\bibnamefont {Diener}}, \ and\ \bibinfo {author}
  {\bibfnamefont {P.}~\bibnamefont {Rodiere}},\ }\bibfield  {title} {\enquote
  {\bibinfo {title} {Superconducting density of states and vortex cores of
  {2H-NbS}$_2$},}\ }\href {\doibase 10.1103/PhysRevLett.101.166407} {\bibfield
  {journal} {\bibinfo  {journal} {Physical Review Letters}\ }\textbf {\bibinfo
  {volume} {101}},\ \bibinfo {pages} {166407} (\bibinfo {year}
  {2008})}\BibitemShut {NoStop}%
\bibitem [{\citenamefont {Soumyanarayanan}\ \emph {et~al.}(2013)\citenamefont
  {Soumyanarayanan}, \citenamefont {Yee}, \citenamefont {He}, \citenamefont
  {Van~Wezel}, \citenamefont {Rahn}, \citenamefont {Rossnagel}, \citenamefont
  {Hudson}, \citenamefont {Norman},\ and\ \citenamefont
  {Hoffman}}]{soumyanarayanan2013quantum}%
  \BibitemOpen
  \bibfield  {author} {\bibinfo {author} {\bibfnamefont {A.}~\bibnamefont
  {Soumyanarayanan}}, \bibinfo {author} {\bibfnamefont {M.~M.}\ \bibnamefont
  {Yee}}, \bibinfo {author} {\bibfnamefont {Y.}~\bibnamefont {He}}, \bibinfo
  {author} {\bibfnamefont {J.}~\bibnamefont {Van~Wezel}}, \bibinfo {author}
  {\bibfnamefont {D.~J.}\ \bibnamefont {Rahn}}, \bibinfo {author}
  {\bibfnamefont {K.}~\bibnamefont {Rossnagel}}, \bibinfo {author}
  {\bibfnamefont {E.}~\bibnamefont {Hudson}}, \bibinfo {author} {\bibfnamefont
  {M.~R.}\ \bibnamefont {Norman}}, \ and\ \bibinfo {author} {\bibfnamefont
  {J.~E.}\ \bibnamefont {Hoffman}},\ }\bibfield  {title} {\enquote {\bibinfo
  {title} {Quantum phase transition from triangular to stripe charge order in
  {NbSe}$_2$},}\ }\href {\doibase 10.1073/pnas.1211387110} {\bibfield
  {journal} {\bibinfo  {journal} {Proceedings of the National Academy of
  Sciences}\ }\textbf {\bibinfo {volume} {110}},\ \bibinfo {pages} {1623--1627}
  (\bibinfo {year} {2013})}\BibitemShut {NoStop}%
\end{thebibliography}

\providecommand{\noopsort}[1]{}\providecommand{\singleletter}[1]{#1}%

\end{document}